\documentclass[a4paper,11pt]{article}

 \usepackage{siunitx}
 \usepackage{url}
 \usepackage{multirow}
 \usepackage{bm}
 \usepackage{hyperref}
 \usepackage{color}
 \usepackage{setspace}
 \usepackage{nicefrac}
\usepackage{lineno}
\usepackage[utf8]{inputenc}
\usepackage{amsthm}
\usepackage{amsfonts}
\usepackage{amssymb}	
\usepackage{amsmath}
\allowdisplaybreaks
\usepackage{mathtools}
\usepackage[english]{babel}
\usepackage{color}
\usepackage{cite}
\usepackage[numbers]{natbib}
\usepackage{slashed}
\usepackage{enumerate}
\usepackage{graphicx}
\usepackage{systeme}
\usepackage{dsfont}
\usepackage{relsize}
\usepackage[margin=0.90in]{geometry}
\usepackage{float}
\usepackage{subcaption}
\usepackage{tikz}
\usepackage[labelformat=simple]{subcaption}

\usepackage{graphicx}
\usepackage{subcaption}
\usepackage{bmpsize}
\usepackage{epstopdf}
\usepackage{bbm}
\usepackage{xcolor,etoolbox}
\usepackage{titling}
\usepackage{authblk}
\newcommand*\samethanks[1][\alph{footnote}]{\footnotemark[#1]}
\usepackage{blindtext,graphicx}
\usepackage[absolute]{textpos}
\usepackage[hang,flushmargin]{footmisc}

\usepackage{xfrac}
\usepackage{nicefrac}
\usepackage{esvect}
\usepackage{cases}
\usepackage{empheq}
\usepackage{stmaryrd}
\usepackage{import} 
\usepackage{transparent}
\usepackage{pifont}
\usepackage{enumitem}

\newcommand{\cmark}{\ding{51}}
\newcommand{\xmark}{\ding{55}}

\newcommand{\mi}{\mathrm{i}} 
\newcommand{\comsol}{\textit{Comsol Multiphysics}\textsuperscript{\textregistered} } 
\newcommand{\matlab}{\textit{Matlab}\textsuperscript{\textregistered} } 
\newcommand{\DivT}{\text{Div}} 
\newcommand{\DivV}{\text{Div}}

\newcommand{\alp}{$\alpha$ }
\newcommand{\bet}{$\beta$ }
\newcommand{\gam}{$\gamma$ }
\newcommand{\del}{$\delta$ }
\newcommand{\conS}{$2\times 3$ }
\newcommand{\conL}{$2\times 4$ }

\title{\vspace{-2cm} Unveiling the key role of Interfaces in the Design of finite-sized Metamaterial Structures}
\author{
\underline{Svenja Hermann}\thanks{Institute of Structural Mechanics, Statics and Dynamics, TU Dortmund University, 44227 Dortmund, Germany}$\;\, ^*$
\, and \, 
Kévin Billon\thanks{Laboratoire de Tribologie et de Dynamique des Systèmes, École Centrale de Lyon, 69134 Ecully, France\\$^*$ Corresponding author, svenja.hermann@tu-dortmund.de} 
\, and \, 
Manuel Collet\samethanks[2] 
\, and \, 
Angela Madeo\samethanks[1]}
\thanksmarkseries{alph}
\date{}

\numberwithin{equation}{section}

\begin{document}
\clearpage          

\maketitle
\setcounter{page}{1}
\vspace*{-20pt}
\begin{abstract}
This paper investigates the influence of interfaces on the performance of finite-sized mechanical metamaterial structures for vibration damping applications. The metamaterial structures are designed in a sandwich configuration in which two homogeneous plates are connected to a metamaterial array. 
We test four different arrays that are obtained from the same metamaterial by differently cutting the metamaterial's unit cell at the metamaterial/plate interface. 
When the four unit cells are periodically repeated in space, they create the same infinitely large metamaterial with an identical mechanical response.
In finite-sized structures, however, the different interfaces between the metamaterial array and the plates~--~called ``material interfaces''~--~and between the metamaterial and the air~--~called ``free interfaces''~--~strongly affect the specimen's vibration transmission characteristics. Using experimental measurements and validated finite-element (FE) models, we demonstrate a significant influence of the different types of interfaces on the global responses and local displacement fields of the structures.
We also demonstrate the presence of a vibroacoustic coupling in the structures which also depends on the type of metamaterial/plate interfaces.
Furthermore, we explore optimization strategies for enhancing the vibration damping performance of the metamaterial structures considering not only the metamaterial array but also the adjacent structures, i.e. the homogeneous plates. A comparison with benchmark cases clearly illustrates the optimization potential that the interfaces' design offers for the vibration damping capability of finite-sized metamaterial structures. We show that optimizing the type of targeted interfaces can shift a given metamaterial's response from underperforming to significantly outperforming compared to classical solutions for noise and vibration damping in civil engineering.

\end{abstract}


\newpage

 \section{Introduction}

	Metamaterials are characterized by an artificially designed microstructure which is the source of the unusual material properties that they show on larger length scales.	
	The microstructure consists of elements, called unit cells, which are periodically repeated in space.
	In mechanics--and particularly in dynamics--metamaterials are used to control the flow of mechanical energy.
	Mechanical metamaterials with a periodic microstructure are also called phononic crystals~\cite{2021_jimenez}.
	Some application examples of phononic crystals are elastic wave guides \cite{2021_laude}, energy harvesting~\cite{2021_hu} or vibration absorbers~\cite{2021_zhou}.
	The underlying unusual property of the metamaterials in these cases is the existence of frequency ranges in which waves cannot freely propagate--the so-called band gaps~\cite{2018_gan}.
	Band gaps are based on two mechanisms: destructive Bragg interference where incident and reflected waves superimpose~\cite{2019_zangeneh} and local resonance where the vibration energy is trapped locally in a part of the structure~\cite{2019_romeroGarcia}.
	Important parameters for the location of the band gaps are the geometry of the unit cell, the base material, the unit cell size and the internal dynamics.

	The exploration of metamaterials in the context of civil engineering is still relatively new. 
	A promising application case is the use of metamaterials for the development of soundproof walls, which is shown by the increasing number of research findings to this topic~\cite{2019_deMelo,2021_riess,2022_kyaw, 2025_chaplain}.
	In residential environments, there is airborne noise originating from acoustic sources (e.g. loud music or airplanes) and structure-borne noise which have a mechanical origin (drilling with a percussion drill or jumping in the building).
	Sound insulation is essential because continuous exposure to noise can lead to serious health problems.
	This is particularly true for low-frequency noise which is difficult to attenuate by common solutions in civil engineering due to the long wave lengths (cf.~\cite{2019_kumar} for example).
	The wavelengths considered in building physics ranges from \SI{100}{\hertz} to \SI{3150}{\hertz}~\cite{1998_fassold}.
	In order to target this frequency range, the unit cell size of metamaterials is generally in the order of centimeters~\cite{2022_holliman}.
 
	The design of metamaterials is based on theoretical methods which are well established for the bulk behavior.
	The design procedure focuses on the unit cell: size, geometry and material properties are modified in parametric studies~\cite{2024_valappil} or in topology optimization studies~\cite{2022_li} such that the band gaps lie in the desired frequency range and/or are as wide as possible.
	To obtain the dispersion curves, Bloch-Floquet boundary conditions are typically applied at the boundaries of a single unit cell. 
	These boundary conditions simulate a metamaterial of infinite size by evaluating the amplitude and phase shift of the waves as they propagate across the single unit cell~\cite{2024_cool}.
	The dispersion curves are obtained as solution of an eigenvalue problem and the band gaps can be easily recognized in the dispersion diagram. 	
	While this approach has become a standard design tool for periodic structures with no or very low damping, there are remaining questions regarding the use of dissipative base materials in metamaterials. The calculation approach becomes non-standard in this case and numerical approaches that allow to compute the dispersion functions for dissipative materials have been presented in~\cite{2011_collet,2018_billon} for example.
In both cases~--~damping taken into account or not~--~the design process is usually completed with the analysis of an infinite metamaterial's properties.

	While the Bloch-Floquet method deals with an infinitely large metamaterial, real engineering applications clearly deal with finite-sized structures.
	To assess the vibration mitigation capacity of these finite-sized structures in practice, a dynamic load is applied to the specimens and the vibration energy propagates across the structure including the metamaterial. 
		For mechanical metamaterials, the load can be applied mechanically (\cite{2016_Dalessandro,2022_demore,2024_hermann}) or in form of an acoustic pressure wave (\cite{2021_gazzola,2022_kyaw}).
	In case of a mechanical load, the vibrations are characterized by the acceleration or the velocity of the structure at the load application position (input side) and positions across the metamaterial (output side).
	In case of an acoustic load, the sound pressure is measured at some distance from the sample.
	The vibration amplitudes on input and output side are compared and band gaps correspond to frequency ranges in which a drastic decrease of the vibrations on the output side is observed.
	Different evaluation metrics are used to quantify the decrease of the vibration amplitude: In mechanics, a drop in the transfer function is observed~\cite{2016_Dalessandro} and in vibroacoustics, the decrease of vibrations causes an increase in the sound transmission loss~\cite{2024_salAnglada}.
	These two metrics are commonly used to evaluate models and experiments.

	In the design of finite sized metamaterials, the influence of the boundary effects occurring at interfaces of a metamaterial with connected structures or interfaces which are unconstrained is usually neglected.
	In general, specimens are designed postulating that the infinite behavior dominates the behavior of the metamaterial so that boundary effects can be neglected. This is true only if the specimens contain many unit cells.
	In mechanical tests, the boundaries of the metamaterial array are most often left unconstrained~\cite{2016_Dalessandro,2022_demore}.	
	In some cases of acoustic tests, correction factors have been applied on results obtained from infinite models to account for the finite size of a specimen~\cite{2021_gazzola}.	
	An exception in which the boundaries are considered is the appearance of so-called edge modes for which vibration energy is transmitted only at the boundaries of metamaterials.
	However, this boundary effect is generally considered as an undesirable side effect which interferes with vibration reduction~\cite{2018_billon,2020_sangiuliano}.
	While the bulk behavior is important for the design of metamaterials, boundary effects must also be taken into account for real application cases since~--~in some cases~--~the influence of boundary effects can be greater than the influence of the bulk effect.
	
Existing studies on boundary effects in metamaterials mainly focus on topological edge modes which are localized vibrations at the materials edges due to specific bulk properties~--~bulk-boundary or bulk-edge correspondence (cf.~\cite{2020_coulais} for example). The energy transport via free boundaries or via the interfaces between metamaterials with different geometric properties is immune to backscattering and energy propagation due to defects~\cite{2019_pal}. It has, therefore, inspired researchers to develop structures that guide waves along specific paths inside a structure which can be useful for energy harvesting or damage detection for example. In finite sizes structures, edge modes can even appear in band gaps of metamaterials. For locally resonant metamaterials, the modes can be controlled by varying the tuning frequency, the number of the resonators or their distribution~\cite{2020_sangiuliano} while in Bragg scattering metamaterials the symmetry of the structure is a determining factor~\cite{2015_hvatov}. Existing studies concentrate on the free boundaries of metamaterials or the interface between different types of metamaterials.
	
	Recent theoretical and numerical works of our group show that constraints at the interfaces are crucial for the vibration damping performance of metamaterials in multi-component structures~\cite{2024_demetriou,2024_hermann,2024_perez}.
	Additional results including experimental work show that constraints at the boundaries in the form of additional structures~\cite{2022_mercer,2024_liang} or imposed conditions~\cite{2020_sangiuliano} significantly modify the behavior of a metamaterial structure.	
	In structures like soundproof walls, metamaterials are coupled with other materials which transfer the vibrations to the metamaterial itself.
	From our perspective it is, therefore, absolutely necessary to include boundaries in the design process of composed metamaterial structures.
	
 	This paper explores the influence of boundary conditions on the performance of finite-sized mechanical metamaterial structures. 
	Using a selected example of a composed metamaterial structure, we 
 \begin{itemize}[noitemsep,topsep=0pt]
     \item prove the influence of material interfaces and free interfaces on the behavior of the targeted metamaterial structure (both, experimentally and in numerical simulations).
     \item show in detail how different types of interfaces influence the wave propagation in the metamaterial. To this aim, we compare the local displacement field for different boundary conditions and assess the resulting global dynamic behavior using transfer functions.    
     \item analyze different factors that influence the behavior of the metamaterial structures: surrounding air, number of unit cells, geometric properties of connected structures.
     \item describe how the design of the interfaces and the aforementioned factors can be used to optimize the vibration damping performance of the metamaterial in the composed structure.
 \end{itemize}


 \section{Materials and Methods}
    \subsection{Specimen Design}
\begin{figure}[!h]
         \centering
         \includegraphics[width=\textwidth]{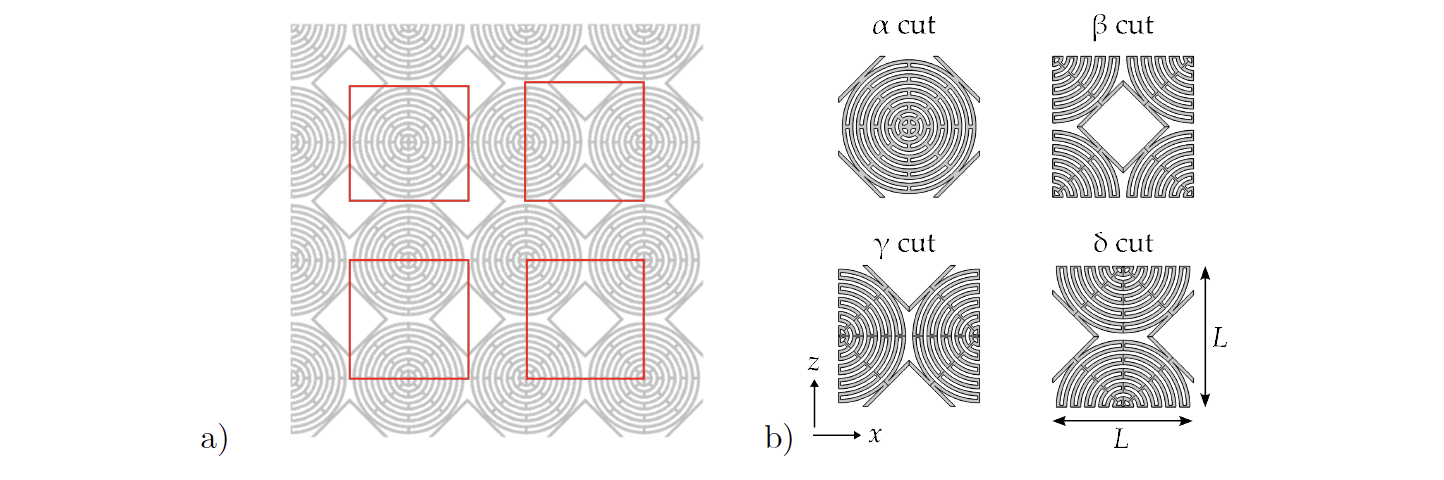}
         \caption{Portion of an infinitely large metamaterial (a) from which four unit cells with similar symmetries are cut (b). $L$~--~unit cell size.}
         \label{fig:unitCellCuts}
     \end{figure}

    This study is based on the labyrinthine metamaterial shown in Figure~\ref{fig:unitCellCuts}a. 
    In a first instance, it is assumed that the metamaterial is infinitely large, which is represented by the dots in the corresponding figure.
    Different types of unit cells can be cut from this metamaterial, as indicated by the red squares in Figure~\ref{fig:unitCellCuts}a, but all of them give rise to the same metamaterial when periodically repeated in space.
    In Figure~\ref{fig:unitCellCuts}b, we show four possible unit cells that are cut from the same labyrinthine metamaterial.
    They are called \alp cut, \bet cut, \gam cut and \del cut in the following. All these cells produce exactly the same dispersion behavior.
    
\begin{figure}[!b]
         \centering
         \includegraphics[trim = 0 0 0 0, clip,width=0.7\textwidth]{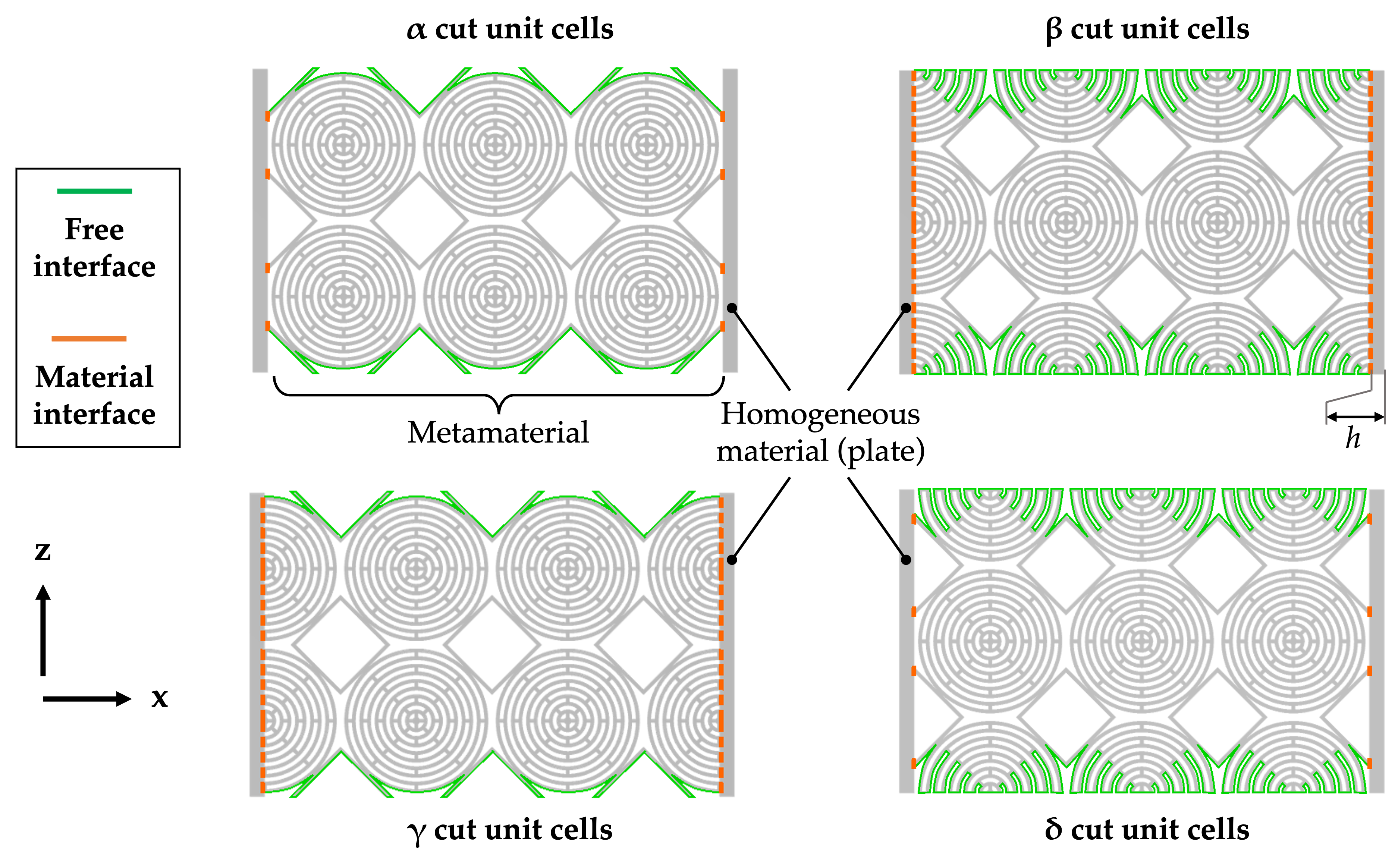}
         \caption{Schematic representation of the sandwich structures comprising \conS unit cells in the four different configurations \alp, \bet, \gam and \del. $h$~--~plate thickness.}
         \label{fig:ints}
     \end{figure}

    The finite-sized specimens of our investigations consist of a metamaterial array which is connected to two homogeneous plates in a sandwich configuration (cf. Figure~\ref{fig:ints}).
    The unit cell size $L$ is \SI{5}{\centi \meter} and the thickness $h$ of the homogeneous plates is \SI{5}{\milli \meter}.      
    In the present study, one specimen with \conS unit cells (as shown in Figure~\ref{fig:ints}) and one specimen with \conL unit cells is analyzed for each of the four cuts shown in Figure~\ref{fig:unitCellCuts}b respectively.
        
\begin{figure}[h!]
         \centering
         \includegraphics[trim = 0 0 0 0, clip,width = 0.9\textwidth]{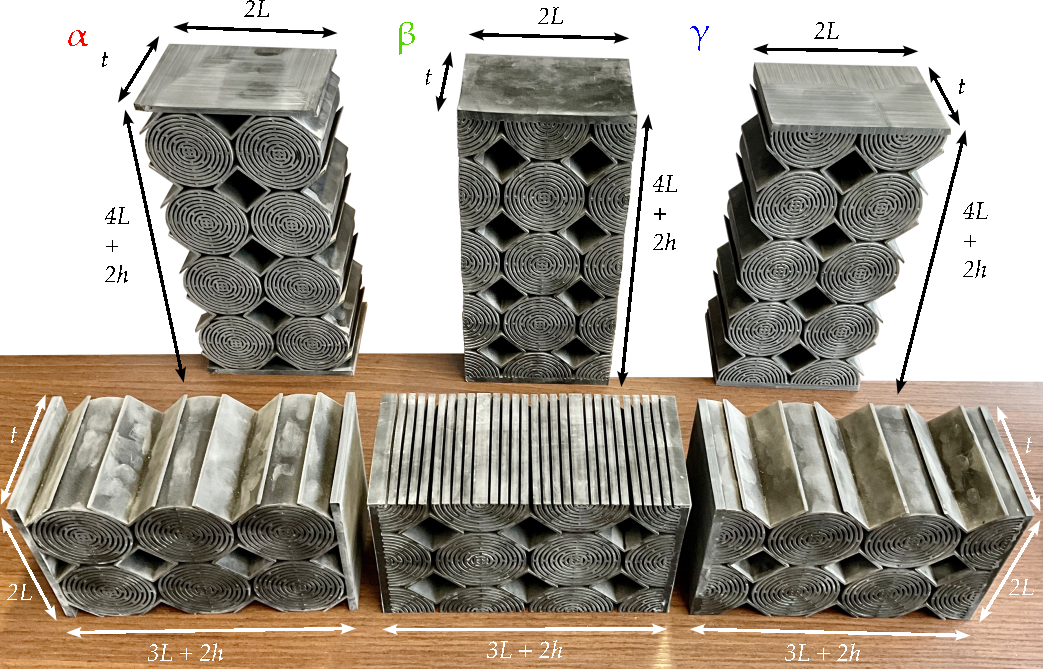}
         \caption{Specimens manufactured for the experiments. Back row from left to right: \conL unit cell specimens with \alp cut, \bet cut and \gam cut metamaterial. Front row from left to right: \conS unit cell specimens with \alp cut, \bet cut and \gam cut metamaterial. $L$~--~unit cell size, $h$~--~plate thickness, $t$~--~specimen thickness.}
         \label{fig:specimens}
     \end{figure}

    As highlighted in Figure~\ref{fig:ints}, the metamaterial has free interfaces at the top and at the bottom (green) and material interfaces where it is connected to the homogeneous material on the left and on the right (orange).
    Extension and topology of the free interfaces and of the material interfaces depend on the unit cell cut and can highly differ between different cuts.
    The configurations \alp, \bet and \gam have been 3D-printed for the experimental testing; a photograph of all printed specimens with \conS and \conL unit cells is shown in Figure~\ref{fig:specimens}.    
The out-of-plane thickness $t$ of the specimen is \SI{7.3}{\centi \meter} so as to optimize the band gap properties~\cite{2024_hermann}.
 
    To manufacture the specimens, the photopolymer¬¬ ``Standard Resin" with black color pigments from the company Anycubic~\cite{2023_anycubic} was used. 
    Being initially liquid, the resin solidifies when exposed to UV light.     
    The metamaterial structures were printed with the Anycubic Photon M3 Premium. 
    This printer works according to the Masked Stereolithography Apparatus (MSLA) principle: a large ultraviolet light source which is masked with an LCD screen is used to generate the different slices. 
    The structure is printed layer by layer; a layer thickness of \SI{0.1}{\milli \meter} was chosen and the exposure time to the light was set to \SI{2.2}{\second}. 
    The printing direction corresponds to the surface normal to the unit cells. 
    Several printing errors are observed on the \bet specimens: the polymer warped during the process and the structure has become wavy on the outer sides. 
    In addition, the specimen was not perfectly straight as shown in Figure~\ref{fig:printingErrors} in the Appendix~\ref{app:A1}. 
    The remaining four specimens show no visible printing errors.
    
 \subsection{Experiments}
    The main goals of the experiments are to validate the numerical models and to show if additional unexpected constraints need to be taken into account in the design process. In the experiments, we assess the specimens' capacity to mitigate vibrations. Therefore, a dynamic load is applied to one side of the structure~--~the ``input'' side~--~and the vibrations on this side of the structure are compared to the vibration on the other~--``output''~--~side of the structure in a specific frequency range. 
    
    \subsubsection{Test setup and measurement procedure}
\begin{figure}[!h]
         \centering
         \includegraphics[width=\textwidth]{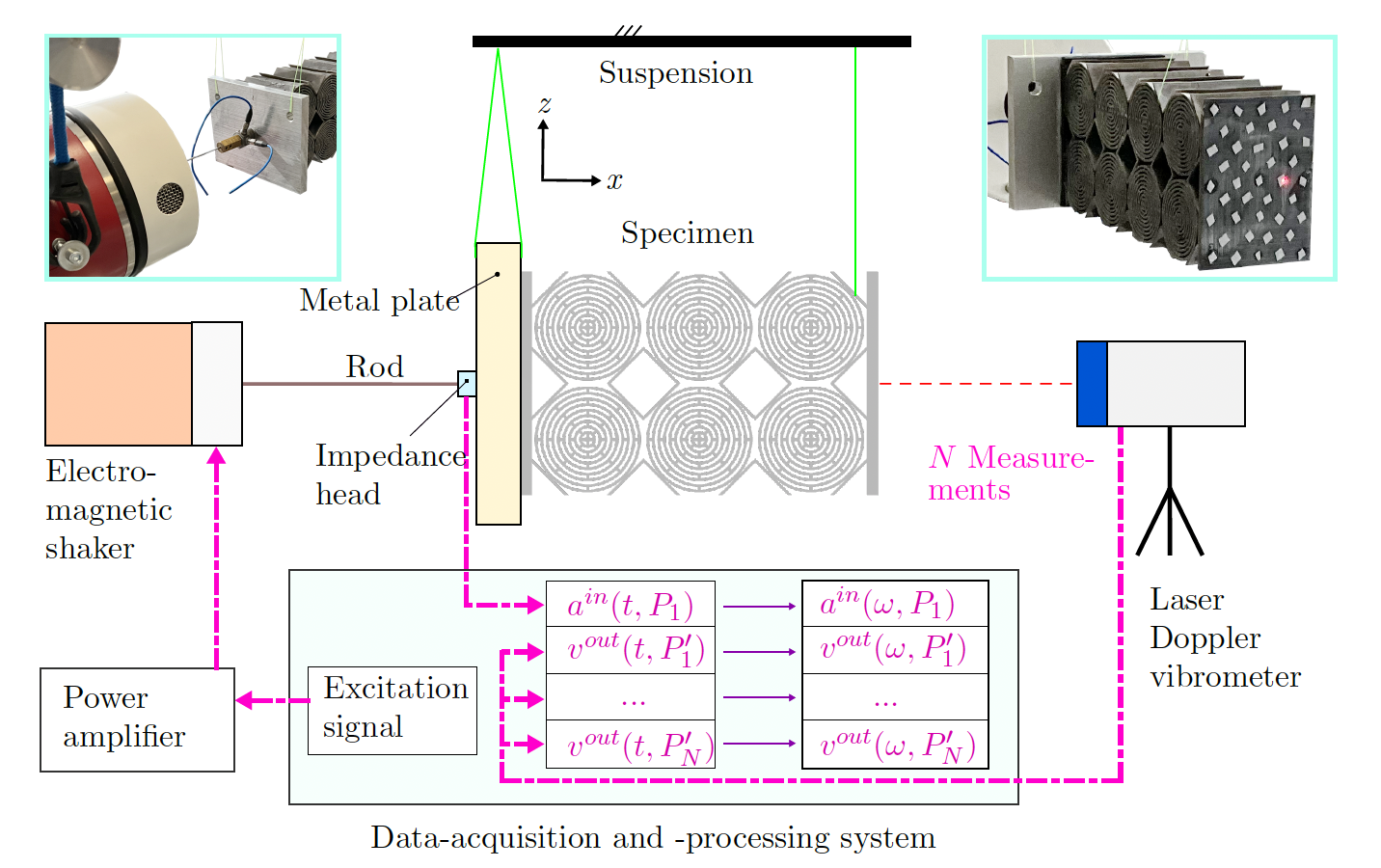}
         \caption{Sketch of the experimental setup. $a(t)$~--~acceleration time signal, $v(t)$~--~velocity time signal, $t$~--~time, $P_1$~--~position of accelerometer on input side, $P'_i$~--~measurement point on output side ($i=1,..,N$), $a(\omega)$~--~acceleration frequency spectrum, $\omega$~--~angular frequency.}
         \label{fig:expScheme}
     \end{figure}

    The experimental setup is shown in Figure~\ref{fig:expScheme}; its individual parts and the measurement procedure will be explained in the following.
    A dynamic mechanical load is applied to the specimen with an electromagnetic shaker. 
    The necessary power supply for the shaker is provided by an amplifier and the excitation signal is provided by a data-acquisition and -processing system from Polytec. 
    The shaker is connected to the specimen via a thin rod which establishes the contact for the load application without significantly constraining the movement of the sample and avoiding the application of a torque. 
    The end of the rod is connected to an impedance head, which is capable of measuring force and acceleration. For the present investigations we only use the acceleration measurements and the measurement direction is perpendicular to the plate's surface.  
    The impedance head is glued on a thick steel plate which is attached to the specimen. Similar steel plates were glued to all of the specimens in advance to make the acceleration of a specimen's input plate as uniform as possible during the tests. Measurements on 	different locations of the metal plate show, that the acceleration is relatively uniform over the nine points on the input plate surface (cf. Appendix~\ref{app:A2}, Figure~\ref{fig:apxAvVsPtw}). 
    The specimen is suspended on a rigid frame by fishing lines, which allows an almost free movement of the structure.
    On the output plate of the specimen, 1D point-wise velocity measurements are performed with a laser Doppler vibrometer from Polytec. The velocity is measured in the direction orthogonal to the plate surface. 
    Due to the matte color of the specimen, reflecting tape snippets are glued on the measurement points to improve the signal received by the vibrometer. 
    41 reflecting tape snippets are glued on each specimen in a similar pattern consisting of consecutive rows (and columns) with 5 and 4 points (cf. Figure~\ref{fig:expScheme}). 
    The laser vibrometer is aligned with the positions on the tape to define the velocity measurement points.
    
    For the excitation, a truncated white noise signal is sent from the data-acquisition and -processing system to the shaker via the amplifier. 
    The signal is filtered by the software such that the frequencies contained in the noise signal are limited to a band ranging from \SI{13.125}{\hertz} to \SI{2}{\kilo \hertz}.
    The sensitivities of the acceleration sensor and the vibrometer are set to \SI{0.00991}{\volt / (\meter / \second}$^2$) and \SI{50}{\milli \meter / \second} respectively. The sample frequency for the measurements is set to \SI{5}{\kilo \hertz} and a Hanning filter is applied. 
    The time signals of the measurements are converted to frequency spectra in the software. 
    The resulting frequency response functions (FRFs) are calculated from the amplitude average over 10 measurements for each measurement point on the specimen. 
    The measurement points on the output plate were scanned one after the other by the laser vibrometer during an excitation sequence.

    \subsubsection{Post-processing} \label{sec:expPostProc}
    From the experimental data-acquisition and -processing system, we obtain the spectral data of the input acceleration $a(\omega,P_1)$ at the position $P_1$ of the impedance head and the output acceleration $a(\omega,P'_i)$ at the positions $P'_i$ of the velocity measurements ($i=1,..,N$ with $N=41$). The output acceleration is obtained from the time derivative of the velocity signal in the frequency domain which is equal to a multiplication with $\mi \omega$.    
    The post processing steps described in the following are performed in \matlab.
    In order to compare the vibrations on the input and output side of the specimen, the transfer function $TF$ is calculated according to
    		\begin{equation}
    			TF(\omega) = \frac{a^{out}(\omega)}{a^{in}(\omega,P_1)} \,,
				\label{eq:avTF}
    		\end{equation}
    		where $a^{in}(\omega,P_1)$ corresponds to the acceleration spectrum at the input measurement point and $a^{out}(\omega)$ correspond to the spatial average of the acceleration on the output side of the specimen respectively.    
    		The spatial average of the output accelerations on the $N=41$ points is obtained from
    		\begin{equation}
    			a^{out}(\omega) = \frac{1}{N} \sum_{i=1}^{N} a^{out}(\omega,P'_i) \, .
    		\end{equation}   
    	The result depends on the number and location of the points over which the average is calculated. 
	We analyzed the influence of a different number of measurement points on the output plate (cf. Appendix~\ref{app:A3}, Figure~\ref{fig:apxNevaluationPoints}) and found that the amplitude of the transfer function converges for $N=41$ measurement points, thus, a good approximation of the average acceleration over the output plate surface is obtained.

 \subsection{Numerical modeling} \label{sec:mAm_sim}

	\subsubsection{Governing equations}
	In the following, the equations that govern the mechanical and vibroacoustic behavior of the specimens and the surrounding air are recalled followed by the description of the vibroacoustic coupling. The modeling is performed in the frequency domain, in which only time-harmonic problems are considered.
	
	The governing equation of the dynamic mechanical behavior in the models stems from linear elasticity:
	\begin{equation}
 		\rho  \omega^2 \bm{u} = \DivT \, \sigma + \bm{f}_v \text{e}^{\mi \phi}  \, ,
 		\label{eq:govEqMech}
	\end{equation} 	
	where $\rho$ corresponds to the mass density of the base polymer material, $\bm{u}$ corresponds to the displacement vector and Div is the divergence operator $\nicefrac{\partial \sigma_{ij}}{\partial x_j} \, \bm{e_i}$ with respect to the base vectors $\bm{e_i}$. 
	The term $\bm{f}_v$e$^{\mi \phi}$ represents a volume force which can show a phase difference $\phi$ with respect to the other terms.
The isotropic stress tensor $\sigma$ is defined by
	\begin{equation}
   		\sigma = \left( \frac{E}{1+\nu}  \left( \text{sym} \left(\nabla \bm{u} \right) 
   		           + \frac{\nu}{(1-2\nu)}  \, \text{tr} \left(\text{sym} \left(\nabla \bm{u} \right) \right) \,  \mathds{1}
   		           \right) \right) \left(1+\mi \eta \right) \, .
		\label{eq:LinElMatLaw}
	\end{equation}
In this definition, $\mathds{1}$ and $\nabla$ correspond to the identity matrix and the nabla operator respectively. 
$\eta$  stands for the isotropic loss factor accounting for the damping by introducing complex terms in the stiffness matrix.
In general, the loss factor introduces a frequency dependence to the problem but we assume here that this dependence can be neglected.
The material properties of the polymer ($E$, $\nu$, $\eta$ and $\rho$) are given in Table~\ref{tab:matProps} at the end of the section.

In the tests we observed that there is an interaction between the air and the specimen. Some of the models, therefore, include the surrounding air as a physical domain -- modeled by an acoustic domain -- and a vibroacoustic coupling between the mechanical and the acoustic domain. The acoustic domain is governed by the Helmholtz equation
	\begin{equation}
    		\DivV \, \left( -\frac{1}{\rho}(\nabla \, p - \bm{q}_d) \right) - \frac{k^2 p}{\rho} = 0 \, .
    		\label{eq:govEqAc}
	\end{equation} 
In this equation, $p$ corresponds to the acoustic pressure, $k= \omega/c$ represents the wavenumber and $\bm{q}_d$ corresponds to a dipole acoustic source term. 

To account for the interaction between the structure and the air in the numerical model, the physical domains are coupled such that the acceleration $a$ of the mechanical domain accelerates the air particles on the one hand and an acoustic pressure variation $p_t$ is applied as a mechanical surface force on the specimen. The coupling is described by the following two equations
\begin{equation}
	\begin{aligned}
    		- \bm{n} \cdot \left( -\frac{1}{\rho_0} \left(\nabla \, p - \bm{q}_d \right) \right) &= - \bm{n} \cdot \bm{a} \,, \\
    		\bm{f}_s &= p_t \,  \bm{n} \,,
	\end{aligned}
    		\label{eq:coupling1}
\end{equation}
	in which $\bm{n}$ is the surface normal of the fluid-solid interface oriented outward the fluid domain, $p_t$ corresponds to the acoustic pressure variation in the air and $\bm{f}_s$ and $\bm{a}$ correspond to a surface force per unit area and the acceleration vector on the solid material's side of the interface.

\subsubsection{Finite-Element (FE) models}
\begin{figure}[!h]
         \centering
         \includegraphics[width=\textwidth]{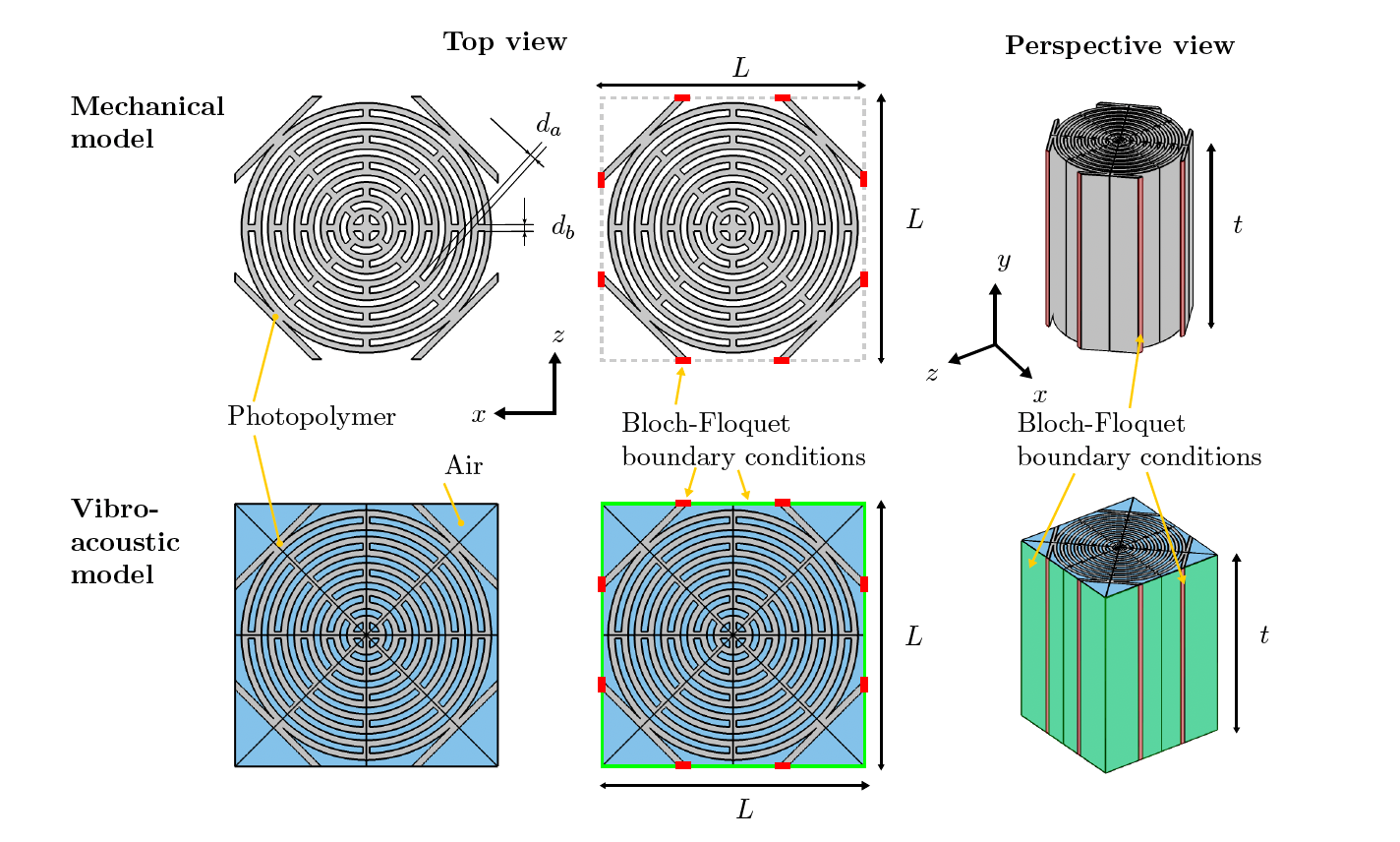}       
         \caption{Models used for Bloch-Floquet analysis: Purely mechanical model (top) and a vibroacoustic model (bottom). $d_a$~--~width of the gap, $d_b$~--~width of the material beams, $L$~--~in-plane unit cell size, $t$~--~out-of-plane unit cell thickness.}
         \label{fig:dispersionDiagramModels}
     \end{figure}

	We start by studying the response of an \textbf{metamaterial with infinitely large extent} to identify its band gap pattern in a dispersion diagram, i.e. we determine the frequency ranges in which waves cannot freely propagate in the metamaterial.
	To this aim, two 3D models of the labyrinthine unit cell shown in Figure~\ref{fig:dispersionDiagramModels} are implemented in \comsol to perform a Bloch-Floquet analysis.
	The results of the study are similar for all four cuts given the infinite extension of the metamaterial (cf. Figure~\ref{fig:unitCellCuts}a). 
	We chose to implement the \alp cut for practical reasons: the boundary conditions are easier to assign to this cut compared to \bet, \gam and \del because the \alp unit cell has the smallest number of outer boundaries. 
	For the unit cell, the thickness of the material in the concentric circles $d_b$ is similar to the thickness of the gaps between the concentric circles $d_a$ and is fixed to $d_a=d_b=$~\SI{1.25}{\milli \meter}.
	In the model, the out-of-plane thickness $t$ of the unit cell is set to \SI{7.3}{\centi \meter} based on an optimization study that aimed to reduce the influence of bending modes on the band gaps~\cite{2024_hermann}.
	A purely mechanical and a vibroacoustic model are implemented; the latter contains an air domain in addition to the unit cell (cf. Figure~\ref{fig:dispersionDiagramModels}). 
	The behavior of the unit cell is governed by Equation~\eqref{eq:govEqMech}.
	The behavior of the air is governed by Equation~\eqref{eq:govEqAc}.

	The vibroacoustic model also includes a coupling between the air and the structure which is described by Equation~\eqref{eq:coupling1}.\\
	To study wave propagation in the infinite metamaterial, in which the unit cell is repeated periodically, Bloch-Floquet boundary conditions are applied on the outer boundaries of the unit cell in the mechanical and acoustic domain (cf. Figure~\ref{fig:dispersionDiagramModels}). 
	The conditions describe the relation between the displacement of corresponding points on two opposite boundaries as a wave propagates across the unit cell:
	\begin{equation}
        		    \bm{u}(\bm{r_L}) = \bm{u}(\bm{r_0}) \, \text{e}^{- \mi \bm{k} (\bm{r}_L-\bm{r}_0)} \; .
        		    \label{eq:BlochFloquet}
			\end{equation}
	In Equation~\eqref{eq:BlochFloquet}, $\bm{k} = (k_x \, k_y \, k_z)$ corresponds to the wave vector and $\bm{r}_L$ and $\bm{r}_0$ correspond to the position vectors of two points on opposing boundaries of the unit cell. Considering the two graphics in the central column of Figure~\ref{fig:dispersionDiagramModels}, $\bm{r}_0$ either points to a point on the boundary on the left or the bottom of the unit cell and $\bm{r}_L$ points to the opposing point either on the right or the top boundary of the unit cell.
	The boundary conditions are applied to surfaces in the $x$-$y$ plane and in the $y$-$z$ plane. 
	Boundaries in the $x$-$z$ plane are not constrained, similar to the experiment.
	An eigenfrequency problem is implemented to obtain the dispersion diagram, i.e. $\omega (k)$.
	
	The dispersion curves are calculated for wave vectors with different components $k_x$ and $k_y = k_z = 0$ which corresponds to the loading in the experiments. 
	Due to the symmetry of wave propagation, it is sufficient to calculate the results in the first Brillouin zone with $k_x \in [0, \; \pi/L]$. 
	In a parametric study, $k_x$ was varied from 0 to $\pi/L$ in 25 steps.
	The eigenfrequencies of the first eleven dispersion curves are obtained with the 3D models.
 	
	To study the dynamic behavior of the \textbf{finite-sized metamaterial structures}, four different types of models are implemented: a purely mechanical model and a vibroacoustic model of the specimens as well as mechanical and vibroacoustic model validation cases. The four model types are described in detail in the following.
	
\begin{figure}[!h]
         \centering
         a) \includegraphics[trim = 0 0 0 0, clip,height = 3.2cm]{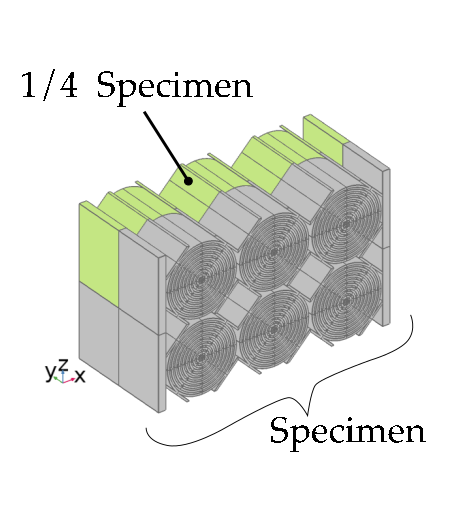}
         b) \includegraphics[trim = 0 0 0 0, clip,height = 3.3cm]{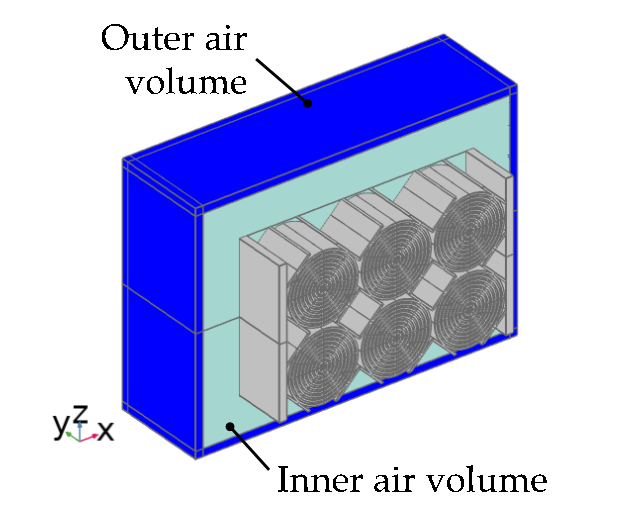}
         c) \includegraphics[trim = 0 -20 0 0, clip,height = 2.8cm]{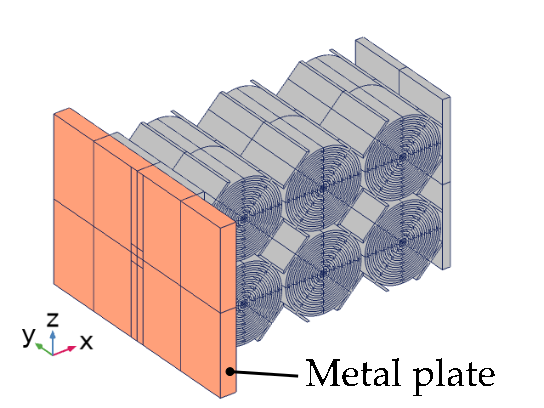}
         d) \includegraphics[trim = 0 0 0 0, clip,height = 3.3cm]{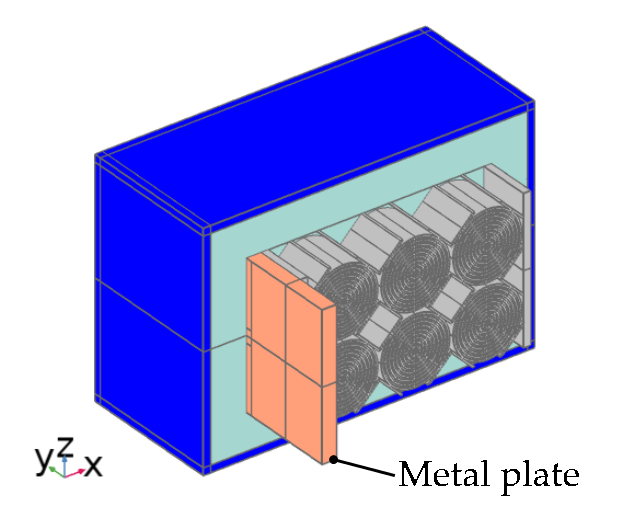}
         \caption{Models of the finite-sized metamaterial specimens for the \alp cut: purely mechanical model (a), vibroacoustic model including two air domains (blue) surrounding the specimen (b) and models including the metal plate (orange) on the input side analogous to the experiment: purely mechanical model (c) and vibroacoustic model (d).}
         \label{fig:modelsFinite}
     \end{figure}

	The \textit{mechanical model} is used to study the purely mechanical dynamic behavior of the specimens. 
	One quarter of the symmetric sample geometry is modeled to save time and computational resources (cf. Figure~\ref{fig:modelsFinite}a).
	The symmetry conditions apply for the displacement at the boundaries where the metamaterial specimen is cut and the normal displacement on these boundaries is set to zero:
	\begin{equation}
		\bm{u} \cdot \bm{n} = 0 \, .
		\label{eq:symMe}
	\end{equation}
	The mechanical behavior of the specimen is governed by Equation~\eqref{eq:govEqMech}.
	A uniform acceleration $\overline{a}^{in}$ is applied in the $x$-direction on the entire surface of the specimen's input plate. The amplitude of acceleration is not a significant parameter in this linear model because we evaluate the ratio of output and input acceleration. For the sake of completeness, we mention here that it is set to \SI{0.3}{\meter / \second}$^2$ which corresponds to the average acceleration amplitude across all frequencies tested in the experiment.

	The \textit{vibroacoustic model} is used to study the coupled vibroacoustic dynamic behavior of the metamaterial specimens. 
	The governing equations for the air domain and the vibroacoustic coupling are stated in Equations~\eqref{eq:govEqAc} and~\eqref{eq:coupling1} respectively.
	The governing equation for the mechanical domain is Equation~\eqref{eq:govEqMech}.
	The air domain surrounding the specimen consists of two parts as shown in Figure~\ref{fig:modelsFinite}b: an inner air volume with an extension of \SI{2.5}{\centi \meter} in the three principal directions around the outer contour of the specimen and an outer volume with an extension of \SI{5}{\milli \meter} in the three principal directions around the inner air volume.
	The outer of the two air domains is defined as a perfectly matched layer (PML) and approximates the large air volume that surrounds the specimen in the laboratory by absorbing radiation of the acoustic waves. Therefore, a layer with a thickness of \SI{5}{\milli \meter} in each spatial direction is added around the air domain. The domains are meshed regularly with 5 elements in the direction of their thickness. A polynomial function is used for the coordinate stretching; the scaling factor and the curvature parameter are set to 1.
	We model a quarter of the entire geometry; therefore, symmetry conditions need to be applied on the surfaces on which the specimen is cut--analogous to the mechanical domain. On these boundaries, the displacement symmetry condition from Equation~\eqref{eq:symMe} applies and the normal velocity of the particles is equal to zero
	\begin{equation}
		\bm{v} \cdot \bm{n} = 0 \, .
		\label{eq:symAc}
	\end{equation}
	In this study, the modeling is carried out exclusively in the frequency domain and problem is solved for the pressure $p$ in the numerical model. The symmetry condition is, therefore, takes the following form:
	\begin{equation}
		- \bm{n} \cdot \left( -\frac{1}{\rho_0} \left(\nabla \, p - \bm{q}_d \right) \right) = 0 \, .
		\label{eq:symAcP}
	\end{equation}
	This boundary condition also applies automatically at the outer limits of the air volume, which is why the perfectly matched layer is required to model the infinite-like air volume around the specimen.
	Two load cases are implemented for the vibroacoustic model: Either, a uniform acceleration $\overline{a}^{in}$~--~similar to the load case in the mechanical model~--~or a uniform pressure loading $\overline{p}^{in}$ is applied on the entire surface of the input plate. In the second case, the pressure amplitude is set to $p=$~\SI{137}{\newton / \meter^2} which corresponds to a total loading force of $F^{in}=$~\SI{1}{\newton}. As explained in the previous paragraph, the loading amplitude is not a significant parameter in the present study and is given for the sake of completeness.	

	The third and fourth type of model of the metamaterial specimens are used for \textit{model validation}.
	These models should replicate the test conditions of the experiment as accurately as possible.
	Therefore, the geometries include the metal plate on the input side  (cf. Figure~\ref{fig:expScheme}). 
	The metal plate is centered on the input plate of the specimen and measures \SI{1}{\centi \meter}~$\times$~\SI{12}{\centi \meter}~$\times$~$2L$ ($x \times y \times z$).
	One of the models is a purely mechanical model~--similar to model type one~--~and the other one a vibroacoustic model~--similar to model type two (cf. Figure~\ref{fig:modelsFinite}~c and~d respectively). Depending on the type of physical domains involved, the corresponding governing equations for the mechanical behavior and the acoustic behavior and vibroacoustic coupling (Equations~\eqref{eq:govEqMech} - \eqref{eq:coupling1}) and symmetry conditions (Equations~\eqref{eq:symMe} and~\eqref{eq:symAcP}) are applied.
	
\newpage
	For the loading, the input acceleration $a^{in}_{S_1}$ obtained from the experimental measurement is applied in a square surface $S_1$ which measures \SI{1}{\centi \meter}~$\times$~\SI{1}{\centi \meter} and which is located in the center of the metal plate's surface. The size and location of $S_1$ correspond to the experimental conditions; the impedance head is glued in the center of the plate surface and the contact area corresponds to \SI{1}{\centi \meter}$^2$. For this model type, we do not only evaluate the ratio of output and input, but also the magnitude of the accelerations. Therefore, we use a loading amplitude similar to the experiment in order to be able to obtain comparable results concerning the acceleration magnitudes.
     
	For all four model types, the specimen geometries are meshed with quadratic Lagrange elements that have a maximum size of \SI{2}{\milli \meter} according to the following strategy: at first, the metamaterial, the homogeneous plates and the surrounding air between the homogeneous plates were meshed with a triangular mesh on the cut surface in the $x$-$z$ plane. The mesh was then swept in $y$-direction. A regular mesh was used for the perfectly matched layer and the steel plate. The remaining air was meshed with a tetrahedral mesh.
	Frequency domain studies have been performed for all four model types. In these studies, the boundary value problems are solved in a frequency range between \SI{20}{\hertz} and \SI{2000}{\hertz} for every $\Delta f=$~\SI{20}{\hertz}.
	In addition, eigenfrequency studies have been performed for both of the models that include the steel plate.

\begin{figure}[b!]
          \centering
          \includegraphics[trim = 0 0 0 0, clip,width=0.75\textwidth]{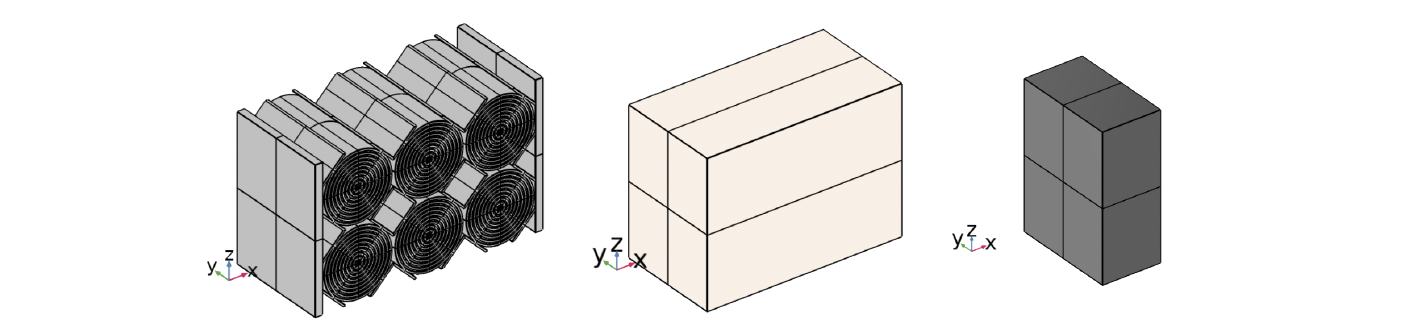}
          \caption{Finite-sized metamaterial structure (left) with \conS unit cells in \alp configuration and corresponding benchmark cases consisting of a block of gypsum (middle) and a block of concrete (right) having the same mass and dimensions in $y$ and $z$-direction as the metamaterial specimen.}
          \label{fig:benchmarks}
      \end{figure}

        Two \textbf{benchmark cases} are implemented to compare the vibroacoustic behavior of the metamaterial specimens to building elements made of commonly used materials in civil engineering. 
        The benchmark cases consist of blocks of gypsum and concrete.
        We use full material blocks as comparison solution because we are interested in solutions that are capable of bearing mechanical loads. 
        As the sound absorption of the solutions is mainly determined by the mass, the benchmark cases are designed such that they have the same mass as the specimens with \conS unit cells (\SI{676}{\gram}) and \conL unit cells (\SI{871.6}{\gram}) respectively. 
         The geometric dimensions of the benchmark cases are determined as follows: in the $y$ and $z$-directions, the blocks have the same dimension as the specimens--$t$ and $2a$ respectively.
         The corresponding thickness in the $x$-direction was obtained from the mass density and the volume of the blocks.
         For the specimen with \conS unit cells, the thickness of the gypsum specimen is \SI{13.62}{\centi \meter} and the thickness of the concrete specimen is \SI{4.07}{\centi \meter}. 
         In Figure~\ref{fig:benchmarks}b and~c, the sizes of the different structures are illustrated exemplarily for the small specimen size.         
         For the specimen with \conL unit cells, the thickness of the gypsum specimen is \SI{17.56}{\centi \meter} and the thickness of the concrete specimen is \SI{5.25}{\centi \meter}. 
         The models are purely mechanical, therefore, Equation~\eqref{eq:govEqMech} applies as governing equation.
         The material properties are summarized in Table~\ref{tab:matProps}.        

         A uniform pressure loading is applied on the input side of the benchmark structures. 
         Analogous to the metamaterial models, a total loading force of $F^{in}=$~\SI{1}{\newton} is applied in $x$-direction; the total force is distributed uniformly over the input surface.
         The maximum mesh size of the benchmark cases is set to \SI{1}{\centi \meter} and contains, thus, larger elements than the metamaterial models. 
         The geometry is much less detailed and we do not consider an interaction with the air in these models and the mesh can, therefore, be coarser.
         Frequency domain studies have been performed for the benchmark cases. In these studies, the boundary value problems are solved in the same frequency range as the metamaterial studies.       
 
     \begin{table}[!h]
         \centering
         \caption{Material properties used in numerical simulations.}
         \label{tab:matProps}
         \begin{tabular}{lccccc}
             \hline
              Material       & $E$ [GPa] & $\nu$  [-] & $\eta$ [-] & $\rho$ [kg/m$^3$]& $c$ [m/s]\\
             \hline
              Photopolymer   & 1.875 & 0.4   & 0.13  & 1220  & -\\
              Steel          & 200   & 0.3   & -     & 7850  & -\\
              Air            & -     & -     & -     & 1.204 & 343\\
              Gypsum			& 2.8	& 0.3	& - 		 & 680  & - \\
              Concrete		& 31.6	& 0.2	& 0.01	 & 2275 & - \\
             \hline
         \end{tabular}
     \end{table}

\subsubsection{Post-processing} 

The information we extract from the FE models for post-processing are acceleration responses and the load input. An overview of the evaluation metrics for the different FE model types is given  in Table~\ref{tab:FEmodelProps}.
For the comparison with the experiment, we extract the acceleration response on measurement points in the FE models which are similar to the experiment. Since we model 1/4 of the geometry, we can define 13 point positions which are directly similar to the experiment. 
As we assume a symmetric response of the specimen, the transfer function can be calculated from the response on these 13 points similarly to the procedures described in experimental post-processing section (Section~\ref{sec:expPostProc}). 

    \begin{table}[!h]
         \centering
         \caption{Overview over the properties of the finite-sized FE models. BC~--~boundary conditions.}
         \label{tab:FEmodelProps}
         \renewcommand{\arraystretch}{1.5}
         \begin{tabular}{|p{2cm} | p{2.1cm} |p{2.5cm} |p{2.3cm} |p{2cm} |p{2cm} |}
             \hline
              Model type & \textbf{Mechanical} & \textbf{Vibroacoustic} & \multicolumn{1}{|c|}{\textbf{Vibroac. + plate}}& \multicolumn{1}{|c|}{\textbf{Mech. + plate}} & \textbf{Benchmark}\\
             \hline
             Geometry &  \multicolumn{2}{|c|}{1/4} &  \multicolumn{2}{|c|}{1/4 or 1/2} &  \multicolumn{1}{|c|}{1/4} \\
             \hline
             Mechanical domain & \multicolumn{5}{|c|}{\cmark} \\
             \hline
             Mechanical BC & \multicolumn{5}{|c|}{Harmonic mechanical loading, symmetry} \\
             \hline
             Acoustic \mbox{domain} & \multicolumn{1}{|c|}{\xmark} & \multicolumn{2}{|c|}{\cmark} & \multicolumn{2}{|c|}{\xmark} \\ 
             \hline
             Acoustic BC & \multicolumn{1}{|c|}{\xmark} & \multicolumn{2}{|c|}{Symmetry, PML} & \multicolumn{2}{|c|}{\xmark} \\             
             \hline
			Loading & \multicolumn{1}{|c|}{$a^{in}$} & \multicolumn{1}{|c|}{$a^{in}$ or $p^{in}$} &  \multicolumn{2}{|c|}{$a^{in}_{S_1}$} & \multicolumn{1}{|c|}{$p^{in}$} \\
             \hline
			Study types & \multicolumn{2}{|c|}{Frequency Domain}&  \multicolumn{1}{|p{2.3cm}|}{Frequency Domain or Eigenfrequency} & \multicolumn{2}{|c|}{Frequency Domain} \\			
             \hline
			Evaluation & \multicolumn{1}{|c|}{$\overline{TF}$ } & \multicolumn{1}{|c|}{$\overline{TF}$, $TL_{\perp}$} &  \multicolumn{2}{|c|}{$TF$} & \multicolumn{1}{|c|}{$\overline{TF}$, $TL_{\perp}$} \\
             \hline
         \end{tabular}
         \renewcommand{\arraystretch}{1.0}
     \end{table}

For the models of the metamaterial structure only, the loading consists of a uniform acceleration or pressure loading. In these study cases, the responses that we obtain from the model correspond to the average accelerations over the input plate and the output plate.
The simulation software allows to calculate a spatial average over the plate surface considering all of the mesh nodes, which allows a straightforward evaluation of the acceleration responses $\overline{a}^{in}$ and $\overline{a}^{out}$. The average transfer function $\overline{TF}(\omega)$ for these models is obtained from 
 
    		\begin{equation}
    			\overline{TF}(\omega) = \frac{\overline{a}^{out}(\omega)}{\overline{a}^{in}(\omega}) \,.
				\label{eq:avTFsim}
    		\end{equation}

In the case of a uniform pressure loading, the sound transmission loss at normal incidence $TL_{\perp}$ is calculated.
This quantity is interesting in the context of sound insulation as it gives an indication about the airborne sound insulation of building components.
	The sound transmission loss at normal incidence is defined as the ratio of the incoming and the transmitted sound power $P$:
	\begin{equation}
		TL_{\perp} = 10 \log_{10} \left( \frac{P^{in}}{P^{out}} \right) \, .
	\end{equation}
        		
It has been shown in~\cite{2024_hermann} that--for a uniform pressure loading--the transmission loss at normal incidence can be obtained from the frequency response function on the input plate,
        		\begin{equation}
        			\overline{FRF}^{in} = \frac{\overline{a}^{in}}{F^{in}} \, ,
        		\end{equation}
        		and the average transfer function (Equation~\eqref{eq:avTFsim}):
        		\begin{equation}
        			TL_{\perp} = 10 \log_{10} \left(  
        				\left| 
        					\frac{1}{2 \, \overline{TF}} 
        					\left(
        						1+ \frac{\mi k}{\overline{FRF}^{in} \, S \rho_0}
        					\right)
        				\right|^2
        				\right) \, .
        		\end{equation}
        		
To allow a quick comparison between specimens and benchmarks, we define a comparison metric that we call performance indicator $Q$. This indicator is based on the $H_2$ norm and compares the sound power ratio $P_r^A = 10^{\sfrac{TL_{\perp}(A)}{10}}$ of the input and output side of a specimen $A$ to the sound power ratio $P_r^B$ of a specimen $B$ in a frequency interval from $f_1$ to $f_2$. Its derivation is shown in Appendix~\ref{app:A5} and it is defined for $N$ consecutive frequencies with the interval $\Delta f$ between two frequencies as
	 \begin{equation}
	 Q(A,B)_{f_1}^{f_2} = \frac{\sqrt{\sum\limits_{n=1}^{N-1} \; \left( |P_r^A(f_n)|^2 + |P_r^A(f_{n+1})|^2 \right) \Delta f}}{\sqrt{\sum\limits_{n=1}^{N-1} \; \left( |P_r^B(f_n)|^2 + |P_r^B(f_{n+1})|^2 \right) \Delta f}} \, .
	 \end{equation}

 \section{Results} \label{sec:results}
 
 In this section, we present experimental and modeling results that illustrate the influence of the interfaces on the dynamic behavior of the finite-size metamaterial specimens.
 At first, we will discuss initial findings of the interfaces' influence and the vibroacoustic coupling that we obtain from the FE models. 
 Afterwards, the results of the experimental measurements are presented and used for the validation of the numerical models.
 The influence of the interfaces of the local displacement fields of the finite-sized specimens is discussed in the subsequent section. 
 Finally, we will show how the vibration damping capacity of the specimens can be improved by modifying the metamaterial or the homogeneous plates and we will compare the performance of the finite-size specimens to two benchmark solutions.
 
 In the description of the finite-sized specimens, the words ``cut" and ``configuration" are used interchangeably to refer to the arrangement (\alp, \bet, \gam or \del) of the unit cells in the metamaterial part of the specimens.
 The average transfer function $\overline{TF}$ (uniform loading), the transfer function $TF$ (experimental loading) and the sound transmission loss at normal incidence $TL_{\perp}$ will be used to characterize the specimen behavior. These functions are complex valued, but for the following analysis we only consider their absolute values (or amplitudes) which are indicated by the symbol $|\cdot |$.

 \subsection{Initial findings on the influence of interfaces}
     In this section we discuss the mechanical and vibroacoustic behavior of both the infinite-size metamaterial and the finite-size sandwich structures. 
     \subsubsection{Infinitely large metamaterial}
\begin{figure}[!h]
          \centering
          \includegraphics[trim = 0 0 0 0, clip,width=\textwidth]{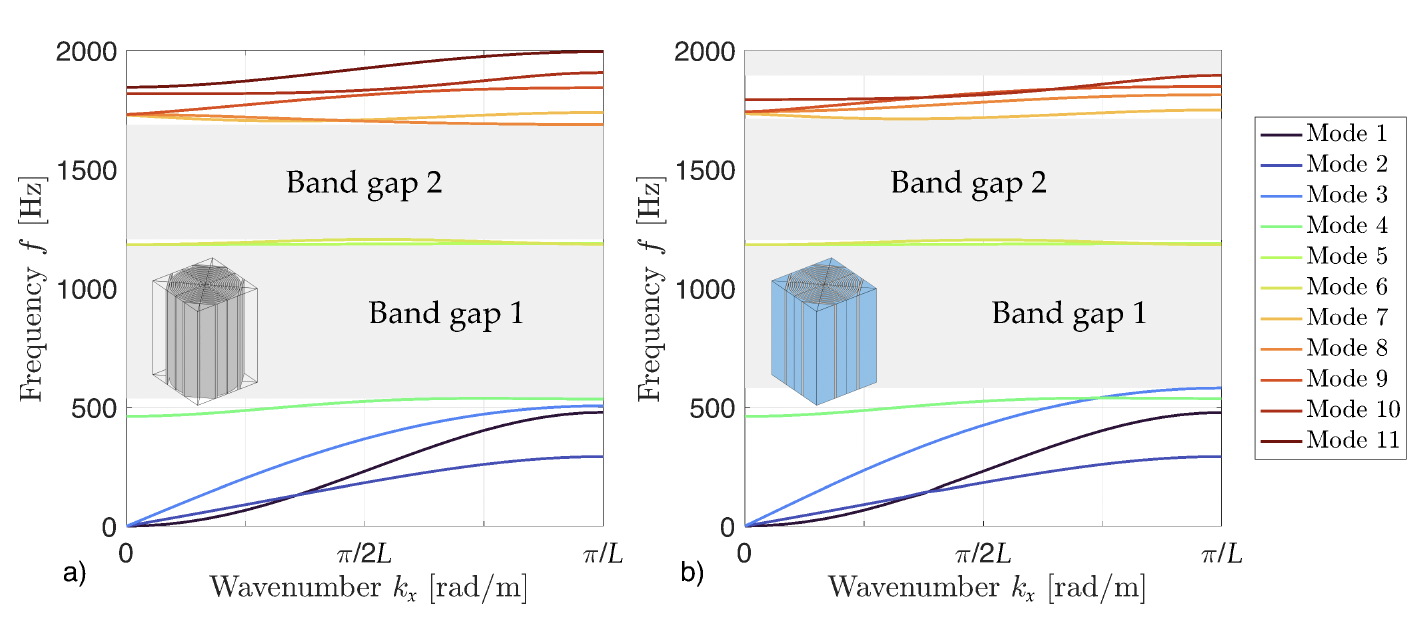}
          \caption{Dispersion curves obtained from 3D simulations from the purely mechanical model~(a) and the vibroacoustic model~(b). $k_x-$wavenumber, $L-$size of the unit cell.}
          \label{fig:dispersionCurves3}
      \end{figure}

     For the dispersion analysis, an infinitely large metamaterial based on the labyrinthine unit cell is considered. 
     The dispersion curves and band gaps obtained from the mechanical model and the vibroacoustic model in a range between \SI{0}{\hertz} and \SI{2000}{\hertz} are shown in  Figure~\ref{fig:dispersionCurves3}a and Figure~\ref{fig:dispersionCurves3}b respectively. 
	Two large band gaps can be observed in both dispersion diagrams.
Bragg scattering is one underlying mechanism generating the band gaps, due to the periodicity of the structure, but there are also contributing local resonances as Eigenfrequency analysis of the structure show. This coexistence of both underlying mechanisms has already been subject of several studies on mechanical metamaterials (cf. [1] for example).	
	 Using the mechanical model, we find the band gaps ranging from \SI{537}{\hertz} to \SI{1183}{\hertz} and from \SI{1205}{\hertz} to \SI{1688}{\hertz}. The band gaps in the vibro-acoustic model have slightly different ranges compared to the mechanical band gaps; they range from \SI{581}{\hertz} to \SI{1182}{\hertz} and from \SI{1203}{\hertz} to \SI{1712}{\hertz}.
     The band gaps are separated by the dispersion curves of modes 5 and 6.
     A comparison with the dispersion curves obtained from a mechanical 2D plane-strain model of the unit cell shows, that these two modes arise due to an out-of-plane movement of the unit cell (cf. Appendix~\ref{app:B1}, Figure~\ref{fig:dispersionCurves2}).
The dispersion curves 5 and 6 are nearly similar for both models; the variation of the frequency range in which waves can propagate is approximately 1190~$\pm$~\SI{10}{\hertz}. The dispersion curve of the eleventh eigenmode is shifted to frequencies higher than \SI{2000}{\hertz} in the vibroacoustic case. Therefore, an additional band gap is opened at \SI{1900}{\hertz} (cf. Figure~\ref{fig:dispersionCurves3}b).
We state that the vibroacoustic coupling has only a small influence on the low frequency band for the first two band gaps of the infinitely large metamaterial.

     \subsubsection{Finite-size sandwich structures}
In this section, we will demonstrate that while the response of an infinite-sized specimen shows only minor changes with or without considering the structure-air coupling, significant changes in response occur in the responses of finite-sized specimens. This indicates that the type of adopted boundary conditions can have a substantial macroscopic effect on the response of finite-sized specimens.    
     The results presented in the following are obtained for the finite-size sandwich structures, in which the metamaterial is integrated with two homogeneous plates.
     
\begin{figure}[h!]
          \centering
          a)\includegraphics[trim = 0 0 400 0, clip,width=0.475\textwidth]{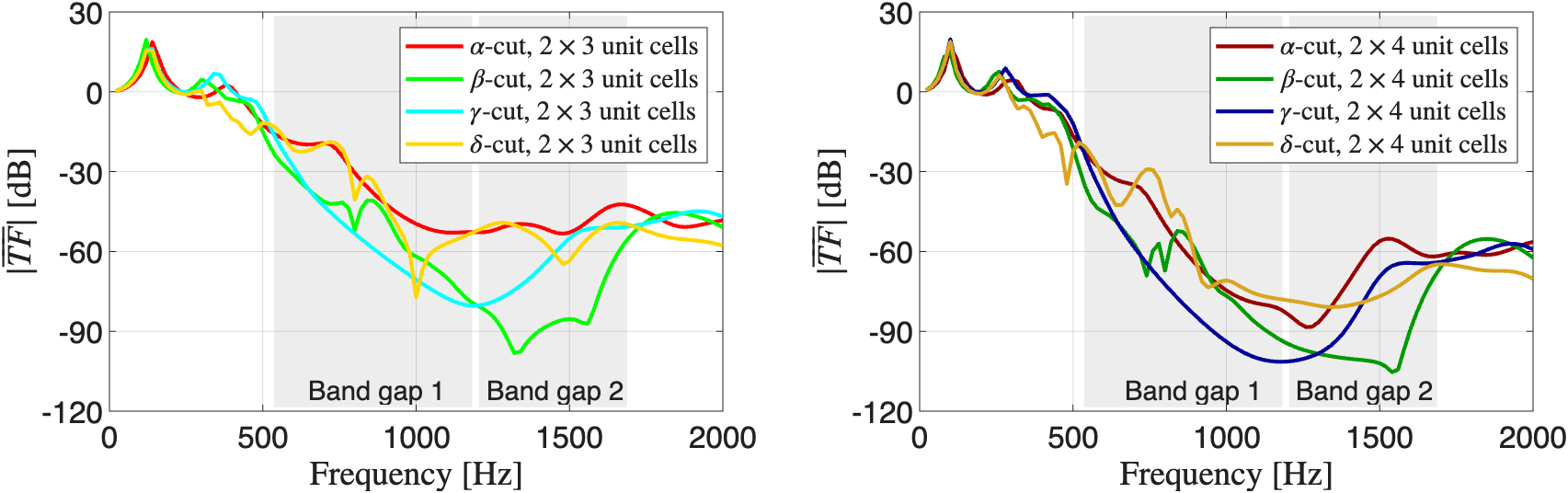}
          b)\includegraphics[trim = 400 0 0 0, clip,width=0.475\textwidth]{figures/mechSim2.png}
          \caption{Amplitude of the average transfer function, $|\overline{TF}|$ obtained for sandwich structures in the four different configurations with \conS unit cells (a) and \conL unit cells (b) from the \textbf{purely mechanical model}.}
          \label{fig:mechSim}
      \end{figure}

     In a first step, the dynamic mechanical behavior of the finite sandwich structures is analyzed. The analysis is based on the average transfer function which is shown in Figure~\ref{fig:mechSim} for the structures with \conS unit cells (a) and \conL unit cells (b). Similar results are obtained for both specimen sizes in the low frequency range (up to \SI{300}{\hertz}). The transfer functions of cuts $\alpha$~-~$\gamma$ continue to evolve in a similar manner while the transfer function for the $\delta$-cut drops and then rejoins the other transfer functions again at \SI{500}{\hertz}. In the very low frequency range, the \del specimen, therefore, shows a superior vibration damping capacity.
     Above \SI{500}{\hertz}, the average transfer functions of the specimens in \bet and \gam configuration drop significantly faster with the frequency and also attain lower values compared to the transfer functions of the \alp and~--~for most of the frequency range~--~also the \del cuts. The vibration damping capacity for this frequency range is, hence, much better for \bet and \gam than for \alp and in most cases also better than for \del. 
     The \alp and \del cuts have the same material interface and \bet and \gam have the same material interface (cf. Figure~\ref{fig:ints}). Therefore, it is clear that the different responses are due to the material interface between metamaterial and homogeneous material.     
     For frequencies above the second band gap, the transfer functions of the different cuts become more similar compared to their differences in the band gap and the \del cut shows a slightly better vibration damping performance than the other three cuts. 
     The transfer function of the \del configuration shows several resonance peaks which are also found for the \alp configuration (near \SI{750}{\hertz},\SI{1300}{\hertz} and \SI{1600}{\hertz}) and other resonance peaks which appear in the transfer function of the \bet configuration (\SI{750}{\hertz},\SI{850}{\hertz}). We will show later in this manuscript that this similarity is due to the similar type of material and free interfaces respectively.

     Based on these first results, configuration \del shows the best vibration damping performance in the lower- and higher-frequency range--we find a difference of up to 10~dB between the transfer functions~--~and configurations \bet and \gam outperform \alp and \del in the medium-frequency range where the band gaps are located--differences up to 40~dB are found. The resonance peaks of the transfer functions are notably attenuated which is due to the introduction of the loss factor $\eta$ in the material model. 
In the present metamaterial, there are, thus, two phenomena that influence the vibration mitigation: the band gap phenomenon (conservative) and the structural damping (dissipative).      
     It should be kept in mind that this dissipative part somewhat counteracts the conservative band gap phenomenon: evanescent waves in materials with dissipative behavior can transfer energy to propagating waves. The peaks in the function are reduced in amplitude and broadened in frequency on the one hand, on the other hand, the drops of the frequency function are attenuated likewise. 
      
\begin{figure}[h!]
          \centering
          a)\includegraphics[trim = 0 0 400 0, clip,width=0.4\textwidth]{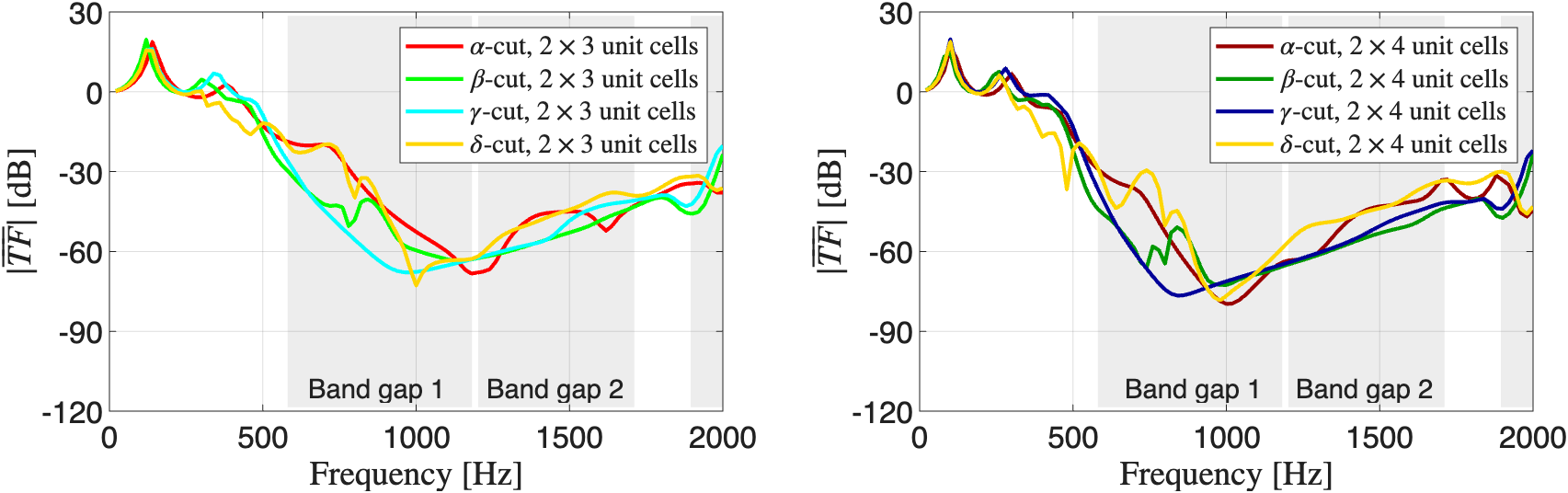}
          b)\includegraphics[trim = 400 0 0 0, clip,width=0.4\textwidth]{figures/vibAcSim3.png}
          \caption{Amplitude of the average transfer function, $|\overline{TF}|$ obtained for sandwich structures in the four different configurations with \conS unit cells (a) and \conL unit cells (b) from the \textbf{vibroacoustic model}.}
          \label{fig:vibracSim}
      \end{figure}

The average transfer function obtained from the vibroacoustic model of the different cuts is shown in Figure~\ref{fig:vibracSim}. In the lower frequency range, we observe similar responses of the structures as for the purely mechanical model. Starting from approximately \SI{1000}{\hertz}, however, the responses of the different cuts become very similar. In addition, the pairs of \alp and \del as well as \bet and \gam show a very high grade of similarity. When the vibroacoustic coupling is taken into account, the performance difference between the specimens becomes much less significant in the higher-frequency range. 
The drop of the transfer functions in the band gap is less in the vibroacoustic case than in the mechanical case, which means that the surrounding air increases the transmission of vibration.
      
\begin{figure}[b!]
          \centering
		\includegraphics[width=\textwidth]{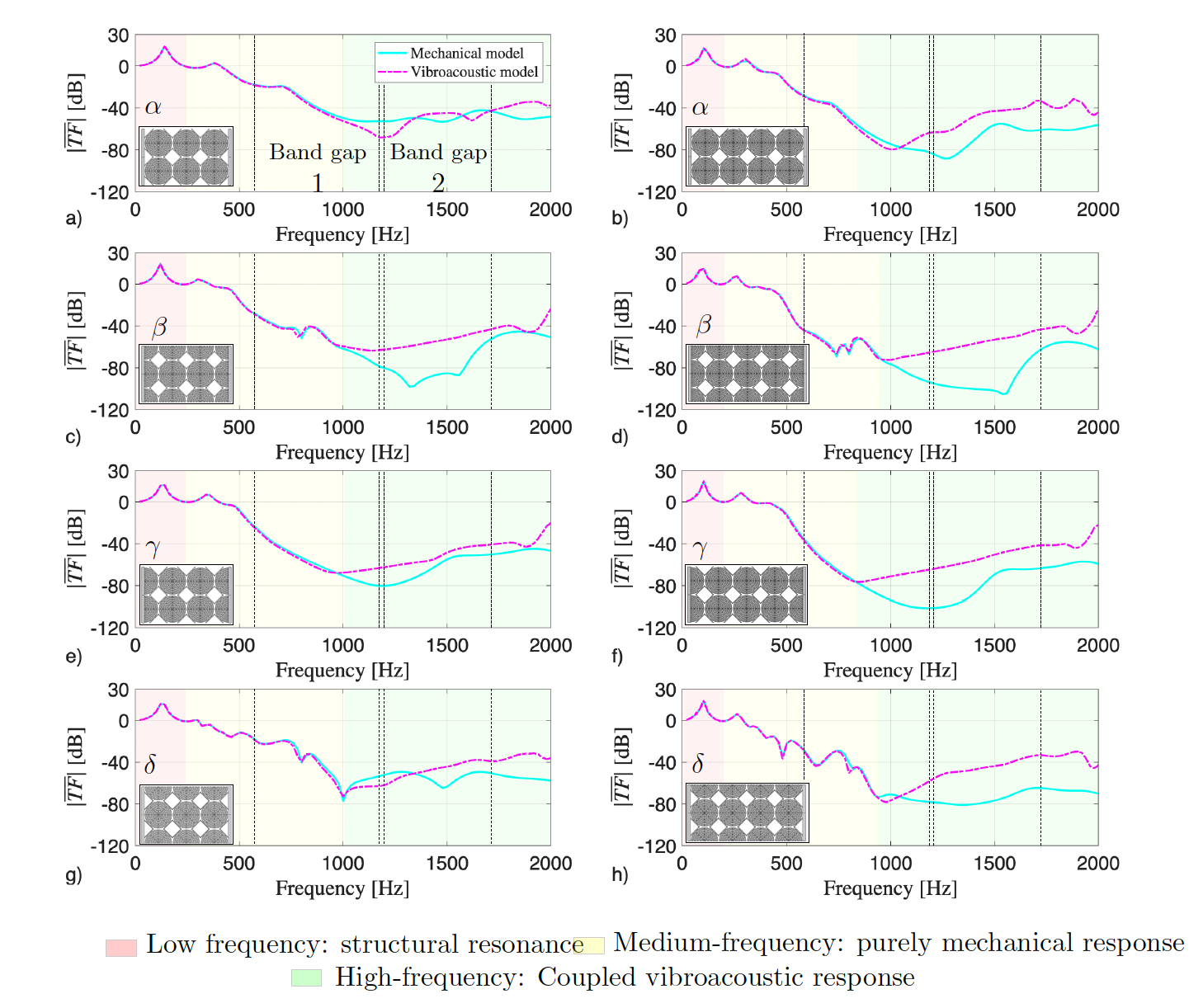}
          \caption{Comparison of the dynamic behavior of all specimens obtained from the mechanical model and the vibroacoustic model. The average transfer function is shown for specimens with \conS unit cells in configurations \alp (a), \bet (c), \gam (e) and \del (g) and for specimens with \conL unit cells in configurations \alp (b), \bet (d), \gam (f) and \del (h).}
          \label{fig:modelAirNoAir}
      \end{figure}

To analyze the influence of the surrounding air on the different cuts more in detail, we compare the average transfer function of the mechanical and the vibroacoustic model for each cut and the two different numbers of unit cells in the following. Figure~\ref{fig:modelAirNoAir} shows the average transfer functions obtained for the four configurations comprising \conS unit cells (a, c, e, g) and \conL unit cells (b, d, f, h). Each graphic depicts the results obtained for a single cut from the mechanical model and the vibroacoustic model. 

      By comparing the graphics of the specimens with \conS unit cells and the specimens with \conL unit cells, we identify a low-frequency ranges up to \SI{250}{\hertz} and \SI{180}{\hertz}, respectively, in which the response of the different configurations is similar. The corresponding range is marked in red in Figure~\ref{fig:modelAirNoAir} and it contains a structural resonance of the metamaterial specimens.
       We also identify a medium-frequency range in which the transfer functions obtained from the mechanical and the vibroacoustic model are similar. For the four smaller specimens its upper limit is \SI{1000}{\hertz}. For the four bigger specimens, the upper limit varies slightly more and lies between \SI{680}{\hertz} and \SI{960}{\hertz}. In this range, we have vibration mitigation via the band gap phenomenon. The different configurations show different responses due to the solid-solid connection at the loaded interface. In his frequency range, \bet and \gam show a better vibration damping capacity than \alp and \del.       
       In the higher-frequency range, the dynamic behavior obtained from the mechanical model is different from the behavior obtained from the vibroacoustic model for each specimen. The vibrations are less damped in the vibroacoustic case than in the purely mechanical case for almost all specimens with the exception of configurations \alp and \del with \conS unit cells. On the one hand, these configurations are the least influenced by the coupling, i.e. the most stable one. On the other hand, they show a lower vibration damping capacity than the \bet and \gam configurations. 
      These results clearly show that there is an important vibroacoustic coupling in the metamaterial specimen for the higher-frequency range: a consistent part of the energy is transmitted by the air even though the metamaterial shows a band gap at that frequency. We can state that, for the considered unit cell, the vibro-acoustic coupling arises for frequencies higher than \SI{1000}{\hertz} which corresponds to a wavelength of \SI{34.3}{\meter} in the air.
      A detailed study of the structure-air coupling in the different cuts will be part of future investigations.

The results show once again that the importance of including the appropriate boundary conditions in the design phase of finite-sized metamaterial structures must not be underestimated. We have shown that that the type of unit cell ``cut'', which defines the boundary conditions, can drastically change the specimens' response both in the purely mechanical and in the vibroacoustic case. More particularly:
\begin{enumerate}
	\item The \del configuration shows the best vibration damping performance in the lower frequency range while the \bet and \gam configurations outperform the other two configurations when a purely mechanical excitation is expected~--~e.g. vibration control.
	\item The \alp configuration can be considered more stable when the air coupling is expected to be important~--~e.g. noise control.	
\end{enumerate}


 \subsection{Model validation}
     In this section, the results of the experimental investigations are analyzed in a first step and are then used to validate the numerical model in a second step. 
      
      \subsubsection{Experimental measurement results}
\begin{figure}[!h]
         \centering
                  \includegraphics[width=\textwidth]{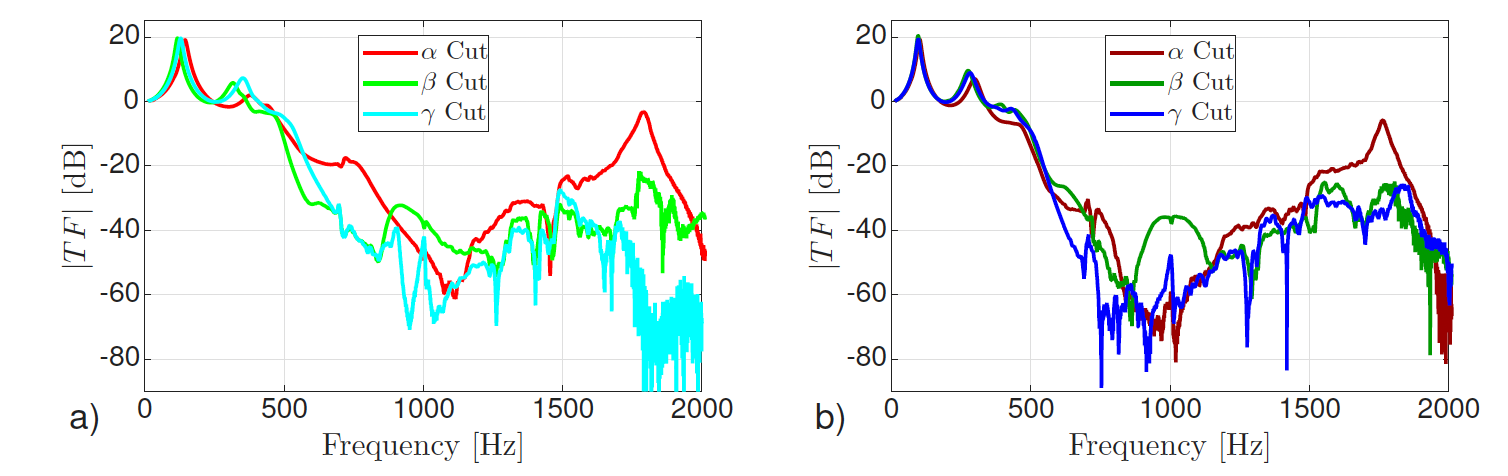}
         \caption{Amplitudes of the experimentally obtained transfer function, $|TF|$, for the three configurations \alp, \bet and \gam for the specimens with \conS unit cells (a) and \conL unit cells (b).}
         \label{fig:expPerNcells}
     \end{figure}

 The experimental results now compared for specimens in different configurations comprising the same number of unit cells. 
     Figure~\ref{fig:expPerNcells}a shows the transfer functions of the tested specimens with \conS unit cells.
     There are several similarities in the transfer functions for all specimens.
     
      In the low frequency range (up to \SI{500}{\hertz}), two resonance peaks can be found: these are the first two structural resonances.     
      In the medium-frequency range (\SI{500}{\hertz} - \SI{1000}{\hertz}), the transfer function drops sharply to values in a range of \SI{-50}{\deci B} \SI{-70}{\deci B}. This corresponds to a vibration reduction by factors ranging from approximately 300 to 3000.
	  Starting from \SI{500}{\hertz}, the \alp specimen is outperformed by the \bet and \gam specimens up to \SI{580}{\hertz} and \SI{850}{\hertz} respectively. 
     The difference of the transfer function reaches up to \SI{22}{\deci B}, which means that vibrations are reduced 12.5 times more by the \bet and \gam configuration than by the \alp configuration. 
     The response is mainly dominated by the structural behavior as it is forecast by the numerical simulations (cf. Figures~\ref{fig:mechSim}a and ~\ref{fig:vibracSim}a).  
      In the high-frequency range ($>$\SI{1000}{\hertz}), the transfer functions of the \bet and \gam cut become very similar up to \SI{1700}{\hertz}, where the transfer function drops sharply for the \gam specimen.
The drop is due to a malfunction of the equipment during the experimental testing, as the low values of the coherence function for this frequency range shows (Figure~\ref{fig:apxCoherence}, Appendix~\ref{app:B2}).     
      the vibration transmission in the \alp configuration is much higher than in the other two configurations due to the presence of a wide resonance peak which is centered around \SI{1790}{\hertz}.
     Specimens \bet and \gam show a similar behavior in the medium-frequency and high-frequency range, besides in the frequency ranges of the disturbances due to the resonances in the \bet specimen (\SI{825}{\hertz} to \SI{1230}{\hertz}) and of the low coherence region for the \gam specimen (above \SI{1680}{\hertz}).
 
     Figure~\ref{fig:expPerNcells}b shows the transfer functions of the tested specimens with \conL unit cells. 
     In the low-frequency and the high-frequency range, the large specimens behave qualitatively in the same way as the small specimens.
     The transfer functions of \bet and \gam are almost similar in the high-frequency range.     
     Besides the large peak at \SI{1750}{\hertz}, the results of the \alp specimen are also similar to the other two configurations in the high-frequency range.
     In the mid-frequency range, the vibration damping capacity is better for \bet and \gam than for \alp~--~from \SI{720}{\hertz} to \SI{820}{\hertz} and from \SI{580}{\hertz} to \SI{830}{\hertz} respectively. The difference of the transfer function with respect to the \alp configuration reach \SI{10}{\deci B} and \SI{30}{\deci B} respectively.
         
     We state, that the specimens in configurations \bet and \gam show a rather similar behavior which is different form the behavior of the specimen in \alp configuration. This finding is in accordance with the results obtained from the numerical modeling and presented in the previous section.	 
    
      \subsubsection{Comparison: numerical vs. experimental results}
\begin{figure}[!h]
             \centering
             \includegraphics[width=\textwidth]{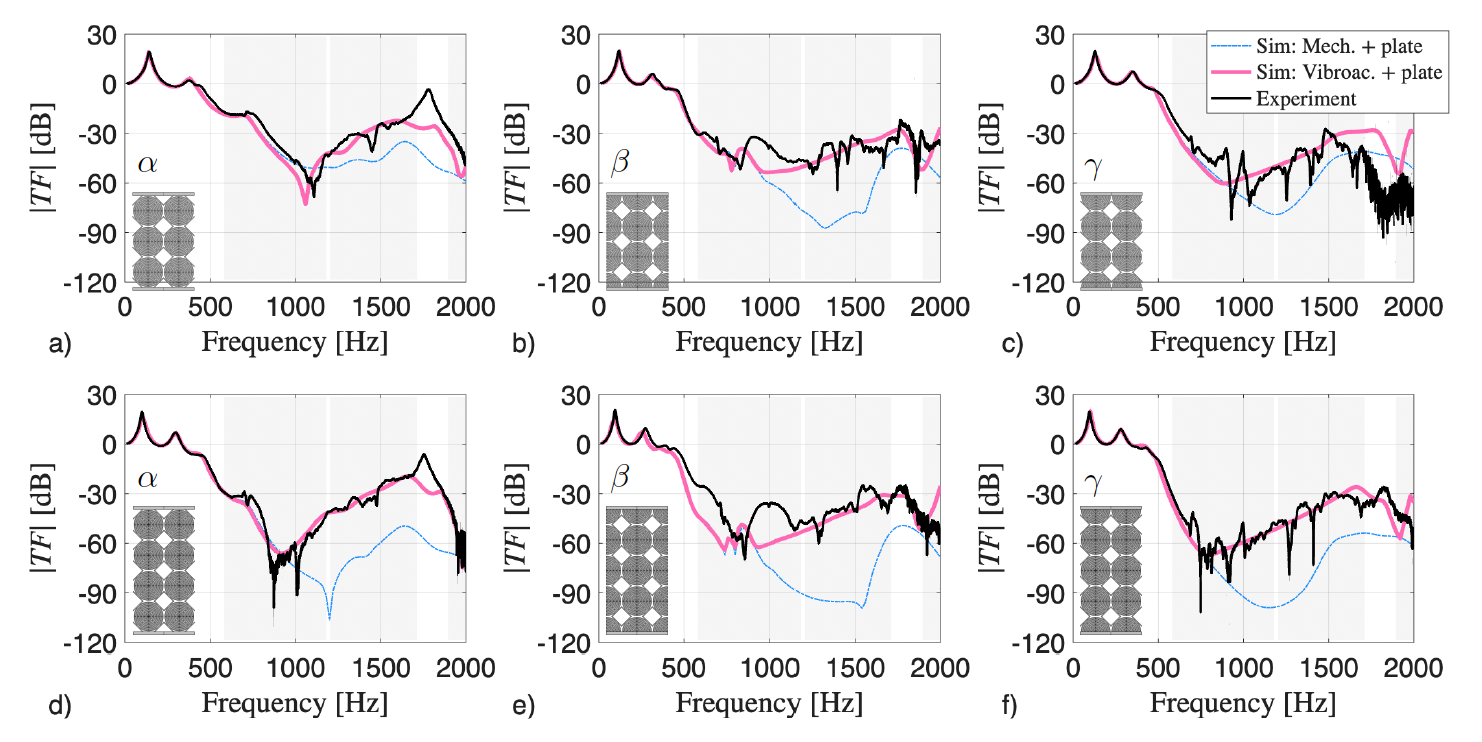}
             \caption{Comparison of experimental and numerical results: amplitude of the average transfer function, $|TF|$, for specimens with \conS unit cells in \alp (a), \bet (b) and \gam (c) configurations and for specimens with \conL unit cells in \alp (d), \bet (e) and \gam (f) configurations.}
             \label{fig:calibCompareModelExp}
         \end{figure}

          The experimental results are used to validate the FE model of the metamaterial structures. 
          Figure~\ref{fig:calibCompareModelExp} shows the average transfer functions $TF$ obtained from the mechanical and vibroacoustic models which include the steel plate and from the experimental measurement results.
          For the low- and medium- frequency range, both model types achieve a good representation of the specimen behavior in the experiment.
          For the higher-frequency range where the vibroacoustic coupling is expected to be present, the vibroacoustic model represents the behavior of the specimens in the experiment much better than the mechanical model. \\
		The comparison shows, that the coupling must be considered in the numerical model to represent the behavior of the specimen in the experiment.
		A comparison of the experimental results and the results obtained from the FE models of the polymer sample without the plate shows that the steel plate has an influence on the dynamic behavior and must, thus, be considered for the model validation (comparison of Figures~\ref{fig:modelAirNoAir} and~\ref{fig:calibCompareModelExp} or cf. Appendix~\ref{app:BN5}).

For three characteristics of the transfer function the model only shows a mediocre agreement with the experiment.
Firstly, the peaks in the transfer function of the \alp cuts (\SI{1780}{\hertz} for \conS unit cells and \SI{1753}{\hertz} for \conL unit cells) are underestimated by the FE models. 
We analyzed several aspects of this difference between experiment and model; the details are given in Appendix and we give a short summary here:
Besides the characteristic peaks, we also have an increased transmission of both \alp specimens in the high frequency range.
The analysis of the acceleration spectra of input and output side show, that there is a drop of the input acceleration in the higher frequency range for all specimens. We attribute this drop to a resonance of the rod connecting the shaker and the specimen (a  detailed explanation is given in Appendix~\ref{app:BN6}). The output acceleration of the \alp configuration is the least affected by this drop.
In our opinion, the reason is a resonance of the structure, which is more sensitive in this regard than configurations \bet and \gam due to the lose solid-solid connections at the interfaces.
An eigenfrequency study shows, that there are resonance modes of the \alp cuts in vicinity of the frequencies for which we observe peaks in the experiments~--at \SI{1792}{\hertz} for the small and at \SI{1774}{\hertz} for the big specimen.
We compared the displacement fields of the output plate obtained from the experiment and the two vibroacoustic models (frequency domain study and eigenfrequency study) and found a good agreement which indicates, that we excite this resonance in the experiment.
However, the displacement amplitude in the simulation is one order of magnitude smaller compared to the experiment.

\begin{figure}[t!]
             \centering
             \includegraphics[width=\textwidth]{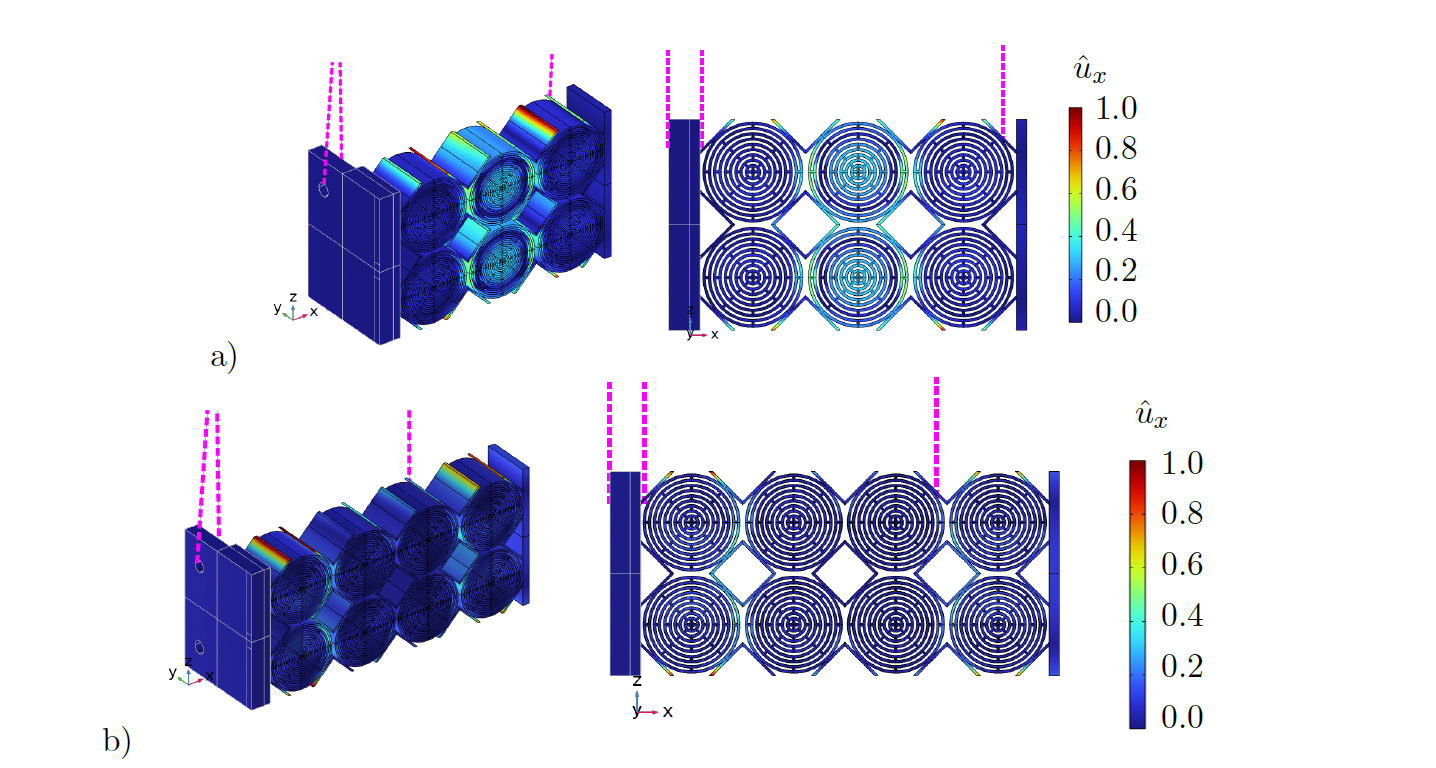}
            \caption{Illustration of resonance modes: Normalized amplitude of the modal displacement field in the $x$-direction, $\hat{u}_x = |u_x|/\text{max}(|u_x|)$ and schematic representation of the suspension lines in the experiment for the \alp cuts with \conS unit cells at \SI{1792}{\hertz} (a) and \conL unit cells at \SI{1774}{\hertz} (b). The results were obtained from an eigenfrequency analysis with the vibroacoustic model.}
             \label{fig:alpResModes}
         \end{figure}

Looking at the mode shapes of the \alp cuts for the specific resonance in Figure~\ref{fig:alpResModes}, we observe that the central unit cells move most for this vibration mode. Considering the location of the suspension lines, it is plausible that this resonance mode will be excited. 
We, thus, conclude that the boundary conditions in the experiment~--~i.e. the resonance of the rod in combination with the suspension lines~--~increase the transmission while this influence is not included in the model. Modeling the influence of the suspension lines is not straightforward, since they are not glued to the specimen and, therefore, the contact is not permanent. Since the influence of the boundary conditions is limited to the upper part of the high frequency range, we consider the behavior of the specimen sufficiently well approximated in most of the frequency range of interest.

         Secondly, the peaks observed in the medium-frequency range of the transfer function for the \bet specimens are underestimated and located at different frequencies in the simulation results compared to the experiment. For these specimens, we clearly see manufacturing inaccuracies at the outer boundaries (cf. Appendix~\ref{app:A1}). These resonance peaks are linked to vibrations that propagate at the boundaries as we will show in the following section. Therefore, it is very likely that this is the origin of the frequency and amplitude shift. Even though these characteristic peaks are not found to be exactly as in the experiment, they are represented by the model.       
         Thirdly, the model of the small specimen with the \gam cut does not represent the large drop of the transfer function in the high frequency range. This drop, however, is due to a malfunction of the equipment during the experimental testing as explained in the previous section and can, thus, not be represented by the model.         
         The fluctuations in the experimental transfer functions are most likely due to the specimen alignment and the excitation, which are not as perfectly symmetric and unidirectional as in the simulation. In addition, the specimens themselves are not completely symmetric due to inaccuracies in the manufacturing and assembly process. The acceleration of the evaluation points that should be (anti-)symmetric and cancel out in the spatial average do, therefore, not cancel out which leading to a fluctuation in the average response.             
              
In this section, experimental results were compared with the results obtained from the vibroacoustic models including the steel plate. Deviations between the model and experimental results were analyzed and can be explained. In summary, the agreement between numerical and experimental results is sufficiently good to consider the model of the finite-sized metamaterial structures as validated.
     
 \subsection{Influence of the interfaces on the local displacement field} \label{sec:cutCharacs}
     In this section, the results obtained from the vibroacoustic FE-models that represent the metamaterial structure are analyzed. The goal is to provide insight into the behavior of the specimen with particular attention to the influence of the material interfaces and free interfaces.
     The models are helpful because they provide local information about the displacement field throughout the specimen.
     We show how the interfaces influence the displacement fields of global modes~--~the entire structure moves~--~and local modes~--~a small part of the structure oscillates locally.

     \subsubsection{Displacement field of global resonance modes}
\begin{figure}[h!]
         \centering
         \includegraphics[width=\textwidth]{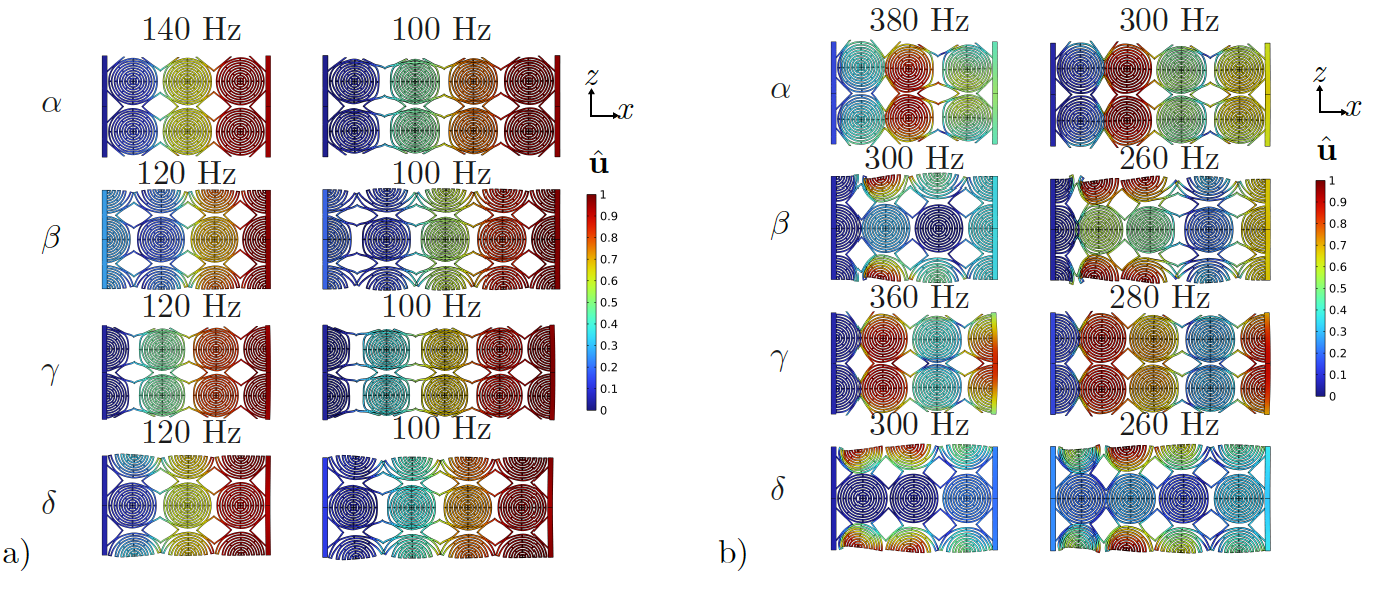}
         \caption{Displacement field $\hat{u}$ of the first resonance mode (a) and the second resonance mode (b) obtained from the numerical vibroacoustic model for the four different cuts with \conS and \conL unit cells respectively.}
         \label{fig:modes12}
     \end{figure}

     In a first step, we study the displacement field of the first resonance mode. Therefore, we show a side view of the specimen in the $x-z$-plane in Figure~\ref{fig:modes12}~a. The color represents the field $\hat{u}$ corresponding to the norm of the displacement field which is normalized with respect to the maximum value of the norm of the displacement field for each configuration and frequency: 
         \begin{equation}
             \hat{u} = \frac{||\vec{u}||}{\text{max}(||\vec{u}||)} \quad \text{with} \quad ||\vec{u}|| = \sqrt{u_x^2+u_y^2+u_z^2} \,.
         \end{equation}
     The mode shape of the first resonance mode corresponds to an extension-compression-motion of the entire structure in the $x$-direction.
     The global displacement field is rather similar for the different configurations and specimen sizes.
     This observation corresponds to the location of the first resonance peak in the transfer function, which is rather similar for the different specimens (cf. Figure~\ref{fig:expPerNcells}).
     Nevertheless, small differences of the displacement field can be observed in the bulk: the fields of \alp and \del, as well as of \bet and \gam, are more similar to each other than to the fields of the other pair.
     This is due to the similar material interfaces and the resulting distribution of similar geometric units in the bulk metamaterial.   
     
     In a second step, the displacement field of the second resonance mode is analyzed for all specimens.
     Figure~\ref{fig:modes12}~b shows $\hat{u}$ for the eight specimens. 
     In contrast to the first resonance mode, the displacement field norm shows bigger differences between the specimens compared to the first resonance mode. 
     However, similar patterns can be observed for the pairs \alp and \gam and \bet and \del. 
	For the first pair, the displacement field propagates in the bulk of the metamaterial and the gradient of $\hat{\textbf{u}}$ in the $z$-direction is small.
     For the \bet and \del configurations, most of the wave propagation happens at the free boundaries. These cells show a rotational movement around the $y$ axis while adjacent cells move upwards and downwards (translation in the $z$-direction).
The differences between the displacement field $\hat{u}$ of the different configurations result from the different free interfaces. 		       
     The analysis of the first two resonance modes illustrates the influence of the boundary conditions on the displacement field of global modes. 
     
     \subsubsection{Displacement field of local resonance modes}
\begin{figure}[h!]
         \centering
         \includegraphics[width=\textwidth]{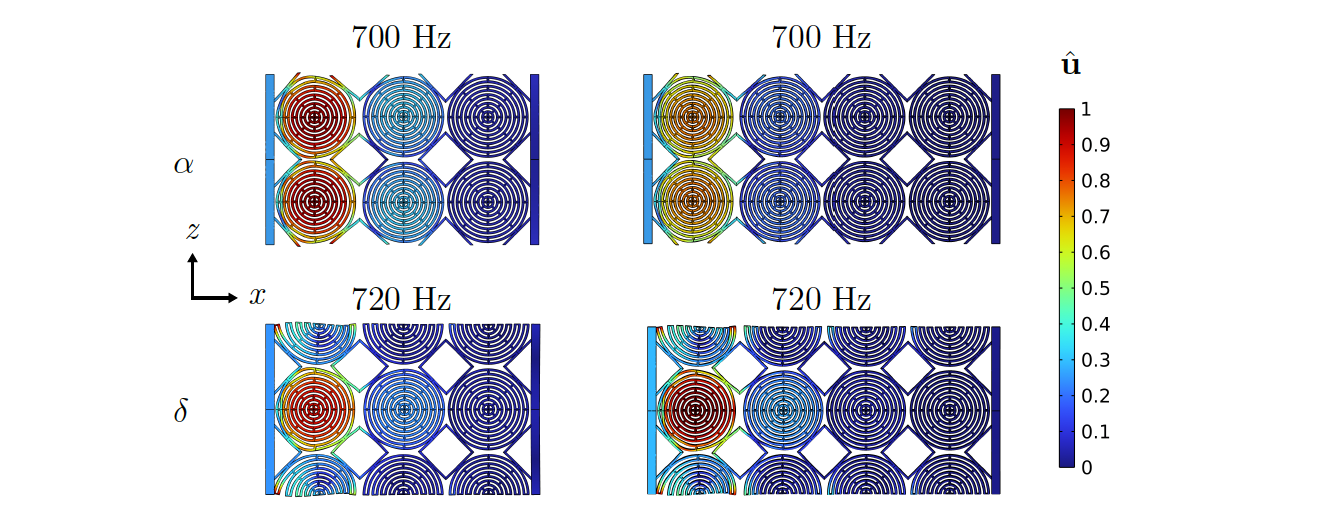}
        \caption{Displacement field $\hat{u}$ of the local resonance mode at \SI{700}{\hertz} in the \alp configuration and at \SI{720}{\hertz} in the \del configuration obtained from the numerical vibroacoustic models.}
         \label{fig:alpha700}
     \end{figure}

     We demonstrate the influence of the boundary conditions on local modes with two examples in the following.
     Firstly, the results in Figure~\ref{fig:vibracSim} show, that there is an additional resonance peak at \SI{700}{\hertz} for the \alp configuration and at \SI{720}{\hertz} for the \del configuration. 
     This peak must have a mechanical origin, because it is also present in the results of the mechanical model.
     The displacement field $\hat{u}$ in Figure~\ref{fig:alpha700} shows, that this peak corresponds to a local resonance of the first column of unit cells connected to the plate. This explains why the resonance peak does not appear in the transfer functions of configurations \bet and \gam: the cells adjacent to the plate are less free to move because the material interface with the plate is larger and the connection to the unit cell is stiffer for \bet and \gam than for \alp and \del. This resonance mode is clearly influenced by the material interface. 
     
     Secondly, the experimental and numerical results in Figure~\ref{fig:vibracSim} show additional resonance peaks in the transfer function of the \bet and \del configuration. 
	\begin{figure}[!b]
         \centering
         \includegraphics[width=\textwidth]{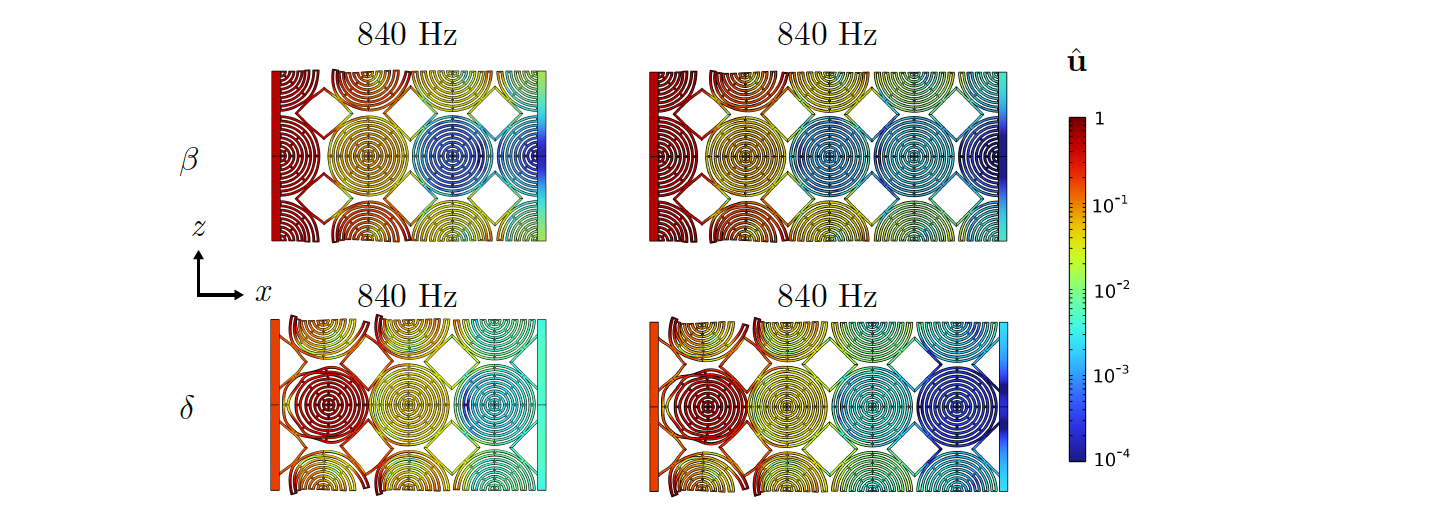}
         \caption{Displacement field $\hat{u}$ of the local resonance modes of the \bet and \del configurations at \SI{840}{\hertz} obtained from the vibroacoustic FE-model. The scale of the color range is logarithmic.}
         \label{fig:betaEdge}
     \end{figure}
     The peaks are located at \SI{840}{\hertz} for both configurations.
     Since the peaks are also found in the purely mechanical simulations, they must have a mechanical origin.
     The corresponding displacement fields $\hat{u}$ are shown in Figure~\ref{fig:betaEdge}.      
     Since the frequencies lie in the band gap, the displacements are very low at the output plate. To better illustrate the displacement field in the structure, the color scale is logarithmic in this graphic.
     Figure~\ref{fig:betaEdge} shows, that most of the displacement is transmitted via the free boundaries of the specimens for the corresponding resonance modes which are so-called edge modes.
     They can be observed in dispersion diagrams when the Floquet-Bloch analysis comprises more than one unit cell and when the periodic boundary conditions are only applied in the direction of wave propagation (cf. Appendix~\ref{app:B4}, Figure~\ref{fig:dispCurves2x3}).
     In the low- and medium-frequency range, the transfer function of the \del specimen shows additional resonance peaks compared to the other configurations. The reason for this is the presence of additional edge modes.
     These resonance modes are clearly influenced by the free interface.

The two examples illustrate the influence of the boundary conditions on the displacement field of local modes.      
The comparison of the displacement fields of the different configurations shows that the specimen \gam combines two geometric assets that lead to better vibration damping performance in the medium-frequency and the high frequency range: a big material interface--no resonance peak at \SI{720}{\hertz}--and a small surface at the free boundaries --no edge modes. On the contrary, the \del specimen combines the opposite characteristics. Its performance in the medium-frequency and the high frequency range obtained from the vibroacoustic model is worse than the performance of the specimens in \alp, \bet and \gam configuration. In the low-frequency range, however, the performance of the specimens in \gam configuration is the best.
     These above results confirm, that material interfaces and free boundaries can be used to tailor the vibration damping performance of finite-sized metamaterial specimens.

 \subsection{Optimization approaches for the finite-sized metamaterial structures}
     In this section, we will show two straightforward methods to improve the vibration damping capacity of the finite-size specimens. 
     The improvement is based on the numerical models comprising only the metamaterial specimen. 
     Firstly, the number of unit cells in the metamaterial is altered and secondly, the geometry of the homogeneous plates is modified.
    
     \subsubsection{Optimizing the metamaterial: number of unit cells}
\begin{figure}[!h]
         \centering
         \includegraphics[trim = 0 0 0 0, clip,width=0.8\textwidth]{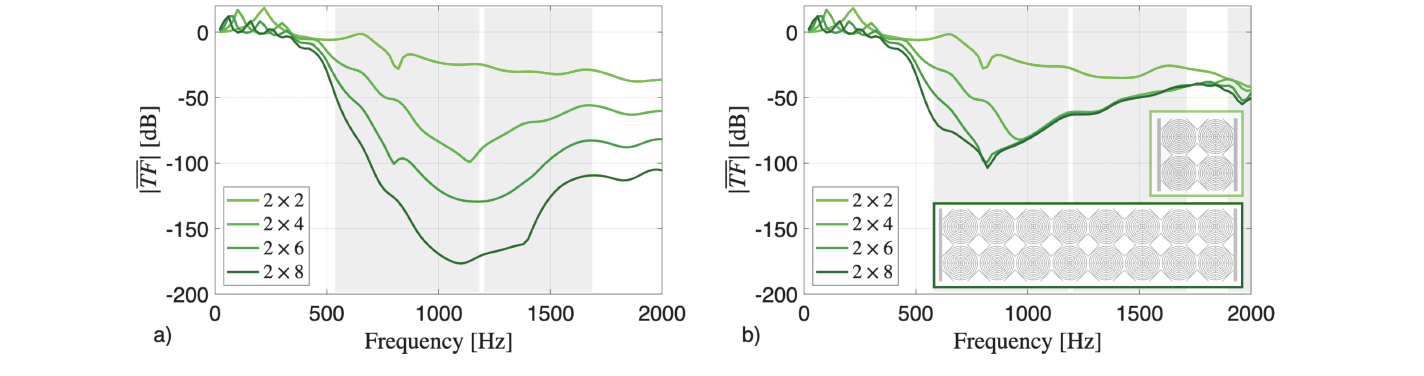}
         \caption{Amplitude of the average transfer function, $|\overline{TF}|$, for increasing number of unit cells in direction of wave propagation ($e_1$-direction) obtained from the mechanical model (a) and the vibroacoustic model (b). The corresponding band gaps for the two models are shown in gray.}
         \label{fig:nCellsHori}
     \end{figure}

     The number of unit cells of the metamaterial is varied in this section; the results are discussed using the example of the \alp configuration. 
     In a first step, we vary the number of unit cells in the direction of wave propagation--the $e_1$-direction.
     Figure~\ref{fig:nCellsHori}a shows the results obtained from the mechanical model. In the medium- and high-frequency range, the transfer function drops with increasing number of unit cells. The transfer function obtained from the vibroacoustic model, shown in Figure~\ref{fig:nCellsHori}b, also drops with increasing number of unit cells, but the trend converges in contrast to the results from the mechanical model. For the high-frequency range, the average transfer function does not change when the number of unit cells in direction $e_1$ is increased above four. In the medium-frequency range, the results start to converge above a unit cell number of six. 
     A convergence of the transfer function is not reached with the mechanical model even with $2\times 30$ unit cells (cf. Appendix~\ref{app:B10}, Figure~\ref{fig:apx_NcellsMec}).
     In the experiments, we observe the same behavior as for the vibroacoustic model. 
     The behavior of specimens with similar configurations but a different number of unit cells is similar in the higher-frequency range (cf. Figure~\ref{fig:expPerCut}).
     We state that adding more unit cells in the metamaterial increases the vibration damping capacity of the finite size structure. In the vibroacoustic case, this effect is limited by the transmission of vibration energy through the air.
     
     \vspace{-0.25cm}
     
     \newpage
    
\begin{figure}[!h]
         \centering
         \includegraphics[trim = 0 0 0 0, clip,width=\textwidth]{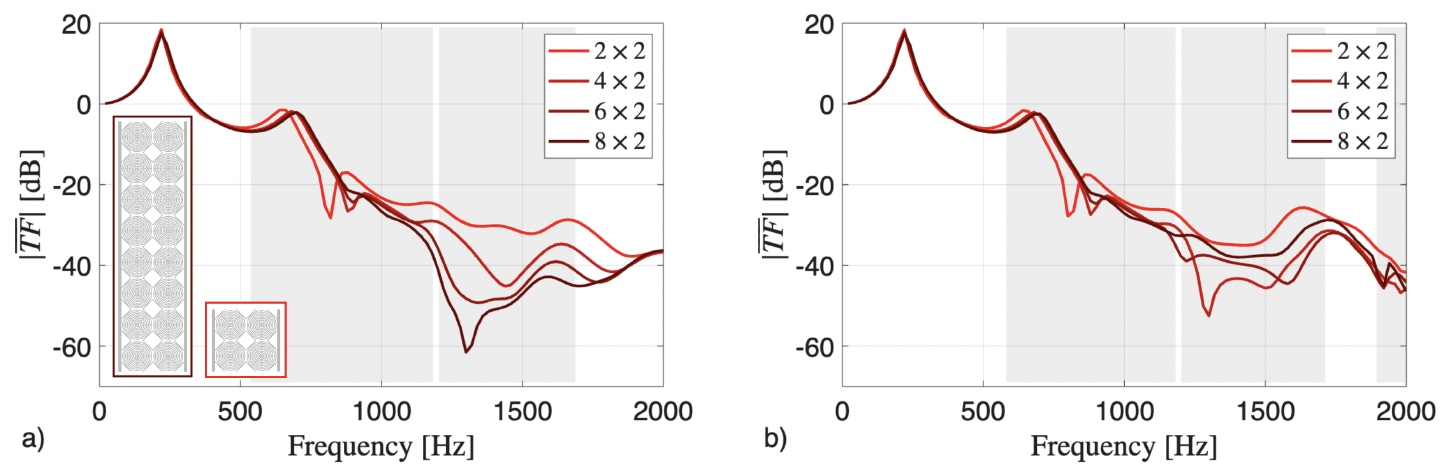}
         \caption{Amplitude of the average transfer function, $|\overline{TF}|$, for increasing number of unit cells perpendicular to the direction of wave propagation ($z$-direction) obtained from the purely mechanical model (a) and the vibroacoustic model (b). The corresponding band gaps for the two models are shown in gray.}
         \label{fig:nCellsVert}
     \end{figure}

     In a second step, the number of unit cells is increased in the transverse direction with respect to the direction of wave propagation--the $e_3$-direction.
     The average transfer functions obtained from the mechanical model and the vibroacoustic model are shown in Figure~\ref{fig:nCellsVert}a and Figure~\ref{fig:nCellsVert}b respectively.
     
     In both cases, the results in the low- and medium-frequency range converge for a number of four unit cells.
     For the purely mechanical model, the average transfer function decreases monotonically in the high-frequency range and the results start to converge. 
     The results obtained from the vibroacoustic simulations do not show a monotonic trend with increasing number of unit cells in vertical direction. The reason for this probably lies in the interaction between the unit cell and the air; this aspect is not analyzed in detail in this paper.
     The air also seems to limit the improvement of the vibration mitigation performance of the specimen, with the number of unit cells in vertical direction.
     By comparing the results in Figure~\ref{fig:nCellsHori} and Figure~\ref{fig:nCellsVert} one can state that  the number of unit cells in the direction of wave propagation has a stronger influence on the transfer function as the number of unit cells perpendicular to the direction of wave propagation.    

     \subsubsection{Optimizing the homogeneous material: plate thickness}
\begin{figure}[!b]
         \centering
		\includegraphics[width=\textwidth]{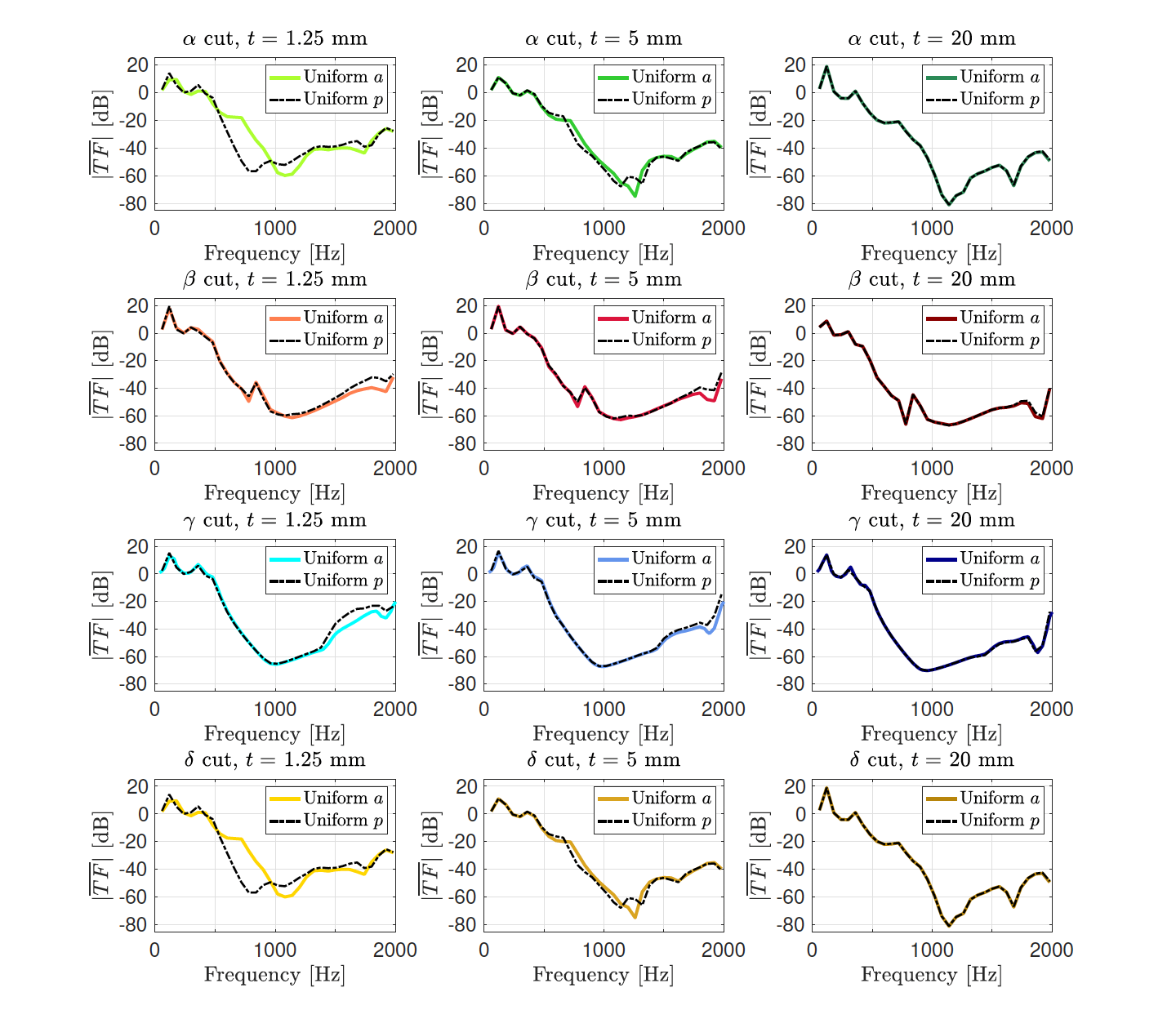}         
         \caption{Influence of loading type (uniform acceleration $a$ or pressure $p$) and plate thickness $t$ on the vibration mitigation: average transfer functions, $|\overline{TF}|$, of specimens in four different configurations (\alp - \del) connected to plates with different thickness $t$.}
         \label{fig:plateThickness4x3}
     \end{figure}

     The results presented in this section are obtained for the smaller specimens (\conS unit cells). We tested three different thicknesses $t$, namely \SI{1.25}{\milli \meter}, \SI{5.00}{\milli \meter} and \SI{20.00}{\milli \meter}. 
     Figure~\ref{fig:plateThickness4x3} shows the average transfer function obtained for the specimens of the four different configurations \alp to \del and three different plate thicknesses respectively. For each specimen, the results of the two different load cases, described in the method section are shown: uniform acceleration and uniform pressure loading. A frequency step of \SI{60}{\hertz} was chosen for the corresponding studies. The larger frequency step allows to save time and computational resources while being still fine enough to analyze the impact of the plate thickness on the specimen behavior.
     
\begin{figure}[!h]
         \centering
         \includegraphics[trim = 0 20 0 0, clip,width=0.9\textwidth]{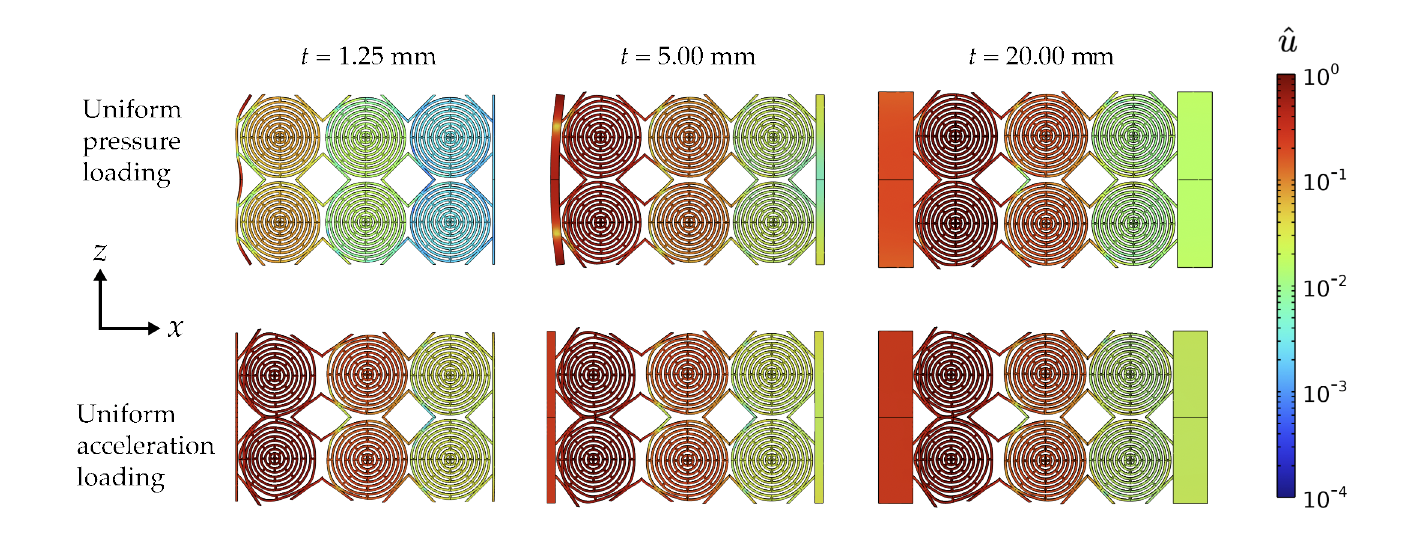}
         \caption{Displacement field $\hat{u}$ of the \alp \conS at \SI{660}{\hertz} obtained with different plate thickness. In the cases represented in the three top pictures, a uniform pressure was applied as a loading; in the cases represented in the three bottom pictures, a uniform acceleration was applied as a loading. The scale of the color range is logarithmic.}
         \label{fig:plateThickness4x3ExA}
     \end{figure}

     A uniform acceleration corresponds to a mechanical-type of loading while the uniform pressure loading can be associated to an acoustic-type of loading in which a plane wave is incident on the specimen surface.
     In general, an increase of the plate thickness improves the vibration mitigation as the transfer function is shifted to lower values.
     This is due to the increased plate mass--more energy is needed to accelerate the plate--and the increased bending stiffness of the plate--high local displacements due to bending are less likely to occur. The effect is most pronounced in the high-frequency range, where the acoustic coupling comes into play (cf. Figure~\ref{fig:modelAirNoAir}) and, therefore, the acoustic energy transmission plays a more important role than in the low- or medium-frequency range.
     The results obtained for the different loading types are similar for almost all cases and become identical for the plate with the highest thickness.
     There are, however, two exceptions: for \alp and \del, the vibration mitigation performance in the medium-frequency range (\SI{540}{\hertz}~-~\SI{960}{\hertz}) is much better for a uniform pressure loading than for a uniform acceleration loading. This is due to the material interface between the plate and the specimen: for a uniform pressure loading, a part of the vibration energy is used to deform the plate and does not enter the specimens.
     This is illustrated in Figure~\ref{fig:plateThickness4x3ExA}, where we show the displacement field $\hat{u}$ of specimens in the \alp configuration for different plate thicknesses and different loading types.
                
      The deformation of the input plate is clearly visible for the smallest plate thickness and uniform pressure loading. In addition, Figure~\ref{fig:plateThickness4x3ExA} illustrates that the plate deforms less and less with increasing thickness; the results of the uniform pressure loading, hence, become similar to the results of the uniform acceleration loading.
     For configurations \bet and \gam, the material interface between metamaterial and plate is bigger than for \alp and \del. A bigger portion of the vibration energy is transmitted to the metamaterial for these configurations because the plate is less free to deform. 
     The graphics also show, that the input plate deforms less and less with increasing plate thickness, since the pressure load chosen for this specific load case becomes too weak to cause a bending. Whether bending occurs depends on the ratio of the pressure load and the plate's bending stiffness. 
     These results show, that the material interface and the plate can be tailored to influence the overall response of the finite-size metamaterial specimen. For a better performance in the low-frequency range, a small material interface and a thin plate are favorable. For a better performance in the medium-frequency and high-frequency range, a big material interface and a plate with high thickness are favorable.

 \subsection{Comparison of vibroacoustic properties to benchmark solutions}
\begin{figure}[!h]
         \centering
		\includegraphics[width=\textwidth]{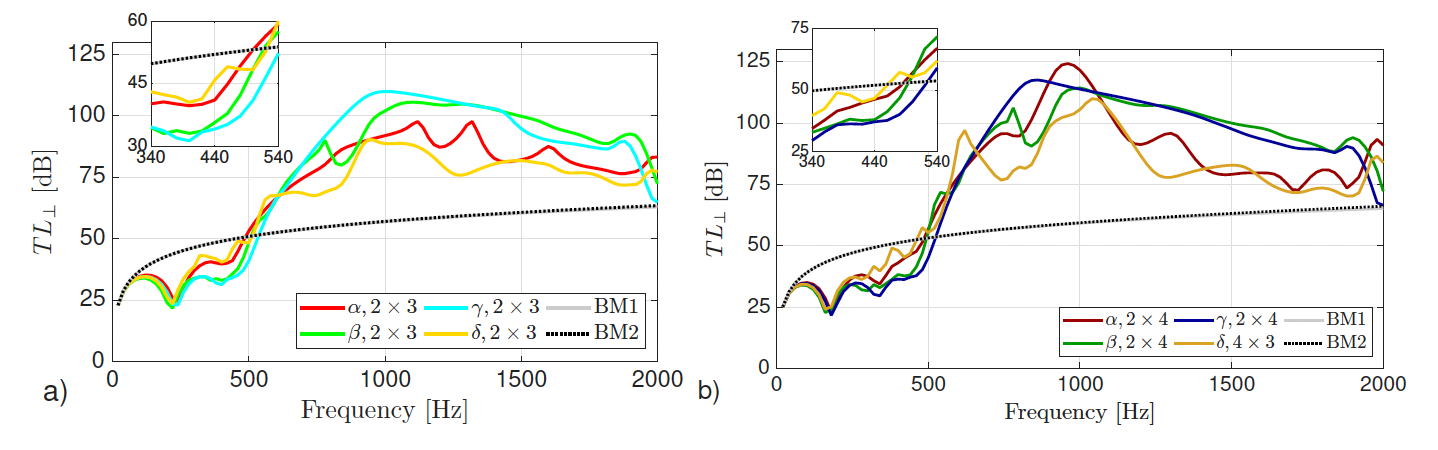}         
         \caption{Comparison to benchmarks: transmission loss at normal incidence $TL_{\perp}$ for specimens comprising \conS unit cells (a) and \conL unit cells (b) and for the benchmark cases consisting of a concrete block (BM1) and a gypsum block (BM2) having the same mass as the specimens. A zoom on the low-frequency range is provided in the upper left box in each graphic.}
         \label{fig:TLbenchmark}
     \end{figure}

     In this section we compare the performance of the small-size metamaterial specimens to the performance of two benchmark cases using the sound transmission loss at normal incidence $TL_{\perp}$. 
     The benchmark cases consist of a gypsum block and a concrete block which have the same mass as the specimen as described in the methods section. 
     A better performance is indicated by a higher value of $TL_{\perp}$. 
     Figure~\ref{fig:TLbenchmark} shows the sound transmission loss at normal incidence for the specimens comprising \conS unit cells (a) and \conL unit cells (b) as well as the results obtained for the benchmark cases with similar mass as the specimens respectively.\\
     A comparison of the specimen's results shows that, in general, the configurations \alp and \del have the best performance in the low-frequency range, and that the configurations \bet and \gam have the best performance in the mid and high-frequency range.
     Deviations from this general statement are the drops of the transmission loss of the \bet specimens due to the edge modes and a peak in the transmission loss of the large specimen (\conL unit cells) around \SI{960}{\hertz}, where a strong bending motion of the plate on the input side is observed.
     The difference in the transmission loss reaches up to \SI{10}{\deci B}, which means that the sound power is reduced 10 times more with specimens in configurations \alp or \del than with specimens in configurations \bet or \gam. 
     The similarities between the transmission loss of the \alp and \del configurations and between the \bet and \gam configurations shows, that the material interface plays a bigger role than the free interface.
  
\begin{figure}[h!]
\centering
\includegraphics[width=\textwidth]{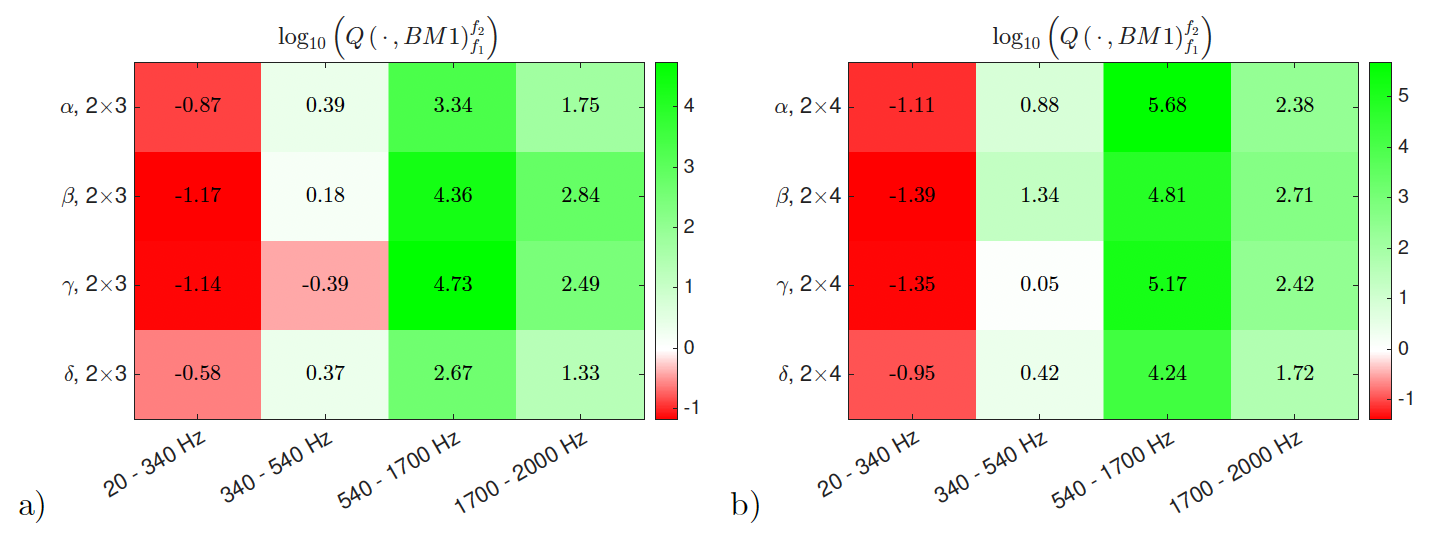}
\caption{Visualization of the performance indicator: The quantity $\log_{10}(Q)$ is represented for different metamaterial structures    with $2\times3$ unit cells (a)  $2\times 4$ unit cells (b) compared to benchmark $BM1$ for frequency ranges ($f_1$ - $f_2$) below the band gap (20~--~340~Hz and 340~--~540~Hz), in the band gap (540~--~1700~Hz) and above the band gap (1700~--~2000~Hz).}
\label{fig:TLmetric}
\end{figure}

     A comparison with the benchmark cases shows, that all configurations largely outperform the gypsum block and the concrete block in the medium- and higher-frequency range.
 In the low frequency range, however, the specimens in \alp and \del configuration perform better than the \bet and \gam configuration in the frequency range between \SI{250}{\hertz} and \SI{500}{\hertz}.
In Figure~\ref{fig:TLmetric}~a and~b, the performance indicator is visualized for different frequency ranges; the higher the value the better the performance. Since the results span several magnitudes, $\log_{10}(Q)$ is shown in the graphics. We can see that far below the band gap, the Benchmark is slightly more effective than the metamaterial which is indicated by a negative value. Closer to the band gap, we can see that the metamaterial becomes more effective than the benchmark. This can also be observed in Figure~23 of the manuscript. In the band gap, the metamaterial is much more effective than the benchmark~--~it should be kept in mind that logarithmic quantities are compared here: a $\log_{10}(Q)=1$ corresponds to $Q = 100$ for example. Above the band gap, this is still the case but it is clearly visible that the most effective frequency range for the metamaterial is the band gap.
       
     These result shows that the choice of the material interface can make a decisive contribution to improving the performance of a structure so that it outperforms another solution.

   \section{Discussion}
    This section discusses the significance of the results and compares them with previous work using relevant references.
   A central point of our analysis is the influence of boundary and interface effects in finite-sized metamaterials which are connected to homogeneous materials in a sandwich configuration. 
In the majority of applied studies on assembled metamaterial structures, the influence of interfaces at the boundaries of metamaterials is not considered. When considered, the studies concentrate on edge effects at free boundaries or interfaces between metamaterials in order to optimize energy transport. In the present study, we concentrate on the influence of the free interfaces of a metamaterial but also on the interfaces between a metamaterial and a connected structure that is generalized as Cauchy continuum. Our results show how designing this interface and the adjacent components can largely influence the vibration transmission and, therefore, can have a large importance in real application cases.      
   If we know how to control these effects it will be possible to design the boundaries to our advantage.
   Our group has shown before in numerical models, that the boundaries and material interfaces can make a big difference for the behavior of the finite sized structure \cite{2024_perez,2024_demetriou}. 
   In the present study, we confirm the influence of boundary effects by experiments for the first time with the metamaterial sandwich structures made from different unit cell cuts. 
   In addition, we establish a link between resonance peaks in the global response of the structure and the local displacement field. By doing this, we are able to understand why some boundary conditions are advantageous compared to others which takes us a step forward in terms of designing an optimal structure.
  
  We also show here for the first time that the metamaterial sandwich structures show a vibroacoustic coupling, i.e. the surrounding air has an influence on the dynamic mechanical behavior of the specimen. 
   The coupling lowers the performance of the different specimens in the upper half of the tested frequency range: for the tested number of unit cells the performance becomes similar for the structures made from different cuts and for an increasing number of unit cells, the performance improvement is bounded. 
   A vibroacoustic coupling in metamaterial sandwich structures is uncommon, in general their behavior is well represented by purely mechanical models (cf. for example~\cite{2016_Dalessandro,2022_demore}).
However, metamaterials that show a vibroacoustic coupling are also demanded since they offer the possibility to combine mechanical and acoustic resonances, thus, broadening the possible spectrum of applications.
   Our analysis shows that the sensitivity of the metamaterial to the coupling is influenced by the metamaterial cut. 
   Strategies to reduce the vibroacoustic coupling or even use effects like acoustic resonances in the mechanical structure are an interesting direction for future research on this type of coupled metamaterial.
  
   To further optimize the performance of the unit cell, we can optimize the bulk metamaterial~--~as opposed to the boundaries~--~and the connecting structure.
   Improving the performance of the bulk metamaterial by adding unit cells is a commonly known solution for improving the metamaterial's performance: boundary effects become less significant and the metamaterial with infinitely large extent is approached with increasing number of unit cells.
   Due to the vibroacoustic coupling, the structures in the present study behave differently in the higher frequency range in presence of the surrounding air.\\
   It must also be taken into consideration, that the available building space is limited in most practical application cases and, therefore, the increase of the unit cell number to a large number is rarely a solution for optimization.
   Another optimization option is to focus on the adjacent structures such that a part of the vibration energy is prevented from entering in the metamaterial. 
   In the present study, we show that a stiffness change of the structure on the input side can lead to the energy deforming the latter instead of entering the metamaterial. 
   One strategy could, thus, be to mitigate vibrations on the input side structure provided this part is not a structural weak point or should not move.
   This approach is efficient in frequency ranges where the metamaterial is not effective, because in the band gap we want the energy to be transferred to the metamaterial for vibration mitigation.
   The adjacent structures should, thus, be included in the optimization process of composed sandwich structures.
   In Table~\ref{tab:OptStra} we summarize the optimization approaches for the different components of an integrated metastructure in two different strategies depending on the frequency range, i.e. depending on whether the frequencies are in a band gap or out of a band gap.
The approaches we give here consider the energy transfer via the specimen. 
     It can also be seen as an impedance mismatch: in the frequency range where the metamaterial can mitigate vibrations energy the goal is to transfer as much vibration energy as possible to the metamaterial. In the frequency ranges where the material is not effective, the goal is to prevent this energy transmission. The design of the interface is dependent on the metamaterial geometry but the impedance mismatch is required in every case. To transmit energy we need to couple our structure better to a vibrating structure (``reinforce connection'') and to prevent transmission we need to uncouple the metamaterial as much as possible from a vibrating structure (``weaken connection''). 
     From our point of view these are macroscopic properties which are not geometry dependent, i.e. the principal work for other geometries including locally resonant metamaterials.
It has already been shown by our group that interfaces affect the vibration response of metamaterial specimens that rely mainly on local resonance~\cite{2024_perez}.
 
\renewcommand{\arraystretch}{1.2} 
\begin{table}[h!]
	\caption{Summary of the optimization strategies for vibration mitigation using the different components of an integrated metamaterial unveiled in this paper.}
	\begin{tabular}{p{2cm}|p{3.75cm}|p{4.5cm}|p{4.5cm}}
	& \multicolumn{1}{|c|}{\textbf{Metamaterial}} & \multicolumn{1}{|c|}{\textbf{Interfaces}} & \multicolumn{1}{|c}{\textbf{Adjacent structure }}\\	
	\hline
	Optimization strategy \textbf{in the band gap}& Increasing the number of unit cells for stringer vibration mitigation in band gap & Reinforce connection between metamaterial and plate for higher energy transmission to metamaterial\\
	\hline
	Optimization strategy \textbf{out of the band gap} && Weaken connection between metamaterial and plate for lower energy transmission to metamaterial & Design adjacent structure such that the vibration energy remains confined in vibrations of adjacent structure and is not transmitted to metamaterial
	\end{tabular}
	\label{tab:OptStra}
\end{table}

Another question that arises from the present investigation concerns the influence of manufacturing constraints and imperfections on the dynamic properties of the structures. During the masked stereo-lithography printing, effects like an uneven exposure to the UV light during the printing or an uneven layer thickness can generate stress gradients and lead to a variation of the mechanical properties or structural deformations. A variation of the geometry can lead to frequency shifts in the structural response as the mass and stiffness distribution changes when the geometry is altered. This is what we see for the $\beta$ specimens (cf. Figure~\ref{fig:calibCompareModelExp}). Another phenomenon which can be caused by imperfections are local modes, i.e. standing waves in a restricted sub-region of the metamaterial that perturb the existence of band gaps. Irregular geometries can lead to localized vibration modes with an amplitude that decreases only slowly with the distance to the localization region (cf. \cite{2012_filoche} for example), thus, transmitting vibration energy even in a band gap. 
Imperfections are inevitable in any manufacturing process; the key question is how strongly they influence the performance of the final structure. To address this point, a model validation is necessary. In the design phase, different interface designs can also be evaluated for their robustness regarding the manufacturing constraints to identify which interfaces (material and free) can be produced with minimal limitations.

   \section{Conclusions}
  
   This article intends to drive forward the design and optimization of finite-sized metamaterial structures. Therefore, we analyze in detail how the following parameters influence the vibroacoustic behavior of an example structure:
  
   \begin{itemize}[noitemsep, topsep=0pt]
       \item \textbf{Boundaries/material interfaces}: in finite-sized structures, the way we cut a unit cell from a metamaterial determines the free boundaries and the material interfaces with adjacent materials. We show the influence of the boundary and interface conditions on the local and global dynamic response of an example structure in experiments and numerical models. A stronger connection between the metamaterial and the adjacent structure, i.e. a larger material interface, improves the energy transmission from the adjacent structure to the metamaterial and avoids local resonances of the adjacent structure, which is useful for vibration damping in the band gap region.
       \item \textbf{Vibroacoustic coupling}: we observe a vibroacoustic coupling in the metamaterial sandwich structure that limits the vibration mitigation performance at higher frequencies in the considered range. 
       A part of the vibration energy can be transmitted by the surrounding air, either directly between the input and output side or such that it amplifies the movement of the structure. As a result of the coupling, the performance differences between the structures with different cuts decreases and the improvement of the vibration damping performance with an increasing number of unit cells is bounded. However, vibroacoustic structures can be desirable in some application cases when acoustic resonances can be used to add additional band gaps or to counteract mechanical resonances by trapping the energy. Structural optimization considering the coupled vibroacoustic behavior will be part of future studies.
       \item \textbf{Adjacent structure}: the structures adjacent to the metamaterial can be designed in order to mitigate a part of the incoming vibration energy. In our example case, a part of the incoming energy is used to deform the structure on the input side instead of being transferred to the metamaterial. In this case, the optimization parameters are the connectivity between metamaterial and adjacent structure at the material interface and the stiffness of the structure on the input side. We find that a weak connection~--~i.e. a small common material interface between metamaterial and adjacent structure~--~and a low stiffness of the structure on the input side lead to higher amount of energy in the input structure and a lower transmission to the metamaterial.
       This mechanism is useful for vibration damping in frequency ranges outside of the band gap region. In some cases, the structure that is needs to be shielded from vibration might not be modifiable. In this case, one could also think of the “adjacent structure” as an additional intermediate structure between the vibrating object and the metamaterial. How to design this third body in an interesting open question for future research.
   \end{itemize}
  As future short-term goals we will further study the vibro-acoustic coupling in our metamaterial structure so that we will be able to also optimize a structure from an acoustic point of view.
   In the long term, we want to up-scale our design procedure towards homogenized large scale metamaterial structures while keeping all of the desirable effects unveiled in this paper. 
These effects should be considered when a non-local behavior~--~originating from internal degrees of freedom~--~is present in the metamaterial. In this case, the finite dimensional response will be dependent on the cell topology and suitable boundary conditions need to be applied.
To achieve this objective, we will employ homogenization models based on enriched continua descriptions. 
More specifically we will use the relaxed micromorphic model~\cite{2022_voss,2022_demore} where the metamaterial is modeled as an equivalent continuum and where boundary effects can be taken into account~\cite{2024_demetriou,2024_perez}.

\vspace{1cm}
  

\noindent  {\small CRediT author statement.} {\footnotesize \textbf{Svenja Hermann}: Conceptualization, Methodology, Investigation, Software, Writing – original draft. 
\textbf{Kévin Billon}: Methodology, Investigation, Formal analysis.
\textbf{Manuel Collet}: Methodology, Supervision, Writing – review \& editing.
\textbf{Angela Madeo}: Funding acquisition, Conceptualization, Supervision, Writing – review \& editing.}

\noindent  {\small Data Accessibility.} {\footnotesize The results obtained from the experiments and from the numerical models can be provided upon request.}

\noindent   {\small Declaration of generative AI use.} {\footnotesize The authors declare that generative AI tools were not used for content or idea generation, data analysis or any other substantial contribution to this work.}

\noindent   {\small Conflict of interests.} {\footnotesize The authors declare no potential conflicts of interest with respect to the research, authorship, and publication of this article.}

\noindent   {\small Acknowledgments.} {\footnotesize This work was supported by the European Commission with the ERC Consolidator Grant META-LEGO [N$^\text{o}$ 101001759].}

\newpage

{
\fontsize{10}{11}\selectfont    
\bibliographystyle{unsrt}
\bibliography{references}

@article{2025_chaplain,
  title={The 2024 Acoustic Metamaterials Roadmap},
  author={Chaplain, G.J. and Langfeldt, F. and Romero-Garc{\'\i}a, V. and Jim{\'e}nez, N. and Meng, Y. and Groby, J.-P. and Pagneux, V. and Moore, D. and Hibbins, A.P. and Sambles, J.R. and Starkey, T.A. and Popa, B.-I. and Zhang, Z. and Christensen, J. and Wen, X. and Li, J. and Fleury, R. and Wallen, S.P. and Haberman, M.R. and Hussein, M.I. and Memoli, G. and Fusaro, G. and D’Orazio, D. and Barbaresi, L. and Garai, M. and Chisari, L. and Mittal, P. and Qi, Z. and Subramanian, S. and Br{\^u}l{\'e}, S. and Enoch, S. and Guenneau, S. and Stoakes, R. and McLean, C. and Gardiner, A.},
  journal={Journal of Physics D: Applied Physics},
  volume={58},
  pages={433001},
  year={2025},
  note ={doi:~\href{{https://doi.org/10.1088/1361-6463/add306}}{10.1088/1361-6463/add306}}
}

@article{2024_cool,
  title={A guide to numerical dispersion curve calculations: Explanation, interpretation and basic Matlab code},
  author={Cool, V. and Deckers, E. and Van Belle, L. and Claeys, C.},
  journal={Mechanical Systems and Signal Processing},
  volume={215},
  pages={111393},
  year={2024},
  publisher={Elsevier},
  note ={doi:~\href{{https://doi.org/10.1016/j.ymssp.2024.111393}}{10.1016/j.ymssp.2024.111393}}  
}

@article{2024_demetriou,
  title={Reduced relaxed micromorphic modeling of harmonically loaded metamaterial plates: investigating boundary effects in finite-size structures},
  author={Demetriou, P. and Rizzi, G. and Madeo, A.},
  journal={Archive of Applied Mechanics},
  volume={94},
  number={1},
  pages={81--98},
  year={2024},
  publisher={Springer},
  note ={doi:~\href{{https://doi.org/10.1007/s00419-023-02509-x}}{10.1007/s00419-023-02509-x}}
}

@article{2024_hermann,
  title={Design and experimental validation of a finite-size labyrinthine metamaterial for vibro-acoustics: enabling upscaling towards large-scale structures},
  author={Hermann, S. and Billon, K. and Parlak, A.M. and Orlowsky, J. and Collet, M. and Madeo, A.},
  journal={Philosophical Transactions A},
  volume={382},
  number={2278},
  pages={20230367},
  year={2024},
  publisher={The Royal Society},
  note ={doi:~\href{{https://doi.org/10.1098/rsta.2023.0367}}{10.1098/rsta.2023.0367}}
}

@article{2024_liang,
  title = {Design of broad quasi-zero stiffness platform metamaterials for vibration isolation},
  author = {Liang, K. and Jing, Y. and Zhang, X.},
  journal = {International Journal of Mechanical Sciences},
  volume = {281},
  pages = {109691},
  year = {2024},
  note ={doi:~\href{{https://doi.org/10.1016/j.ijmecsci.2024.109691}}{10.1016/j.ijmecsci.2024.109691}}
}

@article{2024_perez,
	title = {Effective surface forces and non-coherent interfaces within the reduced relaxed micromorphic modeling of finite-size mechanical metamaterials},
	author = {Perez Ramirez, L.A. and Erel-Demore, F. and Rizzi, G. and Voss, J. and Madeo, A.},
	journal = {Journal of the Mechanics and Physics of Solids},
	volume = {186},
	pages = {105558},
	year = {2024},
	issn = {0022-5096},
	note ={doi:~\href{{https://doi.org/10.1016/j.jmps.2024.105558}}{10.1016/j.jmps.2024.105558}}
}

@article{2024_salAnglada,
  title={Sound transmission loss enhancement through triple-peak coupled resonances acoustic metamaterials},
  author={Sal-Anglada, G. and Yago, D. and Cante, J. and Oliver, J. and Roca, D.},
  journal={International Journal of Mechanical Sciences},
  volume={266},
  pages={108951},
  year={2024},
  publisher={Elsevier},
  note ={doi:~\href{{https://doi.org/10.1016/j.ijmecsci.2023.108951}}{10.1016/j.ijmecsci.2023.108951}}
}

@article{2024_valappil,
  title={Multi-objective design of 3D phononic crystal waveguide by design space trimming},
  author={Valappil, S.V. and Goosen, J.F.L. and Arag{\'o}n, A.M.},
  journal={Materials \& Design},
  volume={237},
  pages={112594},
  year={2024},
  publisher={Elsevier},
  note ={doi:~\href{{https://doi.org/10.1016/j.matdes.2023.112594}}{10.1016/j.matdes.2023.112594}}
}

@misc{2023_anycubic,
  author = {Anycubic},
  title = {{User Guide for Standard Resin}},
  howpublished = "\href{{https://cdn.shopify.com/s/files/1/0245/5519/2380/files/Anycubic_Standard_Resin_User_Manual_V1.0-EN_1.pdf?v=1663574587}}{Anycubic{\_}Standard{\_}Resin{\_}User{\_}Manual{\_}V1.0-EN{\_}1.pdf}",
  year = {2023}, 
  note = {[Online; accessed 13-April-2023]}
}

@article{2022_demore,
  title={{Unfolding engineering metamaterials design: Relaxed micromorphic modeling of large-scale acoustic meta-structures}},
  author={Demore, F. and Rizzi, G. and Collet, M. and Neff, P. and Madeo, A.},
  journal={Journal of the Mechanics and Physics of Solids},
  volume={168},
  pages={104995},
  year={2022},
  publisher={Elsevier},
  note ={doi:~\href{{https://doi.org/10.1016/j.jmps.2022.104995}}{10.1016/j.jmps.2022.104995}}
}

@article{2022_holliman,
  title={Review of foundational concepts and emerging directions in metamaterial research: design, phenomena, and applications},
  author={Holliman, J.E. and Schaef, H.T. and McGrail, B.P. and Miller, Q.R.S.},
  journal={Materials Advances},
  volume={3},
  number={23},
  pages={8390--8406},
  year={2022},
  publisher={Royal Society of Chemistry},
  note ={doi:~\href{{https://doi.org/10.1039/d2ma00497f}}{10.1039/d2ma00497f}}  
}

@article{2022_kyaw,
  title={{A Metamaterial Solution for Soundproofing on Board Ship}},
  author={D’Amore, G.K.O. and Caverni, S. and Biot, M. and Rognoni, G. and D’Alessandro, L.},
  journal={Applied Sciences},
  volume={12},
  number={13},
  pages={6372},
  year={2022},
  publisher={MDPI},
  note ={(doi:~\href{{https://doi.org/10.3390/app12136372}}{10.3390/app12136372})}
}

@article{2022_li,
	title = {Topological design of phononic crystals for multiple wide band gaps},
	author = {Li, Y. and Luo, Y. and Zhang, X.},
	journal = {Journal of Sound and Vibration},
	volume = {529},
	pages = {116962},
	year = {2022},
  	note ={doi:~\href{{https://doi.org/10.1016/j.jsv.2022.116962}}{10.1016/j.jsv.2022.116962}}
}

@article{2022_mercer,
	title = {Effects of geometry and boundary constraint on the stiffness and negative Poisson's ratio behaviour of 	auxetic metamaterials under quasi-static and impact loading},
	author = {Mercer, C. and Speck, T. and Lee, J. and Balint, D.S. and Thielen, M.},
	journal = {International Journal of Impact Engineering},
	volume = {169},
	pages = {104315},
	year = {2022},
  	note ={doi:~\href{{https://doi.org/10.1016/j.ijimpeng.2022.104315}}{10.1016/j.ijimpeng.2022.104315}}
}

@article{2022_voss,
  title={{Modeling a labyrinthine acoustic metamaterial through an inertia-augmented relaxed micromorphic approach}},
  author={Voss, J. and Rizzi, G. and Neff, P. and Madeo, A.},
  journal={Mathematics and Mechanics of Solids},
  pages={10812865221137286},
  year={2022},
  publisher={SAGE Publications Sage UK: London, England},
  note ={doi:~\href{{https://doi.org/10.1177/10812865221137286}}{10.1177/10812865221137286}}
}

@article{2021_gazzola,
  title={{From mechanics to acoustics: Critical assessment of a robust metamaterial for acoustic insulation application}},
  author={Gazzola, C. and Caverni, S. and Corigliano, A.},
  journal={Applied Acoustics},
  volume={183},
  pages={108311},
  year={2021},
  publisher={Elsevier},
  note ={doi:~\href{{https://doi.org/10.1016/j.apacoust.2021.108311}}{10.1016/j.apacoust.2021.108311}}
}

@article{2021_hu,
  title={Acoustic-elastic metamaterials and phononic crystals for energy harvesting: A review},
  author={Hu, G. and Tang, L. and Liang, J. and Lan, C. and Das, R.},
  journal={Smart Materials and Structures},
  volume={30},
  number={8},
  pages={085025},
  year={2021},
  publisher={IOP Publishing},  
  note ={doi:~\href{{https://doi.org/10.1088/1361-665X/ac0cbc}}{10.1088/1361-665X/ac0cbc}}
}

@book{2021_jimenez,
  title={{Acoustic Waves in Periodic Structures, Metamaterials, and Porous Media}},
  author={Jim{\'e}nez, N. and Groby, J.P. and Romero-Garc{\'\i}a, V.},
  year={2021},
  publisher={Springer}  
}

@inproceedings{2021_riess,
  title={{Vibroacoustic Metamaterials for enhanced acoustic Behavior of Vehicle Doors}},
  author={Riess, S. and Droste, M. and Manushyna, D. and Melzer, S. and Druwe, T. and Georgi, T. and Atzrodt, H.},
  booktitle={2021 Fifteenth International Congress on Artificial Materials for Novel Wave Phenomena (Metamaterials)},
  pages={374--376},
  year={2021},
  organization={IEEE},
  note ={(doi:~\href{{https://doi.org/10.1109/Metamaterials52332.2021.9577065}}{10.1109/Metamaterials52332.2021.9577065})}
}

@article{2021_zhou,
  title={A novel hybrid composite phononic crystal plate with multiple vibration band gaps at low frequencies},
  author={Zhou, P. and Wan, S. and Wang, X. and Fu, J. and Zhu, Y.},
  journal={Physica B: Condensed Matter},
  volume={623},
  pages={413366},
  year={2021},
  publisher={Elsevier},  
  note ={doi:~\href{{https://doi.org/10.1088/1361-665X/ac0cbc}}{10.1088/1361-665X/ac0cbc}}
}

@article{2020_sangiuliano,
  title = {Influence of boundary conditions on the stop band effect in finite locally resonant metamaterial beams},
  author = {Sangiuliano, L. and Claeys, C. and Deckers, E. and Desmet, W.},
  journal = {Journal of Sound and Vibration},
  volume = {473},
  pages = {115225},
  year = {2020},
  note ={doi:~\href{{https://doi.org/10.1016/j.jsv.2020.115225}}{10.1016/j.jsv.2020.115225}}
}

@article{2019_kumar,
  title={The present and future role of acoustic metamaterials for architectural and urban noise mitigations},
  author={Kumar, S. and Lee, H.P.},
  journal={Acoustics},
  volume={1},
  number={3},
  pages={590--607},
  year={2019},
  organization={MDPI},
  note ={doi:~\href{{https://doi.org/10.3390/acoustics1030035}}{10.3390/acoustics1030035}}
}

@article{2021_laude,
  title={Principles and properties of phononic crystal waveguides},
  author={Laude, V.},
  journal={Apl Materials},
  volume={9},
  number={8},
  year={2021},
  publisher={AIP Publishing},
  note ={doi:~\href{{https://doi.org/10.1063/5.0059035}}{10.1063/5.0059035}}
}

@article{2020_coulais,
  title={Observation of non-Hermitian topology and its bulk–edge correspondence in an active mechanical metamaterial},
  author={Ghatak, A. and Brandenbourger, M. and Van Wezel,  J. and Coulais, C.},
  journal={Proceedings of the National Academy of Sciences},
  volume={117},
  number={47},
  pages = {29561-29568},
  year={2020},
  note ={doi:~\href{https://doi.org/10.1073/pnas.2010580117}{10.1073/pnas.2010580117}}
}

@article{2019_deMelo,
  title={{Realisation of a thermoformed vibro-acoustic metamaterial for increased STL in acoustic resonance driven environments}},
  author={de Melo Filho, N.G.R and Claeys, C. and Deckers, E. and Desmet, W.},
  journal={Applied Acoustics},
  volume={156},
  pages={78--82},
  year={2019},
  publisher={Elsevier},
  note ={doi:~\href{{https://doi.org/10.1016/j.apacoust.2019.07.007}}{10.1016/j.apacoust.2019.07.007}}
}

@article{2019_pal,
  title={{Topologically protected edge states in mechanical metamaterials}},
  author={Pal, R.K. and Vila, J. and Ruzzene, M.},
  journal={Advances in Applied Mechanics},
  volume={52},
  pages={147-181},
  year={2019},
  note ={doi:~\href{https://doi.org/10.1016/bs.aams.2019.04.001}{10.1016/bs.aams.2019.04.001}}
}

@book{2019_romeroGarcia,
  title={{Fundamentals and applications of acoustic metamaterials: from seismic to radio frequency}},
  author={Romero-Garcia, V. and Hladky-Hennion, A.C.},
  year={2019},
  publisher={John Wiley \& Sons}
}

@article{2019_zangeneh,
  title={{Active times for acoustic metamaterials}},
  author={Zangeneh-Nejad, F. and Fleury, R.},
  journal={Reviews in Physics},
  volume={4},
  pages={100031},
  year={2019},
  publisher={Elsevier},
  note ={doi:~\href{{https://doi.org/10.1016/j.revip.2019.100031}}{10.1016/j.revip.2019.100031}}
}

@book{2018_gan,
  title={{New acoustics based on metamaterials}},
  author={Gan, W.S.},
  year={2018},
  publisher={Springer}
}

@inproceedings{2018_billon,
  title={Design of smart metamaterials for vibration control: extension of Bloch approach to handle finite system boundary conditions},
  author={Billon, K. and Ouisse, M. and Collet, M. and Sadoulet-Reboul, E.},
  booktitle={Health Monitoring of Structural and Biological Systems XII},
  volume={10600},
  pages={384--392},
  year={2018},
  organization={SPIE},
  note ={doi:~\href{{https://doi.org/10.1117/12.2300513}}{10.1117/12.2300513}}
}

@article{2016_Dalessandro,
  title={{Modeling and experimental verification of an ultra-wide bandgap in 3D phononic crystal}},
  author={{D'Alessandro, L. and Belloni, E. and Ardito, R. and Corigliano, A. and Braghin, F.}},
  journal={Applied Physics Letters},
  volume={109},
  number={22},
  pages={221907},
  year={2016},
  note ={doi:~\href{{https://doi.org/10.1063/1.4971290}}{10.1063/1.4971290}}
}

@book{2015_blevins,
  title={Formulas for dynamics, acoustics and vibration},
  author={Blevins, Robert D},
  year={2015},
  publisher={John Wiley \& Sons}
}

@article{2015_hvatov,
  title={{Free vibrations of finite periodic structures in pass-and stop-bands of the counterpart infinite waveguides}},
  author={{Hvatov, A. and Sorokin, S.}},
  journal={Journal of Sound and Vibration},
  volume={347},
  pages={200-217},
  year={2015},
  note ={doi:~\href{https://doi.org/10.1016/j.jsv.2015.03.003}{10.1016/j.jsv.2015.03.003}}
}

@article{2012_filoche,
  title={{Universal mechanism for Anderson and weak localization}},
  author={{Filoche, M. and Mayboroda, S.}},
  journal={Proceedings of the National Academy of Sciences},
  volume={109},
  number={37},
  pages={14761-14766},
  year={2012},
  note ={doi:~\href{https://doi.org/10.1073/pnas.1120432109}{10.1073/pnas.1120432109}}
}

@article{2011_collet,
  title={Floquet--Bloch decomposition for the computation of dispersion of two-dimensional periodic, damped mechanical systems},
  author={Collet, M. and Ouisse, M. and Ruzzene, M. and Ichchou, M.N.},
  journal={International Journal of Solids and Structures},
  volume={48},
  number={20},
  pages={2837--2848},
  year={2011},
  publisher={Elsevier}
}

@book{1998_fassold,
  title={{Schallschutz und Raumakustik in der Praxis, 1}},
  author={Fassold, W. and Veres, E.},
  publisher={Berlin: Verlag f{\"u}r Bauwesen},
  year={1998}
}
}


 \appendix
 
\section*{Appendix}

\renewcommand{\thefigure}{\thesection.\arabic{figure}}
\renewcommand{\thetable}{\thesection.\arabic{table}}

\setcounter{figure}{0}
\setcounter{table}{0}

\section{Materials and Methods} 
\label{app:A}

\vspace{-0.25cm}
\subsection{Printing errors observed on the \bet specimens}
\label{app:A1}
 
\begin{figure}[!h]
    \centering
    a) \includegraphics[trim = 0 0 0 0, clip,width=0.17\textwidth]{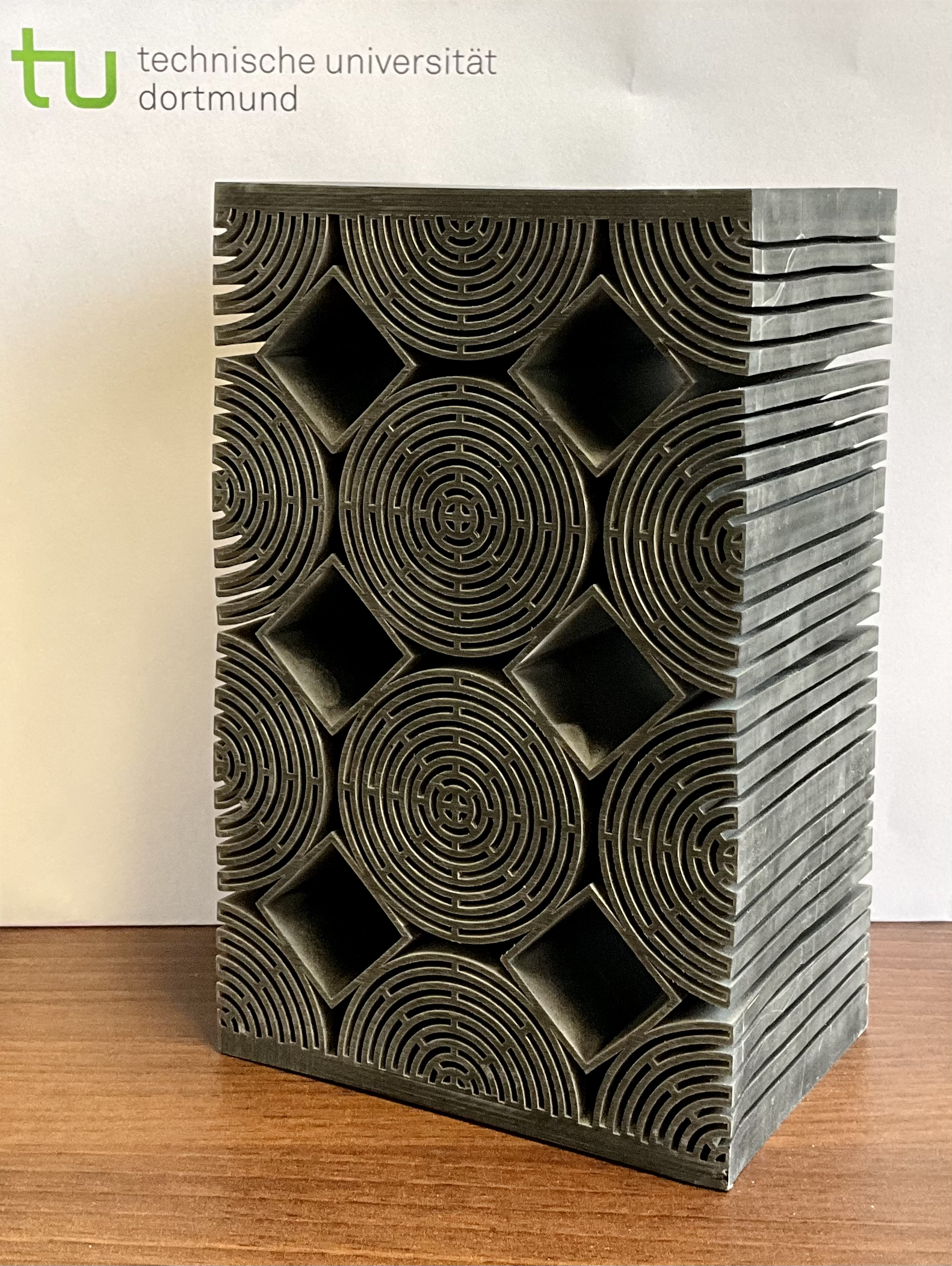} \qquad
    b) \includegraphics[trim = 0 0 0 0, clip,width=0.17\textwidth]{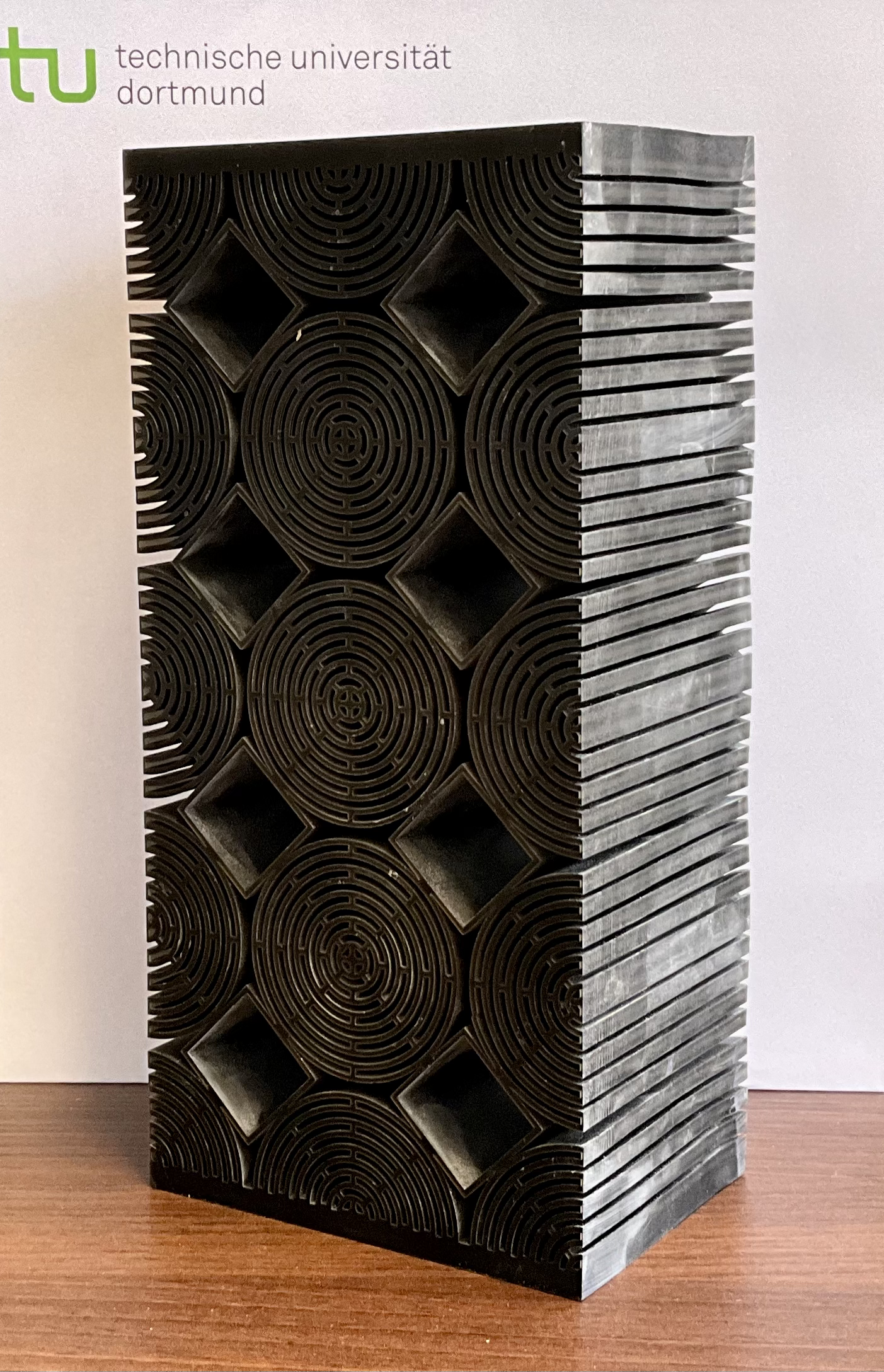}
    \caption{Photograph of the \bet specimens with \conS unit cells (a) and \conL unit cells (b).}
    \label{fig:printingErrors}
\end{figure}

Visible printing errors are observed on the \bet specimens. As it can be seen in Figure~\ref{fig:printingErrors}, the different distances between the lateral bars of both specimens show that the structure has warped during printing.

\vspace{-0.45cm}

\vspace{-0.5cm}
\subsection{Experiment - Uniformity verification of the input acceleration}
\label{app:A2}

\begin{figure}[h!]
    \centering
    \includegraphics[width=\textwidth]{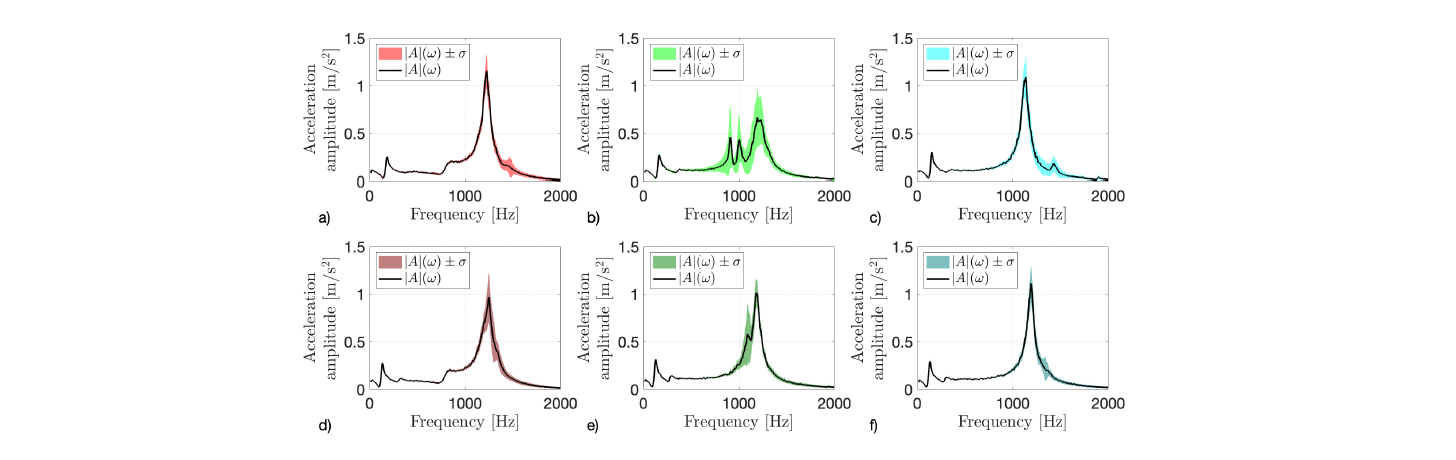}
    \caption{Acceleration amplitudes on the input plate: average measurements results of nine points on the input plate $|A(\omega)|$ and standard deviation of the nine measurements ($\sigma$) obtained for the configuration \alp \conS (a), \bet \conS (b), \gam \conS (c), \alp \conL (d), \bet \conL (e) and \gam \conL (f).}
    \label{fig:inputAccelerationEvaluation}
\end{figure}

The uniformity of the acceleration was tested in the experiment. Therefore, an additional accelerometer was glued on the metal plate on the input side. In eight additional measurements, the acceleration was measured at eight different points in addition to the measurement in the center. The eight positions were chosen such that the points on the outer edge of the input plate of the specimen--not the metal plate-- are covered. Figure~\ref{fig:inputAccelerationEvaluation} shows that the acceleration amplitude varies only slightly over the area of the input plate, especially for the \alp and the \gam specimen. 

\subsection{Experiment - Number of measurement points on the output plate}
\label{app:A3}
\begin{figure}[!h]
    \centering
    \includegraphics[width=\textwidth]{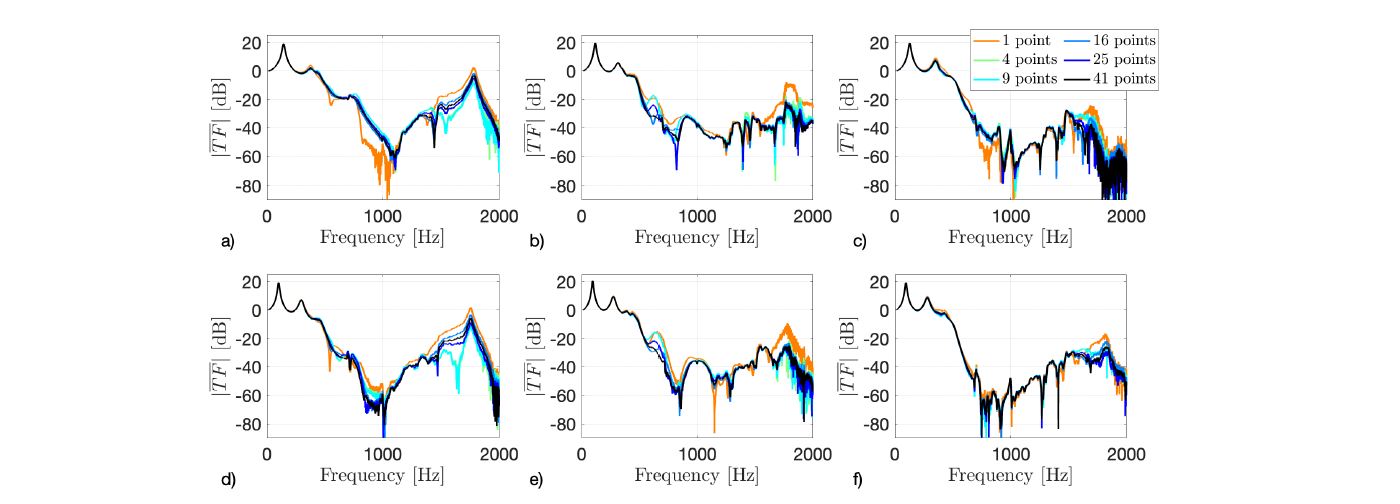}
    \caption{Amplitude of the average transfer function $|\overline{TF}|$ obtained from the experiments for a different number of measurement points on the output plate of the small specimens (\conS unit cells) in configurations \alp (a), \bet (b) and \gam (c) of the big specimens (\conL unit cells) in configurations \alp (d), \bet (e) and \gam (f).}
    \label{fig:apxNevaluationPoints}
\end{figure}

The number of measurement points on the output plate needs to be sufficiently high to approximate the average response of the plate. In Figure~\ref{fig:apxNevaluationPoints}, we show the amplitude of the average transfer function $|\overline{TF}|$ for which the average output acceleration was calculated for six different numbers of measurement points. The result converges with an increasing number of measurement points and the figure shows, that a number of 41 measurement points is sufficiently large.

\vspace{-0.5cm}
\subsection{Modeling - average vs. pointwise results}
\label{app:A4}
\begin{figure}[!h]
    \centering
    \includegraphics[width=\textwidth]{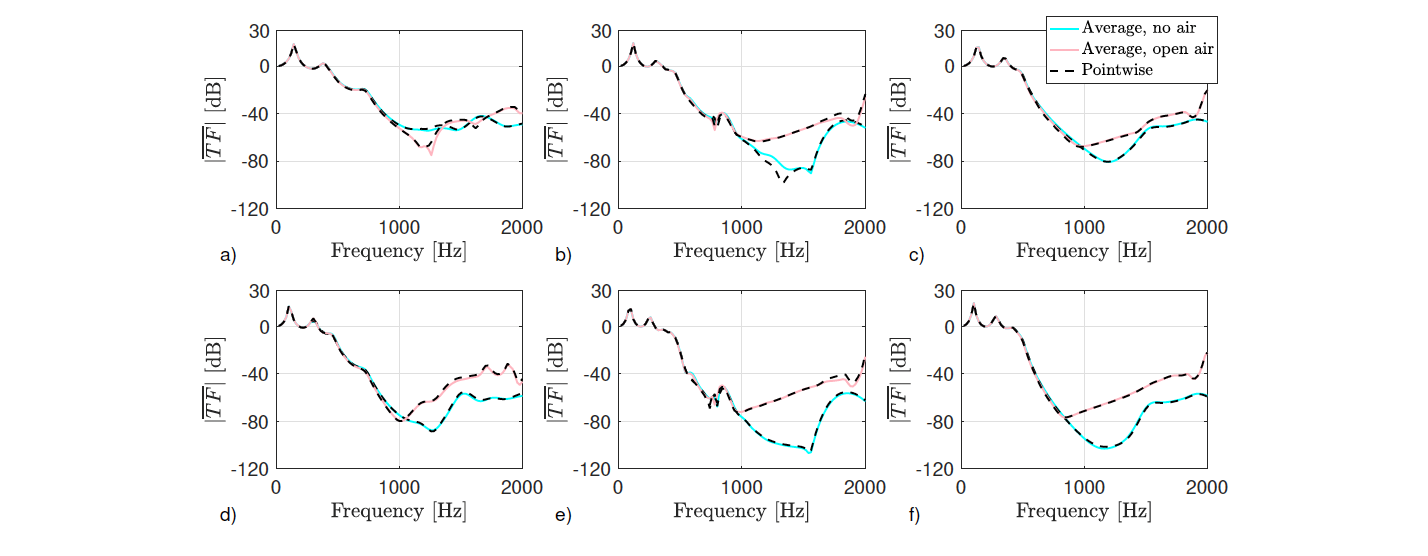}
    \caption{Amplitude of the average transfer function $|\overline{TF}|$ obtained from numerical simulations using the surface average of the acceleration over all mesh points on the output plate (``Average") and for the 41 measurement points used in the experiment (``pointwise"). The results are presented for the small specimens (\conS unit cells) in configurations \alp (a), \bet (b) and \gam (c) of the big specimens (\conL unit cells) in configurations \alp (d), \bet (e) and \gam (f).}
    \label{fig:apxAvVsPtw}
\end{figure}

In the numerical simulations, the spatial average of the acceleration can directly be obtained from the software, while an evaluation on 41 measurement points involves additional post-processing steps. Figure~\ref{fig:apxAvVsPtw} shows, that the difference in the results is insignificant between the two methods. We, therefore, use the average acceleration calculated over the entire plate surface to calculate the average transfer function.

\vspace{-0.5cm}
\subsection{Comparison metric - performance indicator $Q$}
\label{app:A5}

We compare the performances of the specimens regarding the transmission loss with a metric similar to the $H_2$ norm. In frequency domain with angular frequency $\omega$, the $H_2$ norm is defined for a transfer function $TF$ of a system (ratio of output over input quantity) considering symmetry and the relation $\omega = 2 \pi f$ as   
          \begin{equation}
          	||TF||_{H_2} = \sqrt{\frac{1}{2\pi} \int\limits_{- \infty}^{\infty} |TF(\omega)|^2 \;  \text{d}\omega} = \sqrt{2 \int\limits_{0}^{\infty} |TF(f)|^2 \; \text{d}f} \, .
          	\label{eq:metric1}
          \end{equation}
          In the present study, we obtain results between \SI{20}{\hertz} and \SI{2000}{\hertz} in steps of $\Delta f = 20$~Hz. Considering the above equation, two approximations need to be made: we need to approximate the integral in the above equation and limit the integral to the frequency band in which we obtained our results. 
          We assume a linear variation of the function between two neighboring frequencies $f_n$ and $f_{n+1}$ such that the integral becomes 
          \begin{equation}
	          \int\limits_{f_n}^{f_{n+1}} |TF|^2 \approx \frac{|TF(f_n)|^2 + |TF(f_{n+1})|^2}{2} \,  \Delta f 
          	\label{eq:metric3}
          \end{equation}
	To integrate over a frequency band consisting of $N$ consecutive measurement points, we replace the integral in Equation~\eqref{eq:metric1} by a sum over $N$ and obtain a quantity which represents an approximation of the $H_2$ norm for the limited frequency band:
	\begin{equation}
          	||TF||_{H_2} \approx \sqrt{\sum\limits_{n=1}^{N-1} \; \left( |TF(f_n)|^2 + |TF(f_{n+1})|^2 \right) \Delta f}  \; .	
          	\label{eq:metric4}
	\end{equation}
	 We want to use establish a metric with linear quantity that represents the transmission loss $TL_{\perp}$. The transmission loss, however, is a logarithmic quantity as shown in its definition in Equation~(2.14) in the manuscript. Therefore, we calculate the sound power ratio $P_r$ between input and output sound power of a specimen $A$~--~which corresponds to the term in the brackets in (2.14) in the manuscript~--~from the transmission loss as
	 \begin{equation}
	 P_r^A = 10^{\sfrac{TL_{\perp}(A)}{10}} 
          	\label{eq:metric5}
	 \end{equation}
	 We use this expression to define our comparison metric $Q$ between a specimen $A$ and a specimen $B$ on a frequency interval $[f_1 \; f_2]$ by
	 \begin{equation}
	 Q(A,B)_{f_1}^{f_2} = \frac{\sqrt{\sum\limits_{n=1}^{N-1} \; \left( |P_r^A(f_n)|^2 + |P_r^A(f_{n+1})|^2 \right) \Delta f}}{\sqrt{\sum\limits_{n=1}^{N-1} \; \left( |P_r^B(f_n)|^2 + |P_r^B(f_{n+1})|^2 \right) \Delta f}}
	 \end{equation}
	 with $n$ and $N$ being the indices of $f_1$ and $f_2$  in the frequency vector of results respectively.

\section{Results} 
\label{app:B}

\setcounter{figure}{0}
\setcounter{table}{0}

\subsection{Bloch-Floquet analysis - 2D vs. 3D}
\label{app:B1}
\begin{figure}[!h]
     \centering
     \includegraphics[width=\textwidth]{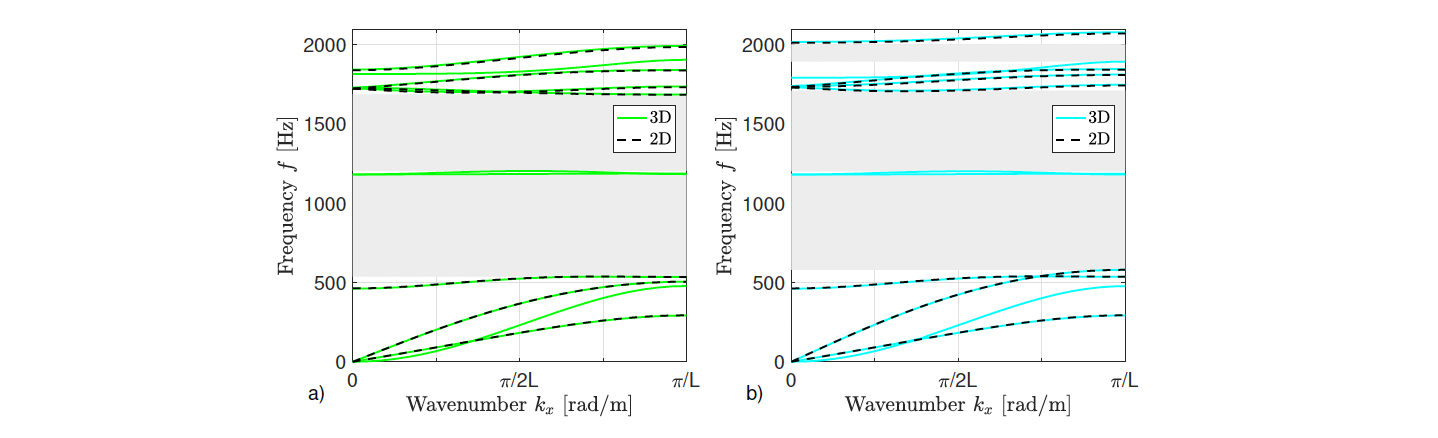}
     \caption{Dispersion curves obtained from numerical simulations with 3D and 2D plane-strain mechanical~(a) and vibroacoustic~(b) models.  $k_x-$wavenumber, $L-$size of the unit cell.}
     \label{fig:dispersionCurves2}
 \end{figure}

The results of the Bloch-Floquet analysis performed with a 2D plane strain and a 3D model of the unit cell are compared in Figure~\ref{fig:dispersionCurves2}. While the in-plane modes are similar, there are four additional dispersion relations in the mechanical and the vibroacoustic model as shown in Figure~\ref{fig:dispersionCurves2}. The corresponding eigenmodes are linked to an out-of-plane movement of the metamaterial in the 3D model.

\subsection{Coherence function obtained for the experimental results}
\label{app:B2}
\begin{figure}[!h]
     \centering
     \includegraphics[width=\textwidth]{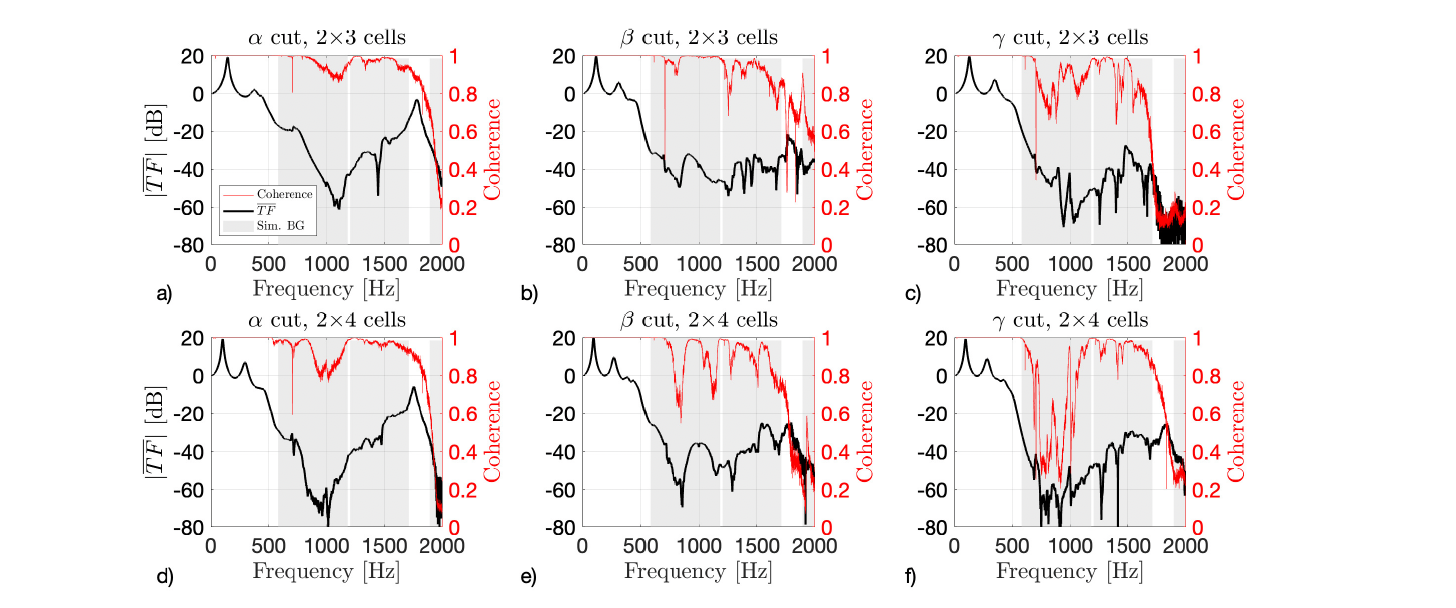}
     \caption{Experimental results: amplitude of the average transfer function $\overline{TF}$ and coherence function between output and input measurement results. The results are shown for the specimens comprising \conS unit cells in configurations \alp (a), \bet (b) and \gam (c), and for specimens comprising \conL unit cells in configurations \alp (d), \bet (e) and \gam (f).}
     \label{fig:apxCoherence}
 \end{figure}

The coherence functions in Figure~\ref{fig:apxCoherence} show, how well the measurement signals on the input side and output side of the specimen are correlated. Values close to 1 indicate a strong correlation and low values indicate, that the signals are either influenced by noise or they indicate the presence of a band gap.

\subsection{Influence of the steel plate on the transfer function obtained from the vibroacoustic models}
\label{app:B3}

\begin{figure}[!h]
     \centering
     \includegraphics[width=\textwidth]{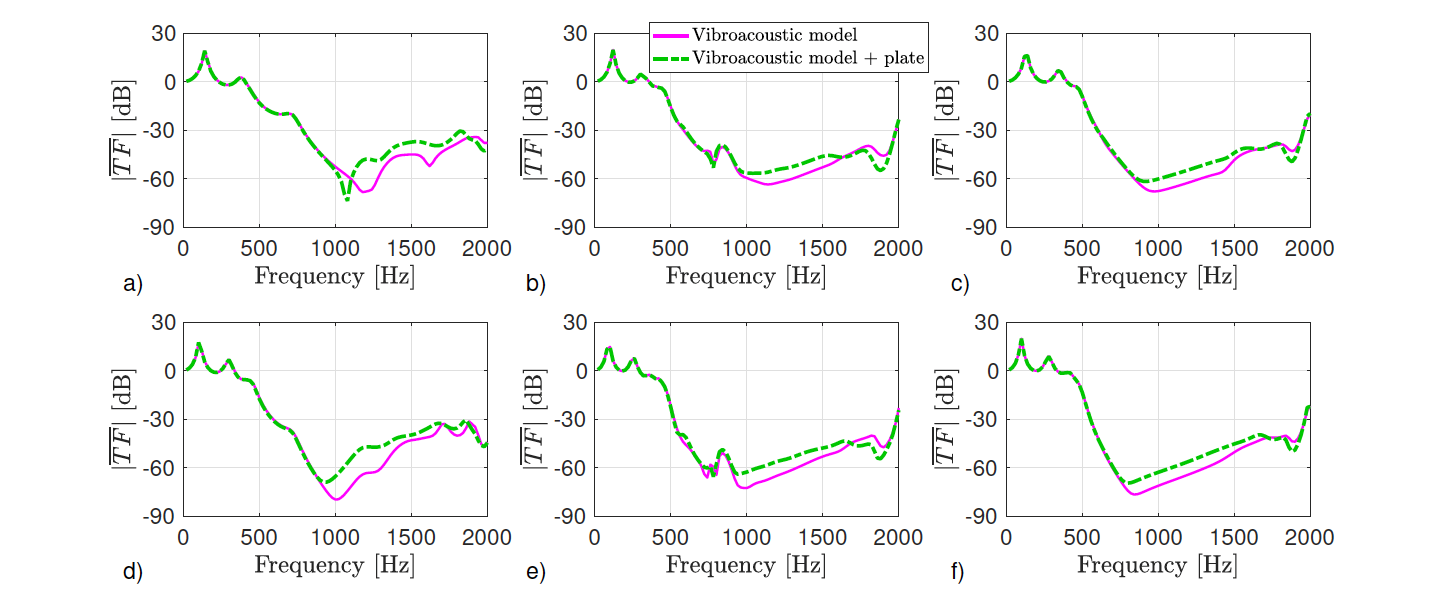}
     \caption{Average transfer function of the metamaterial structures comprising \conS unit cells in configurations \alp (a), \bet (b) and \gam (c), and for specimens comprising \conL unit cells in configurations \alp (d), \bet (e) and \gam (f) obtained from vibroacoustic numerical models.}
     \label{fig:apxPlateNoPlate}
 \end{figure}

Figure~\ref{fig:apxPlateNoPlate} shows the average transfer functions obtained from numerical models including only the specimen and including the additional steel plate used in the experiments. 

\subsection{Experiment - comparison of small and big specimens for each cut}
\label{app:BN4}

\begin{figure}[!h]
         \centering
         \includegraphics[width=\textwidth]{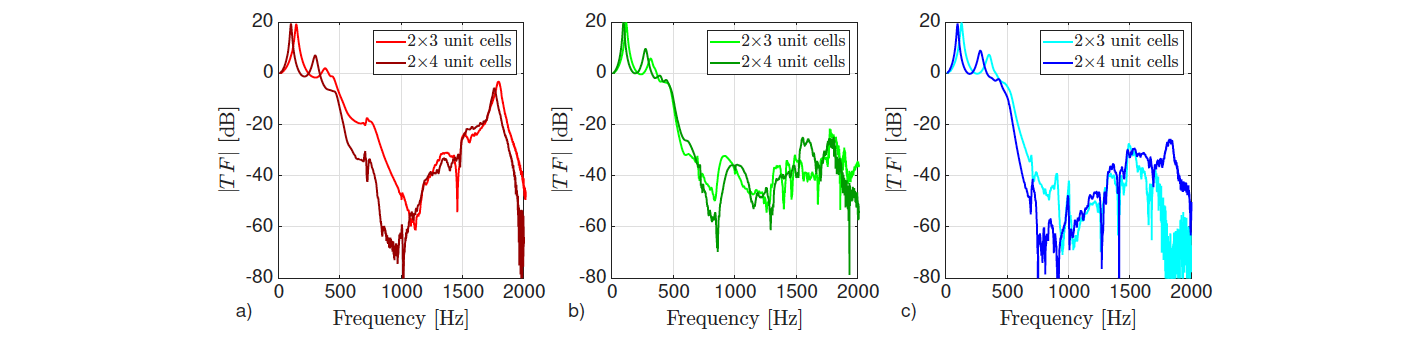}
         \caption{Experimental results: amplitude of the average transfer function, $|TF|$, for the specimen configurations \alp (a), \bet (b) and \gam (c) comprising a different number of unit cells.}
         \label{fig:expPerCut}
     \end{figure}

      In Figure~\ref{fig:expPerCut}, we compare the results of the small and the larger specimens of a single configuration. 
      It shows the transfer functions obtained for specimens comprising \conS and \conL unit cells in configurations \alp (a), \bet (b) and \gam (c) respectively.    
      The specimens comprising \conL unit cells show a higher vibration mitigation capacity~--~i.e. lower values of transfer function~--~in the medium-frequency range. This effect is very pronounced for \alp, less pronounced for \gam and the least pronounced for \bet.
      This is related to the fact that the \alp configuration is the one for which the solid-solid structural connection between the plate and the metamaterial is less effective than for \bet and \gam so that a higher number of unit cells is needed to achieve the structural band gap response.     
      The low-frequency resonances are shifted to even lower frequencies when the number of unit cells increases for a larger number of unit cells.
      In the high-frequency range ($>$\SI{1000}{\hertz}), the responses of the small specimens~--~\conS unit cells~--~and the large specimens~--~\conL unit cells~--~become very similar. 
   
  \subsection{Experiment - comparison of results with models not including plate}
\label{app:BN5}   

\begin{figure}[!h]
             \centering
             \includegraphics[width=\textwidth]{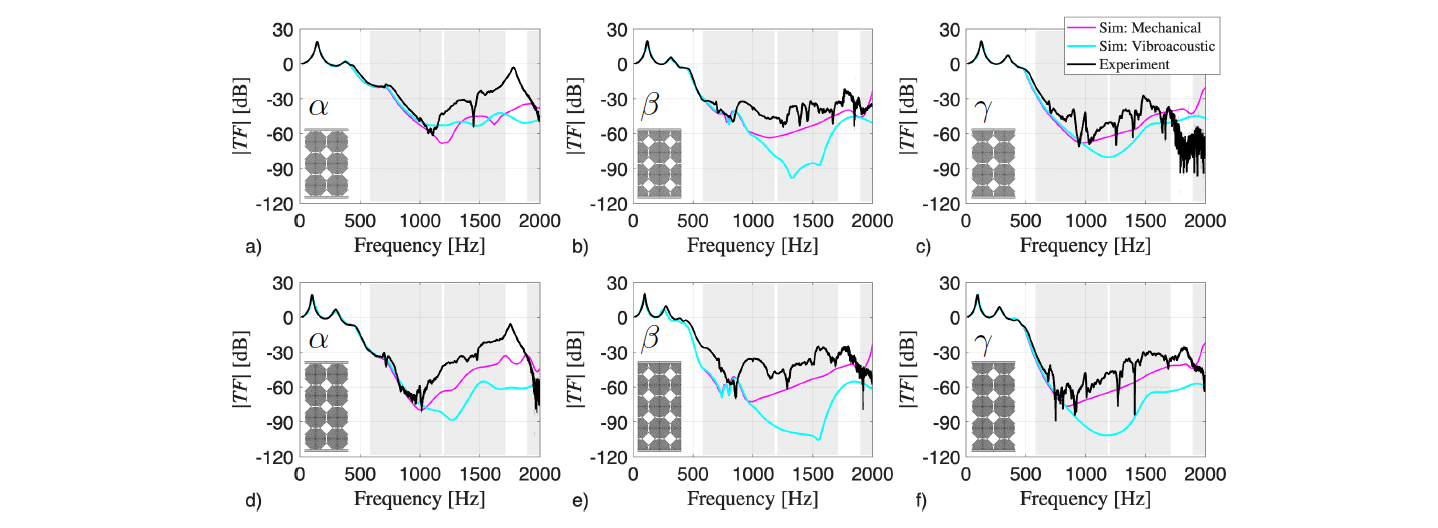}
             \caption{Comparison of experimental and numerical results: amplitude of the average transfer function, $|TF|$, for specimens with \conS unit cells in \alp (a), \bet (b) and \gam (c) configurations and for specimens with \conL unit cells in \alp (d), \bet (e) and \gam (f) configurations.}
             \label{fig:calibCompareModelNoPlate}
         \end{figure}

In Figure~\ref{fig:calibCompareModelNoPlate}, the transfer function obtained from the experimental tests is compared to the average transfer functions obtained with the FE models in which the steel plate is not included and in which we apply a uniform acceleration loading on the input side. The numerical results approximate the response of the structure well in the low- and in parts of the medium frequency range. For the higher frequencies, however, the approximation is not good. This result underlines the importance to include the steel plate for the model validation.

 \subsection{Experiment - Acceleration amplitudes}
\label{app:BN6}   

\begin{figure}[!h]
             \centering
             \includegraphics[width=\textwidth]{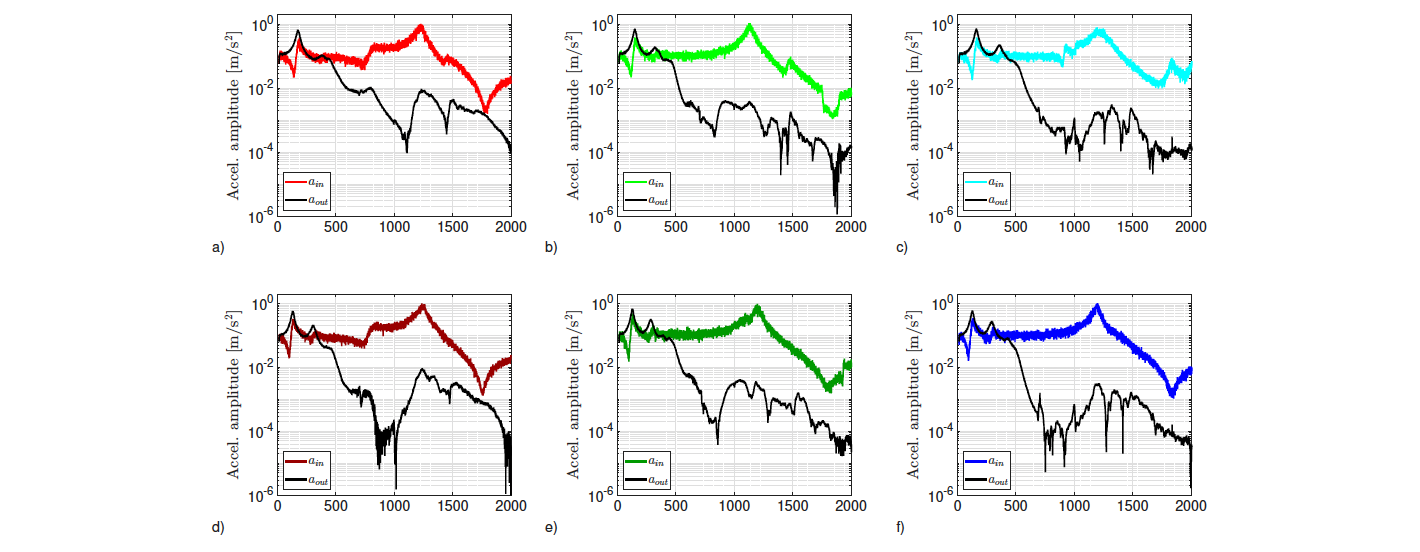}
             \caption{Acceleration amplitude spectra obtained from the experimental measurements: Input acceleration $a_{in}$ (color) and output acceleration $a_{out}$ (black) of the configurations \alp, \bet and \gam obtained for specimens with \conS unit cells (a, b and c respectively) and for specimens with \conL unit cells (d, e and f respectively).}
             \label{fig:accelerationAmplitude}
         \end{figure}

Figure~\ref{fig:accelerationAmplitude} shows the amplitudes of the acceleration spectra obtained for the input side (color) and the output side (black) of the specimen, i.e. with the accelerometer and with the laser Doppler vibrometer. We observe a drop of the input acceleration in the range from \SI{1790}{\hertz} to \SI{1843}{\hertz} for the specimens with \conS unit cells (excluding the \gam configuration for which there were errors in the high frequency measurements) an in a range from \SI{1760}{\hertz} to \SI{1840}{\hertz} for the specimens with \conL unit cells. 

\begin{figure}[!h]
             \centering
             \includegraphics[trim = 10 0 10 0, clip,width=0.85\textwidth]{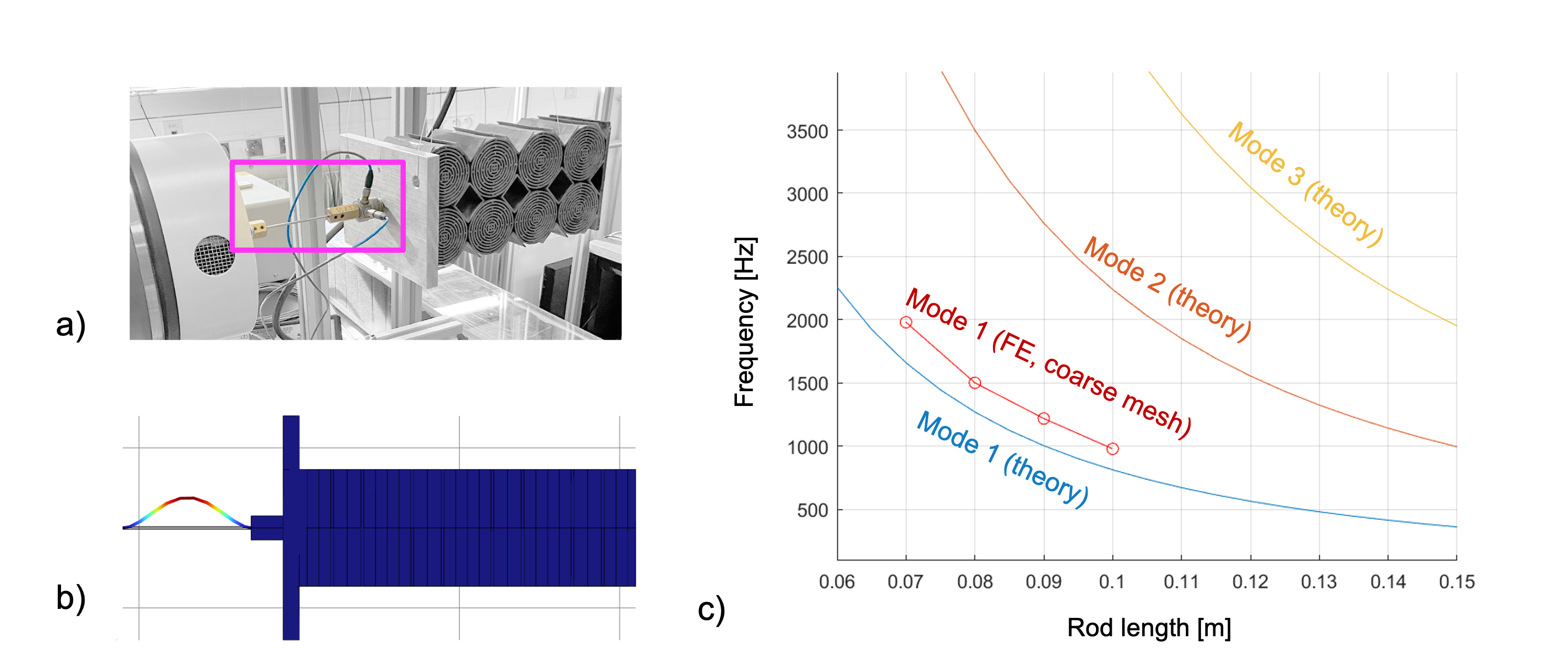}
             \caption{Analysis of the resonances frequencies of the rod connecting the shaker to the specimen in the experiment: Photograph showing the connection and highlighting the rod (a), illustration of the first resonance mode of the rod in a mechanical FE model of the specimen including the rod (b) and resonance frequencies obtained with an analytical model of the rod with clamped-clamped boundary conditions for the first 3 resonance modes and from a FE model with a coarse mesh for the first resonance mode.}
             \label{fig:accelerationAmplitude2}
         \end{figure}

We attribute this drop to the resonance of the rod connecting the shaker to the specimen (cf. Figure~\ref{fig:accelerationAmplitude2}a). When the rod vibrates, less energy is transmitted to the specimen and, hence, the accelerometer at the input side records less acceleration. We analyzed the first three resonance modes of the rod with an analytic model for a beam with clamped-clamped boundary conditions using the relation~\cite{2015_blevins}
\begin{equation}
	f_i = \frac{\lambda_i^2}{2 \pi l^2} \, \sqrt(\frac{E I}{m}) \qquad \text{with } i=1,2,3 \text{ and } \lambda_i = 4.73,7.85,10.99
\end{equation}
where $f$ is the resonance frequency, $\lambda$ is a factor defined for the $i$-th resonance mode, $l$ is the length of the rod, $E$ is the elastic modulus, $I$ is the area moment of inertia and $m=\rho A$ is the nonstructural mass per unit length where $\rho$ is the mass density of the rod and $A$ is its cross sectional area.

The rod length varied in the experiments since the connection was established manually. The results obtained for the resonance frequency with varying rod length are shown in Figure~\ref{fig:accelerationAmplitude2}c. The results obtained for the first resonance mode lie in the range in which we observe a drop of the input acceleration. An illustration of the first resonance mode is given in Figure~\ref{fig:accelerationAmplitude2}b. Results for the resonance frequency obtained from a mechanical model including the specimen and a connected rod (clamped on the left side, cf. Figure~\ref{fig:accelerationAmplitude2}b) are also shown in Figure~\ref{fig:accelerationAmplitude2}c; even though the mesh was very coarse, the numerical results are in good accordance with the analytic results and confirm that a resonance frequency of the rod lies in the considered frequency range.

\subsection{Investigations on the origin of the peak observed in the transfer functions of the \alp configuration}
\label{app:BN7} 
We performed two steps to gain insight into the vibration behavior of the \alp specimen in vicinity of the resonance peak.

In a first step, we performed eigenfrequency studies with the vibroacoustic models of configurations \alp \conS and \alp \conL. The frequencies are complex-valued, since we include structural damping in the numerical model. We only consider the real part of the frequency results since the imaginary part provides information about the damping which is not relevant for this analysis. The real part of the eigenfrequencies that lie in vicinity of the peaks observed in the transfer functions are listed in Table~\ref{tab:resFreqs}.
    
     \begin{table}[!h]
         \centering
         \caption{Eigenfrequencies obtained from the vibroacoustic FE models of the \alp configuration in vicinity of the resonance peak in the transfer function.}
         \label{tab:resFreqs}
         \begin{tabular}{cl}
             \hline
              Metamaterial structure       & Eigenfrequencies [Hz] \\
             \hline
				\alp \conS & 1844, 1851, 1895, 1896, 1913\\
				\alp \conL & 1725, 1752, 1774, 1775, 1778, 1802, 1821\\

             \hline
         \end{tabular}
     \end{table}

In a second step, we compared the displacement fields of the output plate between experiment and model. For the experiments, we used the velocity measurement values obtained on the 41 measurement points. To obtain similar results from the numerical simulation, we created evaluation points on the output plate for the FE models which are located similar to the measurement points in the experiment. 
As described in the post-processing section of the manuscript, the responses of 13 points can be obtained directly from the model (1/4 of the geometry) and the response of the remaining points is obtained from symmetry considerations.
We post-processed the results to obtain a complex-valued displacement vector $\Phi$ containing the displacement of the 41 measurement points such that
\begin{equation}
	\Phi = [u_{x1}, u_{x2}, ..., u_{x41}]^T
\end{equation}
To compare the displacement fields between two different cases $a$ and $b$, we calculate the Modal Assurance Criterion (MAC) via
\begin{equation}
MAC_{ab} = \frac{\Phi_a^T \cdot \Phi_b}{||\Phi_a|| \, ||\Phi_b||}
\end{equation}

\begin{figure}[!h]
             \centering
             \includegraphics[width=\textwidth]{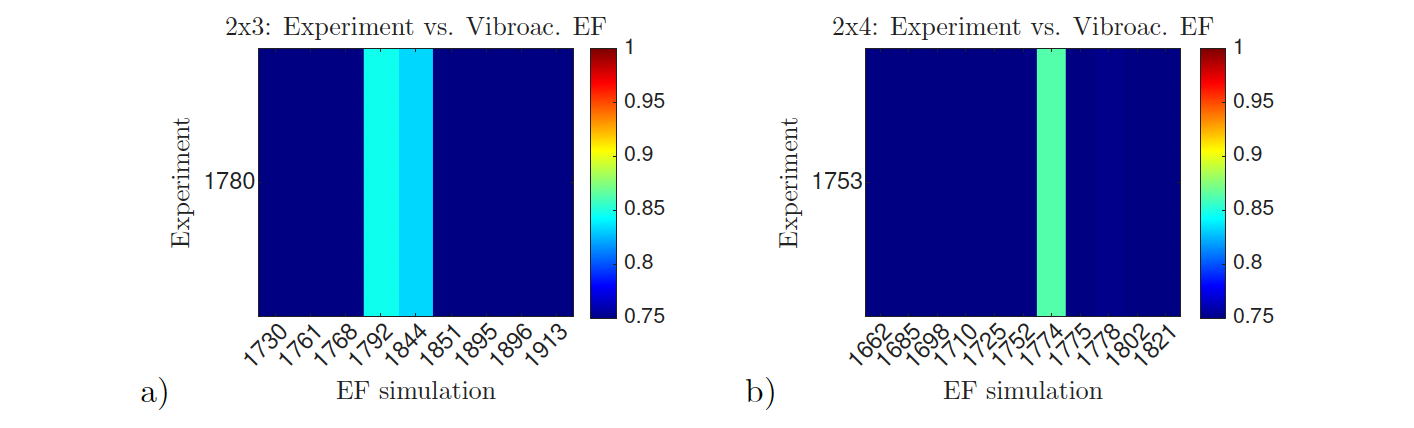}
             \caption{Modal assurance criterion coefficients obtained from the comparison of the displacement field of the output plate obtained from the experiment with the displacement field of the eigenfrequency analysis for the \alp specimens with \conS unit cells (a) and \conL unit cells (b).}
             \label{fig:MAC1}
         \end{figure}

The MAC coefficients represent the correlation between the displacement fields of cases $a$ and $b$. If the coefficient is equal to 1, the compared vectors are co-linear and the displacement fields are identical. If the coefficient is equal to 0, the compared vectors are orthogonal and the displacement fields are very different. The MAC coefficients obtained for a comparison between the displacement fields of the output plate obtained from experiment at the peak frequencies and obtained from the eigenfrequency analysis is shown for the specimens \alp \conS and \alp \conL in Figures~\ref{fig:MAC1}~a and~b respectively. The graphics show, that a good correlation of the displacement fields is obtained for the eigenfrequencies \SI{1792}{\hertz} (\conS unit cells), where $MAC = 0.85$, and \SI{1774}{\hertz} (\conL unit cells) where $MAC = 0.86$. This reinforces the evidence that we have resonance in the experiment.

\begin{figure}[!h]
             \centering
             \includegraphics[width=0.9\textwidth]{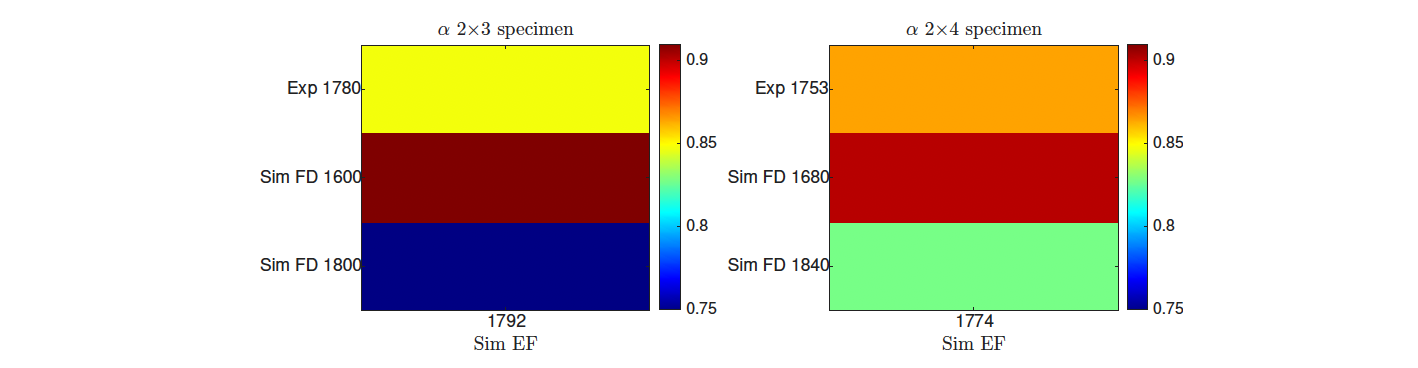}
             \caption{Modal assurance criterion coefficients obtained from the comparison of the displacement fields of the output plate obtained from the experiment (``Exp'') and the frequency domain simulation (``Sim FD'') with the displacement fields of the eigenfrequency analysis (``Sim EF'') for the \alp specimens with \conS unit cells (a) and \conL unit cells (b).}
             \label{fig:MAC2}
         \end{figure}

We also compare the displacement fields obtained at the frequencies of two resonance peaks of the vibroacoustic models in vicinity to the peak from the experiment: \SI{1600}{\hertz} and \SI{1800}{\hertz} for \conS unit cells and \SI{1680}{\hertz} and \SI{1840}{\hertz} for \conL unit cells (cf. Figure~\ref{fig:calibCompareModelExp}). 
The comparison shows similarity of the peaks for the lower frequencies: the MAC coefficients are 0.91 at \SI{1600}{\hertz} for \conS unit cells and 0.90 at \SI{1680}{\hertz} for \conL unit cells. The comparison is illustrated in Figure~\ref{fig:MAC2}. The corresponding frequencies of the resonances in the frequency  domain study are much lower than the eigenfrequencies. The frequencies for which we observe the resonances in the experiments are closer to the eigenfrequencies. This indicates that the boundary conditions in the experiments favor an excitation of the resonance. The MAC coefficients obtained from the comparison of the FE models, on the other hand, are closer to 1 than the MAC coefficient obtained from the comparison between experiment and the eigenfrequency study. The reason is that the boundary conditions in the models, in which the specimen is not mechanically constrained besides the loading, are more similar compared to the experiment, in which a non-permanent contact between the specimen and the suspension lines is present.
    
\vspace{-0.5cm}
\subsection{Bloch-Floquet analysis of different cuts}
\label{app:B4}

\begin{figure}[!h]
     \centering
     \includegraphics[trim = 110 0 60 0, clip,width=0.7\textwidth]{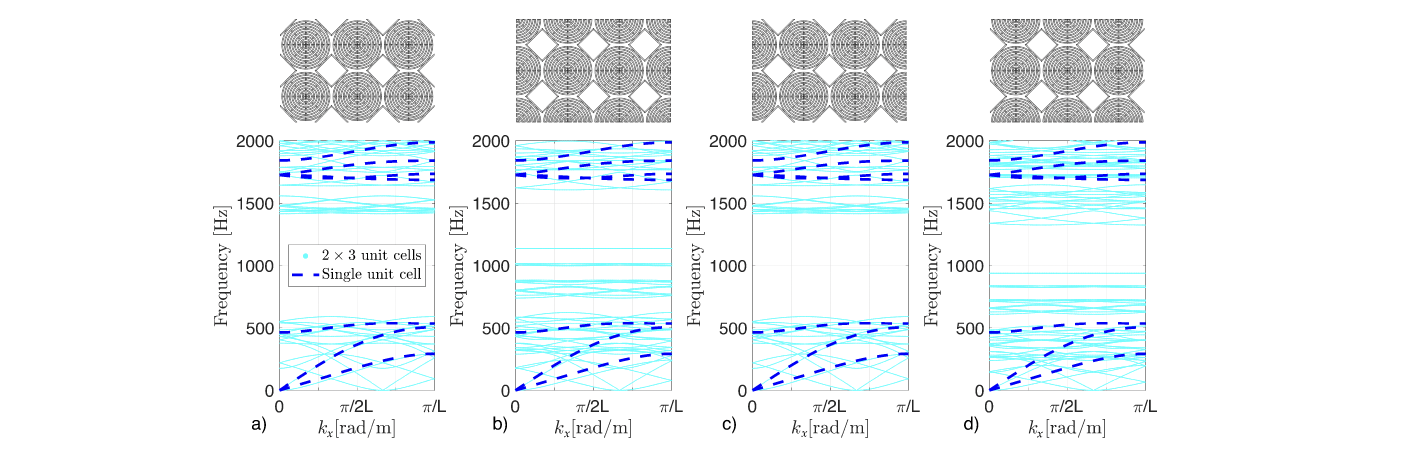}
     \caption{Dispersion curves obtained from 2D mechanical models comprising a single unit cell or $2\times3$ unit cells in configuration \alp (a), \bet (b), \gam (c) and \del (d). $k_x-$wavenumber, $L-$unit cell size}
     \label{fig:dispCurves2x3}
 \end{figure}

Numerical models of the infinitely large metamaterial based on the unit cells shown in Figure~\ref{fig:dispCurves2x3} are performed to illustrate the presence of edge modes in the configurations \bet and \del. The periodic boundary conditions are applied on the outer surfaces of the \conS arrays in these models. Since the edge modes are in-plane modes, we performed numerical simulations with 2D models for this analysis. The resulting dispersion diagrams show that there are propagating waves in a frequency range for which we have a band gap in the configurations \alp and \gam between \SI{600}{\hertz} and \SI{1000}{\hertz}. 

\vspace{-0.5cm}

\vspace{-0.5cm}
\subsection{Transfer function for different number of unit cells}
\label{app:B10}

 \begin{figure}[!h]
     \centering
     \includegraphics[width=\textwidth]{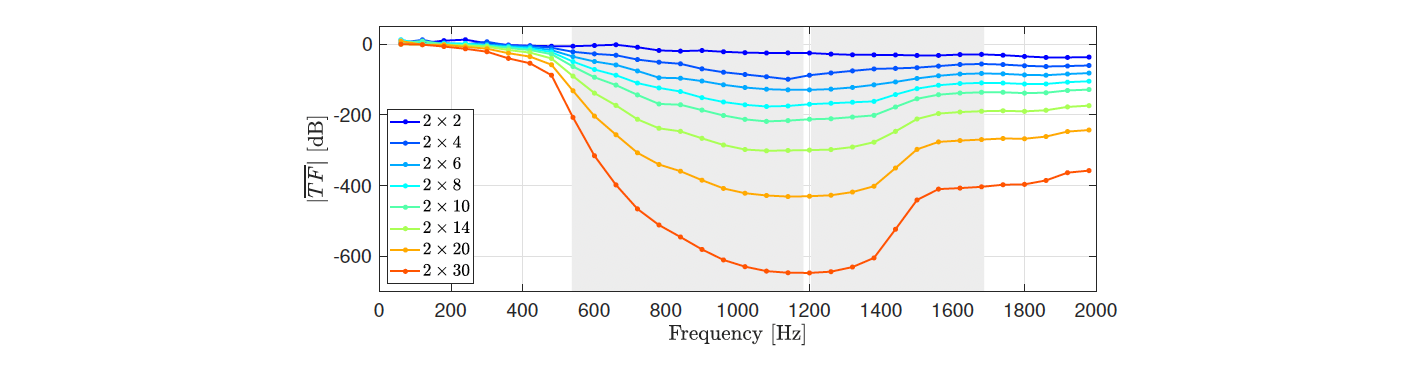}
     \caption{Average transfer function $|\overline{TF}|$ for specimens with different number of unit cells in direction of wave propagation ($e_1$-direction) obtained from the mechanical model.}
     \label{fig:apx_NcellsMec}
 \end{figure}

 \end{document}